\documentclass[]{pasj01}
\usepackage{url}
\usepackage{natbib}
\usepackage{color}
\usepackage{latexsym}
\usepackage{morefloats}
\usepackage{xspace}
\usepackage[usenames, dvipsnames]{xcolor}
\usepackage{colortbl}
\usepackage{multirow}
\usepackage{caption}
\usepackage{graphicx}
\usepackage{pdflscape}
\usepackage{bm}
\DeclareGraphicsExtensions{.pdf,.png,.eps}
\def\btheta{\boldsymbol{\theta}}

\def\etal{et~al.}
\begin{document}
\SetRunningHead{Murata et al.}{HSC CAMIRA cluster splashback radius}
\Received{}
\Accepted{}
%
\title{The splashback radius of optically selected clusters with Subaru HSC Second Public Data Release}
\author{Ryoma~\textsc{Murata}\altaffilmark{1,2}}
\author{Tomomi~\textsc{Sunayama}\altaffilmark{1}}
\author{Masamune~\textsc{Oguri}\altaffilmark{3,2,1}} 
\author{Surhud~\textsc{More}\altaffilmark{4,1}} %
\author{Atsushi~J.~\textsc{Nishizawa}\altaffilmark{5}}
\author{Takahiro~\textsc{Nishimichi}\altaffilmark{6,1}}
\author{Ken~\textsc{Osato}\altaffilmark{7}}
\altaffiltext{1}{Kavli Institute for the Physics and Mathematics of the Universe (WPI), The University of Tokyo Institutes for Advanced Study, The University of Tokyo, Kashiwa, Chiba 277-8583, Japan}
\altaffiltext{2}{Department of Physics, The University of Tokyo, Tokyo 113-0033, Japan}
\altaffiltext{3}{Research Center for the Early Universe, The University of Tokyo, Tokyo 113-0033, Japan}
\altaffiltext{4}{The Inter-University Center for Astronomy and Astrophysics, Post bag 4, Ganeshkhind, Pune, 411007, India}
\altaffiltext{5}{Institute for Advanced Research, Nagoya University Furocho, Chikusa-ku, Nagoya, 464-8602, Japan}
\altaffiltext{6}{Center for Gravitational Physics, Yukawa Institute for Theoretical Physics, Kyoto University, Kyoto 606-8502, Japan}
\altaffiltext{7}{Institut d'Astrophysique de Paris, Sorbonne Universit\'e, CNRS, UMR 7095, 75014 Paris, France}
\email{ryoma.murata@ipmu.jp}
%
\maketitle
\begin{abstract}
Recent constraints on the splashback radius around optically selected galaxy clusters 
from the redMaPPer cluster-finding algorithm in the literature
have shown that the observed splashback radius is $\sim 20\%$ smaller than that predicted by {\it N}-body simulations.
We present analyses on the splashback features around $\sim 3000$ optically selected galaxy clusters
detected by the independent cluster-finding algorithm CAMIRA
over a wide redshift range of $0.1<z_{\rm cl}<1.0$ 
from the second public data release of
the Hyper Suprime-Cam (HSC) Subaru Strategic Program
covering $\sim 427~{\rm deg}^2$ for the cluster catalog.
We detect the splashback feature from the projected cross-correlation measurements 
between the clusters and photometric galaxies
over the wide redshift range, 
including for high redshift clusters at $0.7<z_{\rm cl}<1.0$,
thanks to deep HSC images.
We find that constraints from red galaxy populations only are more precise than 
those without any color cut, leading to
$1\sigma$ precisions of $\sim 15\%$ 
at $0.4<z_{\rm cl}<0.7$ and $0.7<z_{\rm cl}<1.0$.
These constraints at $0.4<z_{\rm cl}<0.7$ and $0.7<z_{\rm cl}<1.0$ are more consistent with the model predictions ($\lesssim 1\sigma$) 
than their $20\%$ smaller values as suggested by the previous studies with the redMaPPer ($\sim 2\sigma$).
We also investigate selection effects of the optical cluster-finding algorithms
on the observed splashback features
by creating mock galaxy catalogs from a halo occupation distribution model,
and find that such effects to be sub-dominant for the CAMIRA cluster-finding algorithm.
We also find that the redMaPPer-like cluster-finding algorithm induces a smaller inferred splashback radius in our mock catalog, especially at lower richness,
which can well explain the smaller splashback radii in the literature.
In contrast, these biases are significantly reduced when increasing its aperture size.
This finding suggests that
aperture sizes of optical cluster finders that are smaller than
splashback feature scales
can induce significant biases on the inferred location of splashback radius.
\end{abstract}
\section{Introduction} \label{sec:intro}
Galaxy clusters are the most massive gravitationally bound structure in the Universe,
which form in dark matter halos around rare high peaks in the initial density field
after interactions between gravitational dynamics and baryonic process related to galaxy formation
\citep[see][for recent reviews]{Allenetal2011, Kravtsov&Borgani2012, Weinbergetal2013, Wechsler&Tinker2018, Prattetal2019, Walkeretal2019, Vogelsbergeretal2019}.
The splashback radius has been proposed as a physical boundary of dark matter halos at the outskirts 
which separates orbiting from accreting materials as a sharp density edge, which is seen 
even after stacking of halos
through numerical simulations
\citep{Diemer&Kravtsov2014, Moreetal2015} 
and 
can be explained with a simple analytical model \citep{Adhikarietal2014}.
At the splashback radius, the logarithmic derivative of density profiles is predicted to drop significantly 
over a narrow range of radius 
in the outskirts
due to the piling up of materials with small radial velocities at their first orbital apocenter after infall onto the halo.
The splashback radius primarily depends on the accretion rate, redshift, and halo mass \citep[e.g.,][]{Diemeretal2017}.
The secondary infall models with the spherical collapse model \citep{Gunn&Gott1972}
predict such a sharp density edge 
analytically \citep[e.g.,][]{Fillmore&Goldreich1984, Bertschinger1985, Adhikarietal2014, Shi2016}.
Furthermore, the splashback features have been investigated from various aspects 
with simulations
\citep{Oguri&Hamana2011, Mansfieldetal2017, Diemer2017, Okumuraetal2018, Sugiuraetal2019, Xhakajetal2019},
including dark energy \citep{Adhikarietal2018}, 
modified gravity \citep{Adhikarietal2018, Contigianietal2019b},
and self-interacting dark matter \citep{Banerjeeetal2019}.

Observationally, the splashback feature has been detected and constrained statistically
for galaxy clusters selected by the optical cluster-finding algorithm 
redMaPPer \citep{Rykoffetal2012, Rykoffetal2014, Rozoetal2014, Rozoetal2015a, Rozoetal2015b, Rykoffetal2016}
through cross-correlation measurements between clusters and galaxies 
\citep{Moreetal2016, Baxteretal2017, Changetal2018, Shinetal2019}
or weak lensing measurements 
\citep{Changetal2018} 
with high precisions thanks to a large number of optically selected clusters.
Less precise measurements have also been done for clusters selected 
by the Sunyaev–Zel’dovich effect \citep{Zurcher&More2019, Shinetal2019}
or by X-ray flux \citep{Umetsu&Diemer2017, Contigianietal2019a}.
We note that the signal-to-noise ratios for weak lensing measurements 
are smaller than those for cluster-galaxy cross-correlation measurements
to constrain the splashback feature.
It is important to 
validate theoretical predictions against observed splashback features
with the help of
weak lensing mass calibrations, 
where weak lensing measurements constrain mass-observable relation mainly from their amplitudes, 
instead of the location of the splashback radius imprinted in lensing measurements.
Interestingly, the measurements of the splashback radius with high precision for the redMaPPer clusters 
\citep{Moreetal2016, Baxteretal2017, Changetal2018}
from the Sloan Digital Sky Survey (SDSS) or the Dark Energy Survey (DES) data
show that 
their observed splashback radius from cluster-galaxy cross-correlation measurements 
is smaller than expectations from {\it N}-body simulations under the standard $\Lambda$ cold dark matter ($\Lambda{\rm CDM}$) cosmology 
at the level of $20 \% \pm 5 \%$.
\cite{Shinetal2019} also show a similar trend at $\sim 2 \sigma$ level for the redMaPPer clusters 
at more massive mass scale.

One of the possible origins of the inconsistency
is some systematics in the optical cluster-finding algorithm,
such as
projection effects in optical cluster finding algorithms 
that are misidentification of non-member galaxies along the line-of-sight direction as member galaxies in optical richness estimation
\citep[e.g.,][]{Cohnetal2007, Zuetal2017, BuschandWhite2017, Costanzietal2018, Sunayama&More2019}.
In particular, \cite{BuschandWhite2017} and \cite{Sunayama&More2019} investigated projection effects on the splashback features
with a simplified redMaPPer-like optical cluster-finding algorithm in the Millennium Simulation \citep{Springeletal2005}.
However, their cluster abundance densities in the simulations is $\sim 3$ times larger than observations as a function of richness
mainly due to the use of a mock galaxy population from a semi-analytical model,
which could lead to the overestimate of the projection effects.
Also, \cite{BuschandWhite2017} and \cite{Sunayama&More2019} did not analyze
their mock data by fitting the model profile proposed in \cite{Diemer&Kravtsov2014}
to the observed projected cross-correlation function,
thus not entirely mimicking the procedure adopted in observational
studies \citep[e.g.,][]{Moreetal2016, Changetal2018}.
Therefore, we can improve this aspect in mock simulation analyses
to investigate possible bias effects from optical cluster-finding algorithms
by more closely resembling the comparison procedure in the observations.
Thus, further independent investigation would be useful to confirm 
whether the inconsistency is caused by artifacts due to optical cluster-finding algorithms
or not.

The distribution of galaxies after magnitude or color cuts in observations
is expected to have the same splashback radius as dark matter distribution
to the first-order approximation 
unless such cuts select particular orbits of recently accreted galaxies in clusters 
due to fast quenching effects for instance.
The difference between dark matter and subhalos, where galaxies should reside,
can be caused by dynamical friction \citep{Chandrasekhar1943}
that affects subhalos with a mass ratio to host halo mass $M_{\rm sub}/M_{\rm host} \gtrsim 0.01$ \citep{Adhikarietal2016}
to 
decrease splashback radii significantly ($\gtrsim 5 \%$) compared to the dark matter.
In observations, fainter galaxies are expected to live in less massive subhalos, and thus dynamical friction effects are smaller.
Hence, fainter galaxies are expected to have splashback radii closer to those for the dark matter.
However, relations between subhalo masses and galaxy luminosities (including scatter between them) have not been fully understood.
Dynamical friction effects on splashback radii are investigated in the literature \citep{Moreetal2016, Changetal2018, Zurcher&More2019}
with different absolute magnitude thresholds.
However, a magnitude dependence of splashback radii for massive clusters has not been detected even with smaller error bars in \cite{Moreetal2016}
at a $\lesssim 5 \%$ precision level,
even though a $\gtrsim 5 \%$ shift of splashback radii 
is expected in simulations especially for more massive subhalos
when selecting subhalos with maximum circular velocities throughout their entire history to reproduce galaxy abundances in observations
\citep{Moreetal2016, Changetal2018}.
The scatter between the maximum circular velocities (or subhalo masses) and galaxy luminosities could wash out dynamical friction effects \citep{Moreetal2016}.
It would be interesting to explore dynamical friction effects with an independent optical cluster-finding algorithm and fainter galaxy samples.

One of the most striking features within galaxy clusters is the presence of a larger number 
of red and elliptical galaxies with little ongoing star formation 
\citep{Oemler1974, Dressler1980, Dressler&Gunn1983, Baloghetal1997, Poggiantietal1999},
exhibiting a tight relation in the color-magnitude diagram \citep[e.g.,][]{Stanfordetal1998}.  
These quenching effects are expected from intra-cluster astrophysical processes
including tidal disruption, harassment, strangulation \citep{Larsonetal1980}, and ram-pressure stripping \citep{Gunn&Gott1972},
or from the age-matching model \citep[e.g.,][]{Hearinetal2014} which employs expectations that galaxies in larger overdensities form earlier.
Quenching of satellite galaxies at the outskirts of galaxy clusters
has been investigated \citep[e.g.,][]{Wetzeletal2013, Fangetal2016, Zingeretal2018, Adhikarietal2019}.
The splashback radius separates orbiting from accreting galaxies as the physical boundary of halos,
and thus investigating relations between the splashback and quenching features around clusters over a wide range of redshift
would be informative to constrain galaxy formation and evolution physics.

The Hyper Suprime-Cam Subaru Strategic Program (HSC-SSP) is a wide-field optical imaging survey 
\citep{Miyazakietal2012, Miyazakietal2015, Miyazakietal2018a, Komiyamaetal2018, Furusawaetal2018, Kawanomotoetal2018}
with a 1.77 ${\rm deg}^2$ field-of-view
camera on the 8.2 m Subaru telescope. The HSC survey is unique in its combination of depth and high-resolution image quality,
which allows us to detect galaxy clusters over a wide redshift range up to $z_{\rm cl}\sim 1$ 
and to investigate their splashback features with photometric galaxies down to fainter apparent magnitudes
compared to other ongoing wide-field surveys, 
although the survey area is smaller than 
those of other imaging surveys
in the literature \citep{Moreetal2016, Changetal2018}.

In this paper, we present constraints on the splashback features of the 
HSC optically-selected clusters in the cluster redshift range of
$0.1 \leq z_{\rm cl} \leq 1.0$ and the richness range of $15 \leq N \leq 200$ detected by the independent cluster-finding algorithm CAMIRA \citep{Oguri2014} 
from the HSC-SSP second public data release \citep{Aiharaetal2019}. 
The main difference between the redMaPPer and CAMIRA cluster-finding algorithms is
that CAMIRA subtracts the background galaxy levels to estimate the richness in cluster selection locally, but redMaPPer does this globally. 
Moreover, CAMIRA employs a larger aperture size ($\simeq 1 h^{-1}{\rm Mpc}$ in physical coordinates)
independent of richness values in cluster selections, 
whereas redMaPPer uses a smaller richness-dependent aperture ($< 1 h^{-1}{\rm Mpc}$ in physical coordinates for almost all clusters).
To constrain the splashback features, we measure the projected cross-correlation function
between the HSC CAMIRA clusters and the HSC photometric galaxies. 
We investigate the dependence of splashback features for the CAMIRA clusters on 
cluster redshift, richness, galaxy magnitude limit, and galaxy color.
We compare results from a model fitting 
to the projected cross-correlation measurements 
with predictions from the halo-matter cross-correlation function
in \textsc{Dark Emulator} \citep{Nishimichietal2019}, which is constructed from a suite of high-resolution $N$-body simulations, 
and the mass-richness relation of the HSC CAMIRA clusters \citep{Murataetal2019} constrained by 
stacked weak gravitational lensing and abundance measurements with a pipeline developed in \cite{Murataetal2018}.
We also investigate how projection effects could bias splashback feature measurements 
using a simplified estimation of the impact of the optical cluster finding algorithms 
partly following approaches presented in \cite{BuschandWhite2017} and \cite{Sunayama&More2019}.
For this purpose, we construct mock galaxy catalogs by populating galaxies in 
$N$-body simulations with a halo occupation distribution \citep[HOD;][]{Zhengetal2005} model
to match cluster abundance densities and lensing profiles to observations approximately.

The structure of this paper is as follows. 
In Section~\ref{sec:data}, we describe the Subaru HSC data and catalogs used in our splashback feature analysis
and the projected cross-correlation function measurements between the HSC clusters and galaxies.
In Section~\ref{sec:model}, we summarize a model fitting method 
with a parametric density profile of \cite{Diemer&Kravtsov2014}
to constrain splashback features from the projected cross-correlation measurements.
In Section~\ref{sec:results}, we show the resulting constraints 
on the splashback features and compare them with model predictions.
We discuss the robustness of our results in Section~\ref{sec:discussion}.
Finally we conclude in Section~\ref{sec:conclusion}.

Throughout this paper, we use natural units where the speed of light is set equal to unity, $c=1$.
We use $M \equiv M_{\rm 200m}=4 \pi (R_{\rm 200m})^3 \bar{\rho}_{\rm m0} \times 200/3$ 
for the halo mass definition, where $R_{\rm 200m}$ 
is the three-dimensional comoving radius 
within which the mean mass density is $200$ times the present-day mean mass density.
In this paper, we use a radius and density in comoving coordinates rather than 
in physical coordinates unless otherwise stated.
We adopt the standard flat $\Lambda{\rm CDM}$ model as the fiducial cosmological model
with cosmological parameters from the {\it Planck15} result \citep{PlanckCollaboration2016}:
$\Omega_{\rm b0}h^2=0.02225$ and $\Omega_{\rm c0}h^2=0.1198$ for the density parameters of baryon and cold dark matter, respectively,
$\Omega_{\Lambda}=0.6844$ for the cosmological constant,
$\sigma_8=0.831$ for the normalization of the matter fluctuation,
and $n_s=0.9645$ for the spectral index.
We employ the convention in \cite{Moreetal2015, Moreetal2016} for the splashback radius, 
where $r_{\rm sp}^{\rm 3D}$ and $R_{\rm sp}^{\rm 2D}$ correspond to
the locations of steepest logarithmic derivative of the radially averaged density profiles after stacking of halos
in the three-dimensional or projected space, respectively, due to its accessibility 
in observations.
%
%
\section{Data and Measurement} \label{sec:data}
%
\subsection{HSC-SSP survey}\label{sec:data:HSCSSP}
HSC is a wide-field prime focus camera with a $1.5$ degree diameter field-of-view
mounted on the 8.2-meter Subaru telescope \citep{Miyazakietal2012, Miyazakietal2015, Miyazakietal2018a, Komiyamaetal2018, Furusawaetal2018, Kawanomotoetal2018}.
With its unique combination of a wide field-of-view, a large aperture of the primary mirror, and excellent image quality, 
HSC can detect relatively higher redshift clusters and fainter galaxies 
compared with other ongoing wide-field surveys.
To fully exploit its survey power, the Hyper Suprime-Cam Subaru Strategic Program \citep[HSC-SSP;][]{Aiharaetal2018a, Aiharaetal2018b, Aiharaetal2019}
conducts a multi-band wide-field imaging survey over six years with 300 nights.
The HSC-SSP survey consists of three layers: Wide, Deep, and UltraDeep.
The Wide layer aims at covering $1400$ ${\rm deg^2}$ of the sky with five broadbands ({\it grizy})
and a $5 \sigma$ point-source depth of ${\it r} \sim 26$.
\cite{Boschetal2018} describes software pipelines to reduce the data. 

In this paper, we use HSC cluster and galaxy catalogs based on the photometric data in the Wide layer with the five broadbands
in the HSC-SSP second Public Data Release \citep[PDR2;][]{Aiharaetal2019}. 
PDR2 includes the datasets from March 2014 through January 2018, about 174 nights in total.
The median seeing for {\it i}-band images is $\sim 0.6$ arcsec
in terms of the point-spread function (PSF) full width at half maximum.
The typical seeing, $5\sigma$ depth for extended sources, and saturation magnitude for {\it z}-band images in the Wide layer
are $0.68$ arcsec, $25.0~ {\rm mag}$, and $17.5~ {\rm mag}$ respectively, 
after averaging over the entire survey area included in PDR2 \citep{Aiharaetal2019}.

\begin{table*}[t]
  \caption{Sample selections for the HSC CAMIRA clusters and their characteristics. $^{*}$ }
  \begin{center}
    \begin{tabular*}{0.995 \textwidth}{ lccccccc } \hline\hline
      \centering
      Selection                  & Full & Low-$z$ & Mid-$z$ & High-$z$ & Low-$N$ & Mid-$N$ & High-$N$ \\ \hline
      $z_{\mathrm{cl, min} }$    & 0.1  & 0.1     & 0.4     & 0.7      & 0.1     & 0.1     & 0.1      \\
      $z_{ \mathrm{cl, max} }$   & 1.0  & 0.4     & 0.7     & 1.0      & 1.0     & 1.0     & 1.0      \\
      $\left<z_{\rm cl} \right>$ & 0.57 & 0.26    & 0.55    & 0.83     & 0.60    & 0.54    & 0.49     \\
      $N_{\mathrm{min} }$        & 15   & 15      & 15      & 15       & 15      & 20      & 30       \\
      $N_{\mathrm{max} }$        & 200  & 200     & 200     & 200      & 20      & 30      & 200      \\
      $\left< N \right>$         & 23   & 25      & 22      & 21       & 17      & 24      & 40       \\
      Number of clusters         & 3316 & 968     & 1140    & 1208     & 1774    & 1067    & 475      \\ \hline
      $\langle M_{\rm 200m} \rangle\ [h^{-1} 10^{14} M_{\odot}]$                               &  $1.72^{+0.08}_{-0.07}$  & $1.71^{+0.07}_{-0.07}$  & $1.93^{+0.09}_{-0.08}$  & $1.51^{+0.13}_{-0.13}$  & $1.23^{+0.06}_{-0.06}$  & $1.78^{+0.08}_{-0.08}$  & $3.32^{+0.14}_{-0.15}$  \\
      $\langle R_{\rm 200m} \rangle\ [h^{-1}{\rm Mpc}]$                                        &  $1.33^{+0.02}_{-0.02}$  & $1.33^{+0.02}_{-0.02}$  & $1.38^{+0.02}_{-0.02}$  & $1.27^{+0.04}_{-0.04}$  & $1.19^{+0.02}_{-0.02}$  & $1.34^{+0.02}_{-0.02}$  & $1.65^{+0.02}_{-0.03}$  \\
      $r_{\rm sp, model}^{\rm 3D}\ [h^{-1}{\rm Mpc}]$                                              &  $1.61^{+0.01}_{-0.01}$  & $1.61^{+0.01}_{-0.01}$  & $1.63^{+0.01}_{-0.01}$  & $1.45^{+0.03}_{-0.02}$  & $1.41^{+0.01}_{-0.01}$  & $1.62^{+0.01}_{-0.01}$  & $1.92^{+0.01}_{-0.01}$  \\
      $\frac{ {\rm d}\ln \xi_{\rm 3D} }{ {\rm d}\ln r }|_{r=r_{\rm sp}^{\rm 3D}, {\rm model}}$ &  $-3.20^{+0.03}_{-0.03}$ & $-3.11^{+0.03}_{-0.03}$ & $-3.25^{+0.03}_{-0.03}$ & $-3.27^{+0.05}_{-0.05}$ & $-3.21^{+0.04}_{-0.04}$ & $-3.28^{+0.03}_{-0.03}$ & $-3.34^{+0.03}_{-0.03}$ \\
      $R_{\rm sp, model}^{\rm 2D}\ [h^{-1}{\rm Mpc}]$                                              &  $1.12^{+0.01}_{-0.02}$  & $1.07^{+0.03}_{-0.02}$  & $1.16^{+0.01}_{-0.01}$  & $1.07^{+0.04}_{-0.03}$  & $0.99^{+0.01}_{-0.01}$  & $1.13^{+0.01}_{-0.02}$  & $1.39^{+0.02}_{-0.01}$  \\
      $\frac{ {\rm d}\ln \xi_{\rm 2D} }{ {\rm d}\ln R }|_{R=R_{\rm sp}^{\rm 2D}, {\rm model}}$ &  $-1.72^{+0.02}_{-0.02}$ & $-1.67^{+0.01}_{-0.01}$ & $-1.75^{+0.01}_{-0.01}$ & $-1.75^{+0.03}_{-0.03}$ & $-1.69^{+0.02}_{-0.02}$ & $-1.75^{+0.02}_{-0.02}$ & $-1.82^{+0.01}_{-0.01}$ \\ \hline
\end{tabular*}
\end{center}
\tabnote{
  $^{*}$ We define each cluster sample by $z_{\rm cl, min}$, $z_{\rm cl, max}$, $N_{\rm min}$, and $N_{\rm max}$,
and $\langle z_{\rm cl} \rangle$ and $\langle N \rangle$ give the mean values of cluster redshift and richness.
We show the constraints for each sample on mean mass and $R_{\rm 200m}$ values, and the model predictions for the splashback feature from the mass-richness relation in \cite{Murataetal2019} and the halo emulator in \cite{Nishimichietal2019} as described in Appendix~\ref{app:modelpredictions}.
We note that we use the halo-matter cross-correlation function for these model predictions and 
we compare our model predictions with model calculation methods in the literature in Appendix~\ref{app:compmodel}. 
We show the median and the 16th and 84th percentiles of the model predictions using the fiducial results of the mass-richness relation
in \cite{Murataetal2019} with the {\it Planck} cosmology.
}
\label{tab:sampleselection}
\end{table*}

\subsection{HSC CAMIRA cluster catalog} \label{sec:data:cluster}
We use the CAMIRA (Cluster finding Algorithm based on Multi-band Identification of Red-sequence gAlaxies)
cluster catalog from the HSC-SSP Wide dataset in PDR2, which was constructed using the CAMIRA algorithm \citep{Oguri2014}.
This catalog is an updated version of the one presented in \cite{Ogurietal2018a} based on the first-year HSC-SSP dataset.
The CAMIRA algorithm is a red-sequence cluster finder based on a stellar population synthesis model \citep{Bruzual&Charlot2003}
to predict colors of red-sequence galaxies at a given redshift and to compute likelihoods of being red-sequence galaxies 
as a function of redshift.
The stellar population synthesis model in the CAMIRA algorithm 
is calibrated with spectroscopic galaxies to improve its accuracy \citep{Ogurietal2018a}.
The richness in the CAMIRA algorithm corresponds to the number of red cluster member galaxies with stellar mass $M_{*}\gtrsim 10^{10.2} M_{\odot}$ (roughly corresponding to the luminosity range of $L \gtrsim 0.2 L_{*}$) within a circular aperture with a radius $R \lesssim 1\ h^{-1} {\rm Mpc}$ in physical coordinates.
Unlike the redMaPPer algorithm \citep{Rykoffetal2012}, the CAMIRA algorithm does not employ a richness-dependent aperture radius to define the richness.
The HSC images are deep enough to detect cluster member galaxies down to $M_{*}\sim 10^{10.2} M_{\odot}$ even at redshift $z_{\rm cl}\sim 1$,
which allows us to reliably detect such high redshift clusters without the richness incompleteness correction.
The algorithm employs a spatially-compensated filter 
that subtracts the background level locally in deriving the richness.
CAMIRA defines cluster centers as locations of the identified brightest cluster galaxies (BCGs).
The bias and scatter in photometric cluster redshifts of $\Delta z_{\rm cl}/(1+z_{\rm cl})$
are shown to be better than $0.005$ and $0.01$ respectively with 4$\sigma$ clipping over most of the redshift range
by using available spectroscopic redshifts of BCGs \citep{Ogurietal2018a}.
We use the cluster catalog after applying bright-star masks for {\it grizy}-bands presented in \cite{Aiharaetal2019}.

The catalog contains 3316 clusters at $0.1 \leq z_{\rm cl} \leq 1.0$ with the richness 
$15 \leq N \leq 200$ with almost uniform completeness and purity over the sky region.
We divide the whole sample into subsamples with different redshift and richness selections
to investigate redshift and richness dependences of splashback features.
The sample selections and characteristics are shown in Table~\ref{tab:sampleselection}. 
The total area for this version of the HSC CAMIRA clusters is estimated to be $427$ ${\rm deg}^2$.
In addition to the cluster catalog, we use a random cluster catalog
with the same redshift and richness distributions as the data in the same footprint.
There are 100 times as many random points as real clusters
to measure projected cross-correlation functions with galaxies accurately.

We also employ the HSC CAMIRA member galaxy catalog to define color cuts to separate red and blue galaxies at each cluster redshift,
following a red-sequence based method in \cite{Nishizawaetal2018},
which we will use to investigate galaxy color dependence of splashback features in Section~\ref{sec:redblue}.
These member galaxies are red-sequence galaxies identified by the CAMIRA algorithm
as described in \cite{Oguri2014}.
We use all the red-sequence member galaxies for clusters with richness $N \ge 10$ to define separation criteria for red and blue galaxies 
\citep[see][]{Nishizawaetal2018}. The total number of member galaxies
are $331,168$ at $0.1 \le z_{\rm cl} \le 1.0$.

We use the richness-mass relation presented in \cite{Murataetal2019} to model splashback features 
for each redshift and richness selection,
as described in Appendix~\ref{app:modelpredictions}.
\cite{Murataetal2019} constrained the richness-mass relation $P(N|M,z)$ 
from the stacked weak gravitational lensing profiles from the HSC-SSP first-year shear catalog \citep{Mandelbaumetal2018a, Mandelbaumetal2018b} 
and the photometric redshift catalog \citep{Tanakaetal2018},
and abundance measurements for the HSC CAMIRA clusters with $0.1 \leq z_{\rm cl} \leq 1.0$ and $15 \leq N \leq 200$.
We note that \cite{Murataetal2019} used the previous version of 
the HSC CAMIRA cluster catalog from the first-year dataset presented in \cite{Ogurietal2018a}.
The richness differences from the updated version for matched clusters in both the versions are negligibly small compared to the scatter values in the mass-richness relation.
\cite{Murataetal2019} also constrained the off-centering between real cluster centers and the identified BCGs for the HSC CAMIRA clusters from the stacked lensing profiles. 
In Section~\ref{sec:model}, we use this result as inputs to our prior distributions 
of off-centering parameters 
to marginalize over its effect on the projected cross-correlation measurements 
between the HSC CAMIRA clusters and photometric galaxies.

\subsection{HSC photometric galaxy catalog} \label{sec:data:galaxy}
We use the photometric galaxy catalog from PDR2 in \cite{Aiharaetal2019}
to measure projected cross-correlation functions with the clusters.
We employ $\texttt{cmodel}$ magnitudes \cite[see][for more details]{Boschetal2018} 
for total {\it z}-band magnitudes in absolute magnitude cuts \citep{Nishizawaetal2018}
since the {\it z}-band magnitude is more suitable for measurements of brightnesses of high-redshift galaxies at $z_{\rm cl}\sim 1.0$ 
than the {\it i}-band magnitude used in the literature for lower redshifts \citep[e.g.,][]{Moreetal2016}
given the wavelength of 4000${\rm \AA}$ break at such high redshifts.
We define magnitudes in the other broadbands 
such that we derive colors of individual galaxies with afterburner aperture photometries 
\citep{Boschetal2018, Ogurietal2018a, Nishizawaetal2018}
to avoid inaccurate photometries in crowded regions 
including cluster centers due to a failure of the deblender in such regions.
The HSC pipeline measures the afterburner aperture photometries after the pipeline matches the PSF sizes
on the undeblended images between all five broadbands to have accurate colors even in crowded regions.
We use the target PSF size of $1.3$ arcsec and the aperture size of $1.5$ arcsec in diameter 
($\texttt{undeblended\_convolvedflux\_3\_15}$)
for PDR2 \citep{Aiharaetal2019}.

We apply the following quality cuts to select galaxies for the measurements.
First, we select galaxies observed in all five broadbands by imposing cuts 
in the number of exposure images used to create a coadd image for each galaxy, as described below.
While the nominal definition of the full depth are four exposures for {\it gr} bands and six for {\it izy} bands
in the Wide layer of PDR2, we adopt these relaxed conditions 
to avoid gaps in the galaxy distribution due to CCD gaps.
\begin{itemize}
\item $\texttt{[gr]\_inputcount\_value}\ge 2$ 
\item $\texttt{[izy]\_inputcount\_value}\ge 4$
\end{itemize}
Second, we apply the following basic flag cuts for all five broadbands
to remove galaxies that can be affected by bad pixels or poor photometric measurements.
We also focus on unique detections only \citep[see][]{Aiharaetal2018b, Aiharaetal2019}.
\begin{itemize}
\item $\texttt{[grizy]\_pixelflags\_edge}$=\texttt{False}
\item $\texttt{[grizy]\_pixelflags\_interpolatedcenter}$=\texttt{False}
\item $\texttt{[grizy]\_pixelflags\_crcenter}$=\texttt{False}
\item $\texttt{[grizy]\_cmodel\_flag}$=\texttt{False}
\item $\texttt{[grizy]\_undeblended\_convolvedflux\_3\_15\_flag}$=\texttt{False}
\item $\texttt{isprimary}$=\texttt{True}
\end{itemize}
Third, we use galaxies with {\it z}-band \texttt{cmodel} magnitude brighter 
than $24.5~{\rm mag}$ after dust extinction corrections ($\texttt{a\_z}$),
its flux signal-to-noise ratio larger than $5$, and 
{\it i}-band star-galaxy separation parameter indicating extended sources. 
We use the {\it i}-band star-galaxy separation because {\it i}-band images are taken in better seeing conditions on average. 
We also place weak constraints on the {\it ri}-band magnitudes
as ${\it r}<28~{\rm mag}$ and ${\it i}<28~{\rm mag}$ to remove artifacts.
\begin{itemize}
\item $\texttt{z\_cmodel\_mag - a\_z}<24.5$ 
\item $\texttt{z\_cmodel\_flux} > 5\times \texttt{z\_cmodel\_fluxsigma}$ 
\item $\texttt{i\_extendedness\_value}=1$
\end{itemize}
We then apply bright-star mask cuts for all five broadbands.
\begin{itemize}
\item $\texttt{[grizy]\_mask\_s18a\_bright\_objectcenter}$=\texttt{False}
\end{itemize}
We use angular positions ($\texttt{ra}$ and $\texttt{dec}$) of the galaxies 
for the measurements.
The catalog after these quality cuts includes 48882039 galaxies 
for analyses in Sections~\ref{subsec:result:fid} and \ref{subsec:dynamical}.
Also, we apply absolute magnitude cuts for additional galaxy selections 
in each cluster redshift bin, as described in Section~\ref{sec:measure}.

We note that these quality cuts for the galaxy catalog
are the same as those used for constructing the cluster catalog from PDR2 in Section~\ref{sec:data:cluster}
except that $\texttt{z\_cmodel\_mag - a\_z}<24.0$ and $\texttt{z\_cmodel\_magsigma}<0.1$ 
are used for the galaxies to construct the cluster catalog 
instead of $\texttt{z\_cmodel\_mag - a\_z}<24.5$ and $\texttt{z\_cmodel\_flux} > 5\times \texttt{z\_cmodel\_fluxsigma}$ above. 
In the former case, the catalog includes 31719028 galaxies.
These cuts are almost the same as those used for constructing the cluster catalog from the HSC first-year dataset
presented in \cite{Ogurietal2018a}.
\cite{Nishizawaetal2018} also used similar quality cuts for the HSC first-year datasets.
We employ the catalog with these cuts for analyses with red and blue galaxies in Section~\ref{sec:redblue}
since we define red and blue galaxies by the CAMIRA member galaxy catalog 
with the same galaxy quality cuts.

We also employ a random catalog for the photometric galaxies from PDR2 with the same quality cuts
when related quantities are available for the random catalog in PDR2.  
Specifically, we do not apply 
the $\texttt{cmodel\_flag}$, $\texttt{undeblended\_convolvedflux\_3\_15\_flag}$,
$\texttt{cmodel\_mag}$, $\texttt{cmodel\_fluxsigma}$, $\texttt{cmodel\_magsigma}$, and $\texttt{extendedness\_value}$ cuts described above for the random catalog. 
The input number density of random points is $100~ {\rm points}/{\rm arcmin^2}$ when constructing the random catalog \citep{Aiharaetal2019}.
Since we use the conservative magnitude and flux signal-to-noise ratio (or magnitude error) cuts for the galaxy catalogs compared with the detection limits for the {\it z}-band described in Section~\ref{sec:data:HSCSSP}, 
the footprint of the random catalog follows the data footprint. 
We use this random galaxy catalog for measurements of projected cross-correlation functions in Section~\ref{sec:measure}.

\subsection{Projected cross-correlation measurement} \label{sec:measure}
We measure projected cross-correlation functions between the HSC CAMIRA clusters and galaxies for each cluster and galaxy selection with the Landy-Szalay estimator \citep{Landy&Szalay1993} similarly to methods in the literature \citep{Baxteretal2017, Changetal2018, Shinetal2019}.
To obtain accurate measurements of projected cross-correlation functions, we divide each sample into subsamples with small redshift intervals, measure signals in each subsample, and combine these measurements with appropriate weighting. 
Specifically, we estimate the projected cross-correlation function 
for each subsample 
defined by $z_{i} \leq z_{\rm cl} \leq z_{i+1}$ with a cluster redshift bin width $\Delta z_{\rm cl}=z_{i+1}-z_{i}=0.05$ as
\begin{eqnarray}
&&\widehat{\xi}_{ {\rm 2D}, i }(R; z_{i} \leq z_{\rm cl} \leq z_{i+1}) \nonumber \\
&=&\left.\frac{D_{\rm c} D_{\rm g} - D_{\rm c} R_{\rm g} -R_{\rm c} D_{\rm g}+ R_{\rm c} R_{\rm g}  }{ R_{\rm c} R_{\rm g} }
\right|_{R=\chi_{\rm c}|\btheta_{\rm c}-\btheta_{\rm g}|, z_{\rm cl}= z_{ {\rm cl}, i} },
\label{eq:meas_estimator}
\end{eqnarray}
where $D_{\rm c} D_{\rm g}$, $D_{\rm c} R_{\rm g}$, $R_{\rm c} D_{\rm g}$, and $R_{\rm c} R_{\rm g}$ 
are the normalized pair counts between the clusters and galaxies after selections in the data and random catalogs,
$\chi_{\rm c}$ is the comoving distance to clusters, 
and $\btheta_{\rm c}$ and $\btheta_{\rm g}$ are angular positions of the clusters and the photometric galaxies, respectively.
We employ 15 radial bins with equal spacing in logarithmic space from $0.3~h^{-1}{\rm Mpc}$ to $5.4~h^{-1}{\rm Mpc}$ in comoving coordinates.
We use a mean redshift value $z_{ {\rm cl}, i}=(z_{i}+z_{i+1})/2$
for all clusters in each subsample
to calculate the comoving distance $\chi_{\rm c}$ and galaxy selections.
We then estimate the projected cross-correlation function for each cluster and galaxy selection 
by averaging over all the subsamples
using redshift-dependent weights
from the comoving volume per unit redshift interval and unit steradian, ${\rm d}^2 V/{\rm d}z{\rm d}\Omega = \chi^2(z)/H(z)$
at $z=z_{ {\rm cl}, i}$ as
\begin{eqnarray}
&&\widehat{\xi}_{ {\rm 2D} }(R; z_{\rm cl, min} \leq z_{\rm cl} \leq z_{\rm cl, max}) \nonumber \\
&=&\frac{ \displaystyle \sum_{i} \widehat{\xi}_{ {\rm 2D}, i }(R; z_{i} \leq z_{\rm cl} \leq z_{i+1}) \frac{ \chi^2( z=z_{ {\rm cl}, i} ) }{ H( z=z_{ {\rm cl}, i} ) }  }
        { \displaystyle \sum_{i}   \frac{ \chi^2( z=z_{ {\rm cl}, i} ) }{ H( z=z_{ {\rm cl}, i} ) }  }.
\label{eq:meas_estimator_allz}
\end{eqnarray}
We employ these redshift-dependent weights
since we calculate model predictions for splashback features from the mass-richness relation and the halo emulator in Appendix~\ref{app:modelpredictions}
with the same redshift-dependent weights. 

We apply an absolute magnitude cut $M_{z}-5 \log_{10} h< -18.8$ 
for our fiducial galaxy selection in Section~\ref{subsec:result:fid}. 
This absolute magnitude limit corresponds to an apparent magnitude limit of {\it z}-${\rm mag}=24.5$ at $z_{\rm cl}=1.0$.
We use the mean redshift value $z_{ {\rm cl}, i}$ for each subsample to convert the absolute magnitude to each apparent magnitude limit. 
We employ absolute magnitude cuts in order to make fair comparisons
between results at different redshifts.
\cite{Nishizawaetal2018} employed a similar absolute magnitude cut on {\it z}-band magnitude with the first-year HSC-SSP datasets.
In Section~\ref{subsec:dynamical}, we also use several different absolute magnitude cuts to investigate 
how constraints on the splashback feature change
mainly to investigate possible dynamical friction effects.
Furthermore, in Section~\ref{sec:redblue}, we apply 
a red-sequence based red and blue separation cut in \cite{Nishizawaetal2018} for the photometric galaxies 
with a slightly brighter absolute magnitude cut $M_{z}-5 \log_{10} h< -19.3$ 
to check the dependence of splashback features on galaxy colors. 

We then estimate the covariance describing the statistical uncertainties of the projected cross-correlation measurements 
from the jackknife method \citep[e.g.,][]{Norbergetal2009}
following the literature \citep{Moreetal2016, Baxteretal2017, Changetal2018, Nishizawaetal2018, Zurcher&More2019, Shinetal2019}.
We divide the cluster footprint into $200$ rectangular subregions with almost equal areas, and we estimate the jackknife covariance by repeating the measurements by excluding each subregion for $200$ times. 
The side length for each jackknife subregion corresponds to $\sim 10~h^{-1}{\rm Mpc}$ for clusters at $z_{\rm cl}=0.1$, 
which is larger than the maximum radial distance of our cross-correlation measurements.
We calculate the signal-to-noise ratio of the measurements 
for each cluster and galaxy selection and list these values in the tables below.
For the full cluster sample with the fiducial galaxy selection, the signal-to-noise ratio is $59.2$,
whereas the signal-to-noise ratio in \cite{Moreetal2016} is $\sim 250$ with measurements from the larger survey area.

We note that the projected cross-correlation function is related
to surface galaxy density profiles $\Sigma_{g}$ after subtracting mean surface galaxy densities
$\langle \Sigma_{g} \rangle$ as
$\Sigma_{g}(R) = \langle \Sigma_{g} \rangle \xi_{\rm 2D}(R)$.
We estimate the mean surface galaxy density for each cluster and galaxy selection, 
which we also list in tables below for reference.
First, we estimate mean surface density values in cluster redshift subsamples
by employing the number of galaxies used for measurements after galaxy selections 
and survey areas in comoving coordinates from the median cluster redshift values. 
We then measure the averaged surface galaxy density profile $\widehat{\Sigma}_{ {g} }(R)$
in a manner similar to equation~(\ref{eq:meas_estimator_allz}) 
to estimate the mean surface density values for each cluster and galaxy selection as 
$\langle \Sigma_{g} \rangle={\rm Median}[\widehat{\Sigma}_{ {g} }(R)/\widehat{\xi}_{ {\rm 2D} }(R)]$ 
where we use a median value from the radial bins for simplicity.
Since the mean surface density values do not affect constraints on the splashback feature,
we employ the projected cross-correlation function for our analyses following \cite{Zurcher&More2019}.
We use the mean surface galaxy density values to estimate three-dimensional galaxy density profiles 
in Section~\ref{subsec:model}.   

\begin{table}[t]   \caption{ Model parameters and their prior distribution. $^{*}$ }
  \begin{center}
    \begin{tabular*}{0.45 \textwidth}{ ll } \hline\hline
      \centering
 Parameter &  Prior \\ \hline
 $\log_{10}{\rho_s}$ & ${\rm flat}[-3, 5]$        \\
 $\log_{10}{\alpha}$ & ${\rm Gauss}(\log_{10}(0.2), 0.6)$   \\
 $\log_{10}{r_s}$    & ${\rm flat}[\log_{10}(0.1), \log_{10}(5.0)]$      \\
 $\log_{10}{\rho_o}$ & ${\rm flat}[-5, 5]$        \\
 $S_e$               & ${\rm flat}[0.1, 10]$        \\
 $\log_{10}{r_t}$    & ${\rm flat}[\log_{10}(0.5), \log_{10}(5.0)]$      \\
 $\log_{10}{\beta}$  & ${\rm Gauss}( \log_{10}(6.0), 0.2)$   \\
 $\log_{10}{\gamma}$ & ${\rm Gauss}( \log_{10}(4.0), 0.2)$    \\
 $f_{\rm cen}$       & ${\rm Gauss}(0.65, 0.06)$ for Full and Mid-$z$\\
                     & ${\rm Gauss}(0.71, 0.08)$ for Low-$z$ \\
                     & ${\rm Gauss}(0.60, 0.09)$ for High-$z$ \\
                     & ${\rm Gauss}(0.55, 0.07)$ for Low-$N$ \\
                     & ${\rm Gauss}(0.67, 0.06)$ for Mid-$N$ \\
                     & ${\rm Gauss}(0.85, 0.06)$ for High-$N$ \\
 $R_{\rm off}$       & ${\rm Gauss}(0.56, 0.10)$ for Full, Mid-$z$, Low-$N$, \\
                     & Mid-$N$, and High-$N$ \\
                     & ${\rm Gauss}(0.40, 0.09)$ for Low-$z$ \\
                     & ${\rm Gauss}(0.59, 0.17)$ for High-$z$ \\ \hline
\end{tabular*}
\end{center}
\tabnote{
  $^{*}$
  ${\rm flat}[x,y]$ denotes a flat prior with a region between $x$ and $y$.
  ${\rm Gauss}(\mu, \sigma)$ shows a Gaussian prior
  with a mean $\mu$ and a standard deviation $\sigma$.
  The unit in $R_{\rm off}$ is $h^{-1}{\rm Mpc}$.
  In addition to the Gaussian priors above, we limit the ranges of $f_{\rm cen}$ and $R_{\rm off}$ as $0<f_{\rm cen}<1$ and $10^{-3}<R_{\rm off}<1.0$.
}
\label{tab:paramspriors}
\end{table}

\section{Model} \label{sec:model}
\subsection{Profile model for projected cross-correlation measurement}\label{subsec:model}
We model the projected cross-correlation measurements 
by employing a 
parametric model of a three-dimensional radial profile proposed in \cite{Diemer&Kravtsov2014} (hereafter DK14 profile)
that is also used in previous analyses of splashback features
\citep{Moreetal2016, Baxteretal2017, Changetal2018, Nishizawaetal2018, Zurcher&More2019, Shinetal2019}.
The DK14 profile consists of an inner Einasto profile and an outer power-law profile 
with a smooth transition function as
\begin{eqnarray}
&&\xi_{\rm 3D}(r)=\rho_{\rm in}(r)f_{\rm trans}(r) + \rho_{\rm out}(r),\\
&&\rho_{\rm in}(r)=\rho_s \exp\left( -\frac{2}{\alpha}\left[ \left(\frac{r}{r_s}\right)^{\alpha} -1 \right] \right),\\
&&\rho_{\rm out}(r)=\rho_o \left( \frac{r}{r_{\rm out}} \right)^{-S_e},\\
&&f_{\rm trans}(r)=\left( 1+\left( \frac{r}{r_t} \right)^{\beta} \right)^{-\gamma/\beta}, 
\end{eqnarray}
where $r$ denotes the three-dimensional radial distance 
from the cluster center and we fix $r_{\rm out}=1.5~h^{-1}{\rm Mpc}$.
The inner profile describes materials after passing through at least one orbital pericenter around the cluster (a multi-stream region), 
whereas the outer profile denotes materials falling toward the cluster before an orbital pericenter (a single-stream region at large scales)
and is modeled by a pure power-law function \citep{Bertschinger1985}. 
We note that nonlinear dynamics at inner radii
may modify profiles for the infalling materials and 
thus we need to calibrate a precise form for the infalling materials 
or to verify this assumption with simulations for further details 
as mentioned in \cite{Baxteretal2017}.
However, we assume this form for simplicity following the literature.   
The transition term $f_{\rm trans}$ controls the steepening of the profile
around a truncation radius $r_t$.
The DK14 model profile is flexible enough 
to reproduce simulation results for dark matter and subhalos 
as well as power-law profiles without any splashback features \citep[e.g., appendix in][]{Shinetal2019}.

We then model the projected cross-correlation function for well-centered clusters
by integrating the DK14 profile along the line-of-sight direction, following a convention in \cite{Zurcher&More2019} for a normalization as
\begin{equation} 
\xi_{\rm 2D}^{\rm cen}(R)=\frac{1}{R_{\rm max}} \int_0^{R_{\rm max}}~ {\rm d}\chi~ \xi_{\rm 3D}(r=\sqrt{R^2+\chi^2} ),
\label{eq:3Dto2D}
\end{equation}
where $R$ is the projected distance to the cluster center 
and we adopt $R_{\rm max}=40~h^{-1}{\rm Mpc}$ for the maximum projection length.
Note that three-dimensional galaxy density profiles after subtracting the mean density can be obtained as
$\rho_{g}(r)=\langle \Sigma_{g} \rangle \xi_{\rm 3D}(r)/(2 R_{\rm max})$ in this convention.
In practice, however, identified BCGs as cluster centers can be off-centered from the true halo centers 
\citep[e.g.,][]{Linetal2004, Rozoetal2014, Oguri2014, Ogurietal2018a, Zhangetal2019}.
Following the method in the literature \citep{Baxteretal2017, Changetal2018, Zurcher&More2019, Shinetal2019},
we include the effect of off-centered clusters on the projected cross-correlation measurements as 
\begin{equation}
\xi_{\rm 2D}(R; f_{\rm cen}, R_{\rm off})=f_{\rm cen} \xi_{\rm 2D}^{\rm cen}(R) + (1-f_{\rm cen}) \xi_{\rm 2D}^{\rm off}(R; R_{\rm off}),
\end{equation}
where $f_{\rm cen}$ is a parameter to model the fraction of centered clusters, 
$(1-f_{\rm cen})$ is the fraction of off-centered clusters, and $R_{\rm off}$ is a typical width for an offset distribution
in equation~(\ref{eq:Poffdist}) shown below.
The projected cross-correlation function for off-centered clusters is given 
by averaging over an offset distribution $P_{\rm off}(\tilde{R}; R_{\rm off})$ as
\begin{equation}
\xi_{\rm 2D}^{\rm off}(R; R_{\rm off})=\int_0^{\infty} {\rm d}\tilde{R}~ P_{\rm off}(\tilde{R}; R_{\rm off}) \xi_{\rm 2D}^{\rm off}(R|\tilde{R}),
\end{equation}
where $\xi_{\rm 2D}^{\rm off}(R|\tilde{R})$ is the projected cross-correlation function with a fixed offset value $\tilde{R}$ below.
For the offset distribution, we assume a Rayleigh distribution
for the normalized one-dimensional radial distribution for detected centers
with respect to true halo centers in the projected two-dimensional space
as
\begin{equation}
P_{\rm off}(\tilde{R}; R_{\rm off})=\frac{\tilde{R}}{R^{2}_{\rm off}} \exp\left( - \frac{ \tilde{R}^2 }{ 2 R_{\rm off}^2 }\right).
\label{eq:Poffdist}
\end{equation}
We calculate the projected cross-correlation function with an offset value $\tilde{R}$ 
by azimuthally averaging over the projected cross-correlation function without the off-centering effects as
\begin{equation}
 \xi_{\rm 2D}^{\rm off}(R|\tilde{R})=\int_0^{2 \pi} \frac{ {\rm d} \theta}{2 \pi} \xi_{\rm 2D}^{\rm cen}(\sqrt{R^2+\tilde{R}^2+2R \tilde{R} \cos{\theta} } ).
\end{equation}
We note that we apply a modification for the projected cross-correlation model without the off-centering effects 
such that $\xi_{\rm 2D}^{\rm cen}(R \leq 0.1\ h^{-1}{\rm Mpc})=\xi_{\rm 2D}^{\rm cen}(R =0.1\ h^{-1}{\rm Mpc})$ 
to avoid numerical instability 
especially when the outer power-law profile has very large amplitudes at inner radii. 
This modification effectively alters the DK14 profile only at $r<0.1 h^{-1}{\rm Mpc}$ and therefore is not important 
for our analyses of splashback features.

\begin{table*}[t]
   \caption{ Model parameter constraints and derived constraints on splashback features with the fiducial absolute magnitude cut. $^{*}$
            }
  \begin{center}
    \begin{tabular*}{0.99 \textwidth}{ lccccccccccccc } \hline\hline
      \centering
  Parameter                    & Full & Low-$z$ & Mid-$z$ & High-$z$ & Low-$N$ & Mid-$N$ & High-$N$                  \\ \hline
  $\log_{10}\rho_s$ & $1.97^{+0.58}_{-0.43}$   &  $1.31^{+0.63}_{-0.91}$   &  $1.83^{+0.68}_{-0.72}$   &  $2.10^{+0.56}_{-0.70}$   &  $2.13^{+0.55}_{-0.70}$   &  $1.28^{+0.64}_{-0.92}$   &  $1.87^{+0.71}_{-0.69}$   \\
  $\log_{10}\alpha$ & $-0.68^{+0.31}_{-0.36}$   &  $-0.75^{+0.54}_{-0.56}$   &  $-0.80^{+0.39}_{-0.50}$   &  $-0.73^{+0.30}_{-0.49}$   &  $-0.62^{+0.26}_{-0.46}$   &  $-0.71^{+0.52}_{-0.57}$   &  $-1.02^{+0.41}_{-0.46}$   \\
  $\log_{10}r_s$    & $-0.58^{+0.20}_{-0.27}$   &  $-0.21^{+0.58}_{-0.32}$   &  $-0.48^{+0.36}_{-0.32}$   &  $-0.64^{+0.32}_{-0.24}$   &  $-0.66^{+0.31}_{-0.23}$   &  $-0.22^{+0.60}_{-0.33}$   &  $-0.43^{+0.33}_{-0.34}$   \\
  $\log_{10}\rho_o$ & $-0.48^{+0.14}_{-0.21}$   &  $-0.21^{+0.15}_{-0.24}$   &  $-0.75^{+0.31}_{-0.44}$   &  $-0.63^{+0.28}_{-0.47}$   &  $-0.44^{+0.17}_{-0.26}$   &  $-0.42^{+0.21}_{-0.36}$   &  $-0.74^{+0.44}_{-0.67}$   \\
  $S_e$             & $1.30^{+0.21}_{-0.26}$   &  $1.40^{+0.22}_{-0.29}$   &  $0.98^{+0.40}_{-0.48}$   &  $1.25^{+0.40}_{-0.54}$   &  $1.38^{+0.25}_{-0.33}$   &  $1.25^{+0.29}_{-0.43}$   &  $1.17^{+0.77}_{-0.64}$   \\
  $\log_{10}r_t$    & $0.16^{+0.11}_{-0.09}$   &  $0.02^{+0.18}_{-0.15}$   &  $0.13^{+0.16}_{-0.15}$   &  $0.16^{+0.20}_{-0.18}$   &  $0.16^{+0.29}_{-0.26}$   &  $0.01^{+0.21}_{-0.17}$   &  $0.13^{+0.07}_{-0.06}$   \\
  $\log_{10}\beta$  & $0.77^{+0.22}_{-0.23}$   &  $0.72^{+0.19}_{-0.18}$   &  $0.70^{+0.21}_{-0.19}$   &  $0.78^{+0.22}_{-0.23}$   &  $0.73^{+0.22}_{-0.22}$   &  $0.72^{+0.20}_{-0.18}$   &  $0.81^{+0.19}_{-0.18}$   \\
  $\log_{10}\gamma$ & $0.64^{+0.19}_{-0.20}$   &  $0.56^{+0.20}_{-0.19}$   &  $0.58^{+0.20}_{-0.19}$   &  $0.61^{+0.20}_{-0.20}$   &  $0.58^{+0.20}_{-0.20}$   &  $0.54^{+0.20}_{-0.18}$   &  $0.67^{+0.18}_{-0.18}$   \\
  $f_{\rm cen}$     & $0.65^{+0.06}_{-0.06}$   &  $0.69^{+0.08}_{-0.08}$   &  $0.64^{+0.06}_{-0.06}$   &  $0.59^{+0.11}_{-0.12}$   &  $0.53^{+0.08}_{-0.08}$   &  $0.66^{+0.06}_{-0.06}$   &  $0.85^{+0.06}_{-0.06}$   \\
  $R_{\rm off}$     & $0.56^{+0.07}_{-0.08}$   &  $0.40^{+0.09}_{-0.09}$   &  $0.56^{+0.08}_{-0.09}$   &  $0.61^{+0.08}_{-0.10}$   &  $0.59^{+0.06}_{-0.06}$   &  $0.53^{+0.09}_{-0.09}$   &  $0.55^{+0.10}_{-0.10}$   \\ \hline
  $r_{\rm sp}^{\rm 3D}\ [h^{-1}{\rm Mpc}]$ & $1.51^{+0.19}_{-0.18}$   &  $1.24^{+0.21}_{-0.19}$   &  $1.58^{+0.28}_{-0.25}$   &  $1.52^{+0.31}_{-0.45}$   &  $1.23^{+0.35}_{-0.33}$   &  $1.26^{+0.25}_{-0.21}$   &  $1.68^{+0.25}_{-0.19}$   \\
  $\frac{ {\rm d}\ln \xi_{\rm 3D} }{ {\rm d}\ln r }|_{r=r_{\rm sp}^{\rm 3D}}$ &$-4.04^{+0.44}_{-0.65}$   &  $-3.69^{+0.34}_{-0.46}$   &  $-4.29^{+0.55}_{-0.75}$   &  $-4.12^{+0.63}_{-0.93}$   &  $-3.59^{+0.40}_{-0.58}$   &  $-3.89^{+0.41}_{-0.56}$   &  $-5.18^{+0.79}_{-1.13}$   \\
  $\frac{ {\rm d}\ln (\rho_{\rm in} f_{\rm trans}) }{ {\rm d}\ln r }|_{r=r_{\rm sp}^{\rm 3D}}$  & $-5.51^{+0.95}_{-1.25}$   &  $-4.83^{+0.73}_{-0.92}$   &  $-5.28^{+0.85}_{-1.10}$   &  $-5.32^{+1.04}_{-1.38}$   &  $-4.70^{+0.80}_{-1.14}$   &  $-4.89^{+0.79}_{-1.01}$   &  $-6.24^{+1.04}_{-1.44}$   \\
  $\frac{ \rho_{\rm in}f_{\rm trans} }{ \rho_{\rm in}f_{\rm trans}+\rho_{\rm out} }|_{ r=r_{\rm sp}^{\rm 3D} }$ & $0.66^{+0.11}_{-0.10}$ & $0.67^{+0.13}_{-0.11}$ & $0.78^{+0.10}_{-0.12}$ & $0.73^{+0.12}_{-0.15}$ &  $0.68^{+0.12}_{-0.13}$ & $0.74^{+0.13}_{-0.13}$ & $0.81^{+0.10}_{-0.15}$\\
  $R_{\rm sp}^{\rm 2D}\ [h^{-1}{\rm Mpc}]$ & $1.15^{+0.17}_{-0.18}$   &  $0.95^{+0.14}_{-0.12}$   &  $1.13^{+0.17}_{-0.15}$   &  $1.12^{+0.26}_{-0.36}$   &  $0.83^{+0.26}_{-0.19}$   &  $0.95^{+0.14}_{-0.13}$   &  $1.36^{+0.14}_{-0.12}$   \\
  $\frac{ {\rm d}\ln \xi_{\rm 2D} }{ {\rm d}\ln R }|_{R=R_{\rm sp}^{\rm 2D}}$ & $-2.10^{+0.17}_{-0.22}$   &  $-2.01^{+0.17}_{-0.21}$   &  $-2.28^{+0.25}_{-0.32}$   &  $-2.24^{+0.27}_{-0.40}$   &  $-1.97^{+0.20}_{-0.29}$   &  $-2.14^{+0.21}_{-0.26}$   &  $-2.97^{+0.42}_{-0.63}$   \\ \hline
  $S/N$ & $59.2$   &    $45.4$   &    $39.5$   &    $44.8$   &    $44.4$   &    $39.2$   &    $30.2$   \\
   $\langle \Sigma_{g} \rangle\ [h^{2}{\rm Mpc}^{-2}]$   &  76.8     &       60.8     &      73.4     &     84.0       &     77.2       &     77.3       &      75.4     \\
  $\chi_{\rm min}^2/{\rm dof}$                         & $11.5/10$   &    $19.6/10$   &    $6.3/10$   &    $19.2/10$   &    $11.3/10$   &    $15.6/10$   &    $8.9/10$   \\ \hline
\end{tabular*}
\end{center}
\tabnote{
  $^{*}$ 
  We show the median and the 16th and 84th percentiles of the posterior distribution of the model parameters for each cluster sample 
  in Table~\ref{tab:sampleselection} with the fiducial absolute magnitude cut $M_{z}-5 \log_{10} h< -18.8$.
  We present the derived constraints from MCMC chains on splashback features for each cluster sample. 
  $S/N$ denotes the signal-to-noise ratio within the radial bins for the fitting analyses,
  and $\langle \Sigma_{g} \rangle$ value is the mean galaxy surface density 
  of galaxies after the galaxy selection for each cluster sample
  (see Section~\ref{sec:measure} for more details).
  We also show the minimum chi-square ($\chi_{\rm min}^{2}$) with the number of degrees of freedom (dof) in the bottom row.
  Since the model parameters with informative Gaussian priors ($\alpha$, $\beta$, $\gamma$, $f_{\rm cen}$, and $R_{\rm off}$) 
  are determined strongly by the priors in Table~\ref{tab:paramspriors}, we do not include them as free parameters when calculating dof
  (i.e., dof=10=15-5, where 15 is the total number of data points and 5 is the total number of model parameters without informative priors).
  In Appendix~\ref{appendix:MCMCcontour} we show correlations among the parameters for the Full sample. 
}
\label{tab:mainresults}
\end{table*}

We employ the constraints on the off-centering parameters 
$f_{\rm cen}$ and $R_{\rm off}$ in \cite{Murataetal2019} from stacked lensing profiles around the HSC CAMIRA clusters as the prior distributions in our splashback analysis. 
Note that \cite{Murataetal2019} 
assumed the same offset distribution 
as given by equation~(\ref{eq:Poffdist}) when expressed in the projected two-dimensional space.
\cite{Murataetal2019} used 
an empirical parametric model for $f_{\rm cen}$  with mean richness and redshift dependences 
in each cluster sample, and one $R_{\rm off}$ parameter
in each redshift bin with the same redshift binning 
as our Low-$z$, Mid-$z$, and High-$z$ samples.
For the prior distribution in our analyses, 
we employ the posterior distribution in \cite{Murataetal2019} 
to calculate the constraints of $f_{\rm cen}$ 
from the mean richness and redshift values in Table~\ref{tab:sampleselection},
and those of $R_{\rm off}$ using the redshift bin in \cite{Murataetal2019} 
with the closest mean redshift value 
for each cluster selection.
Gaussian distributions can well approximate the constraints, and we derive their mean and standard deviation values for each cluster sample.
We show these values in Table~\ref{tab:paramspriors}. 
We ignore a weak correlation between the constraints of $f_{\rm cen}$ and $R_{\rm off}$ in \cite{Murataetal2019} for simplicity.

To summarize, our model of cross-correlation signals contains ten parameters, including two parameters for off-centering effects.
The model parameters and their prior distributions are given in Table~\ref{tab:paramspriors}.
The literature used Gaussian distributions for the prior distributions of $\log_{10}\alpha$, $\log_{10}\beta$, 
and $\log_{10}\gamma$,  
which were determined conservatively compared to the simulation results in \cite{Gaoetal2008} and \cite{Diemer&Kravtsov2014}.
For the Gaussian priors of $\log_{10}\alpha$, $\log_{10}\beta$, and $\log_{10}\gamma$, 
we follow moderate assumptions made by \cite{Moreetal2016}\footnote{ \cite{Moreetal2016} has a typographic error in the priors for $\log_{\
10}\beta$ and $\log_{10}\gamma$. The priors used in \cite{Moreetal2016} for $\log_{10}\beta$ and $\log_{10}\gamma$ are the same as those in \cite{Shinetal2019} and ours.} and \cite{Shinetal2019}. We use flat priors for the other model parameters in the DK14 profile
with conservatively large widths as shown in Table~\ref{tab:paramspriors}.
We employ a constant value for the mean of $\log_{10}\alpha$ parameter in Table~\ref{tab:paramspriors} for simplicity ignoring mass and redshift dependence
in simulation calibrations for dark matter in \cite{Gaoetal2008}.
We have confirmed that our results do not change significantly ($\lesssim 1{\rm -}2 \%$ for splashback locations and derivatives at these locations) 
even when using mean values for $\log_{10}\alpha$ 
from \cite{Gaoetal2008} with mean mass and redshift values 
for each sample with a fixed standard deviation.

\begin{figure*}
  \begin{center}
    {
      \includegraphics[width=4.2cm]{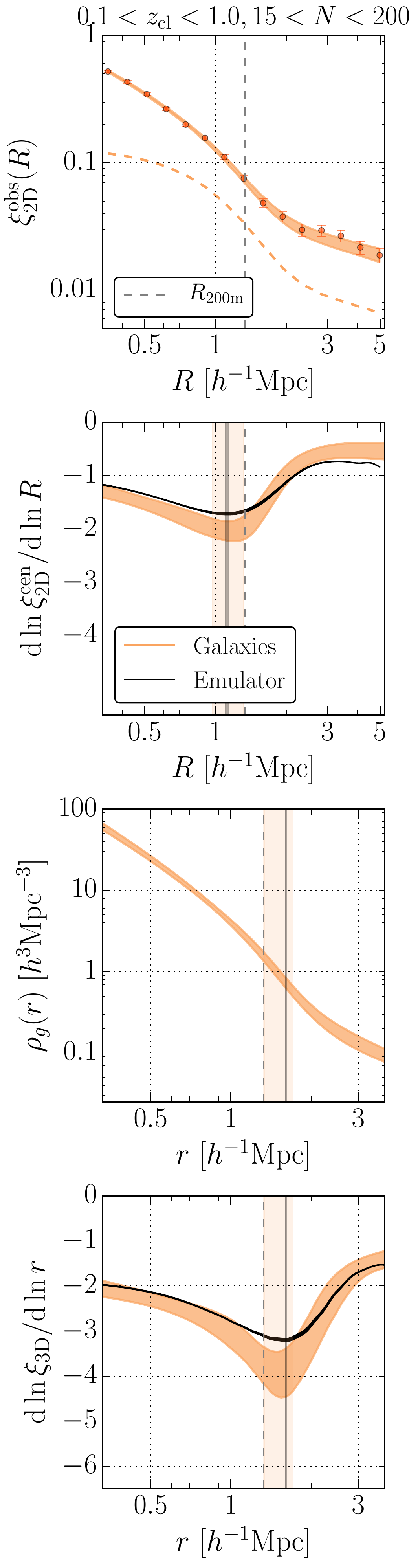}
      \includegraphics[width=4.2cm]{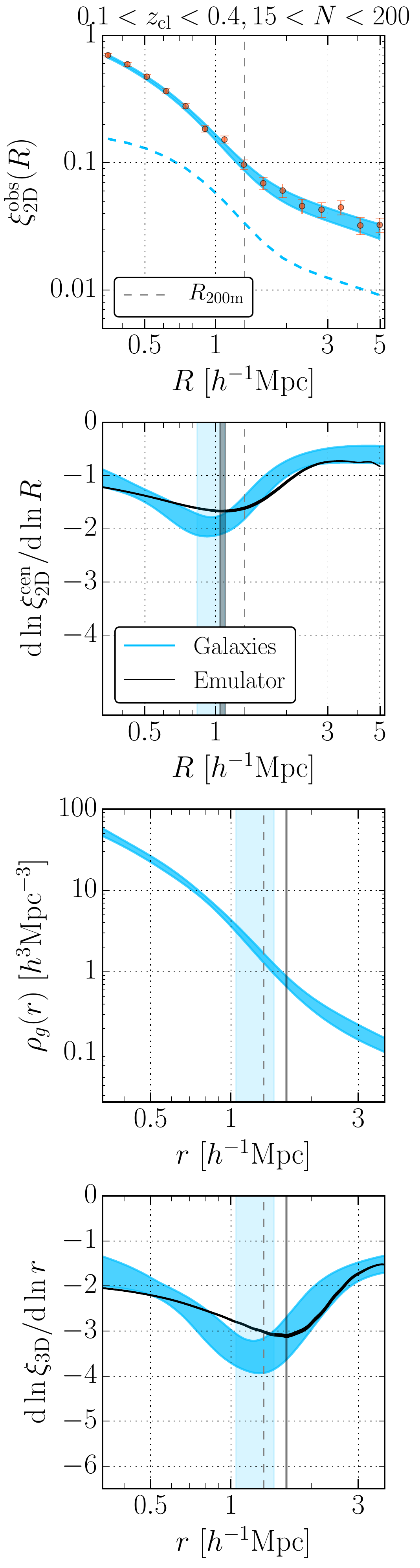}
      \includegraphics[width=4.2cm]{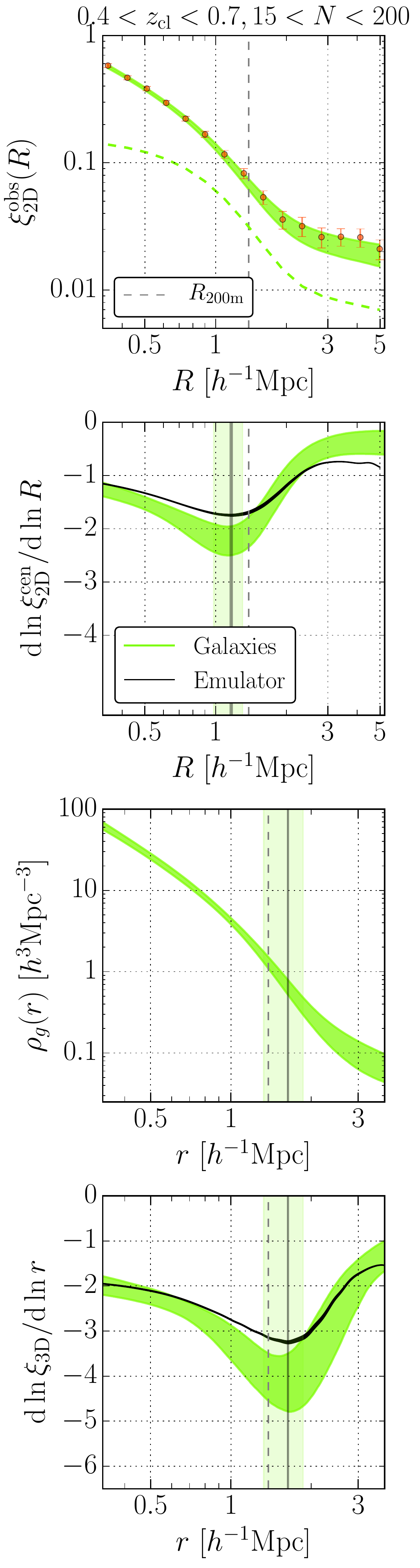}
      \includegraphics[width=4.2cm]{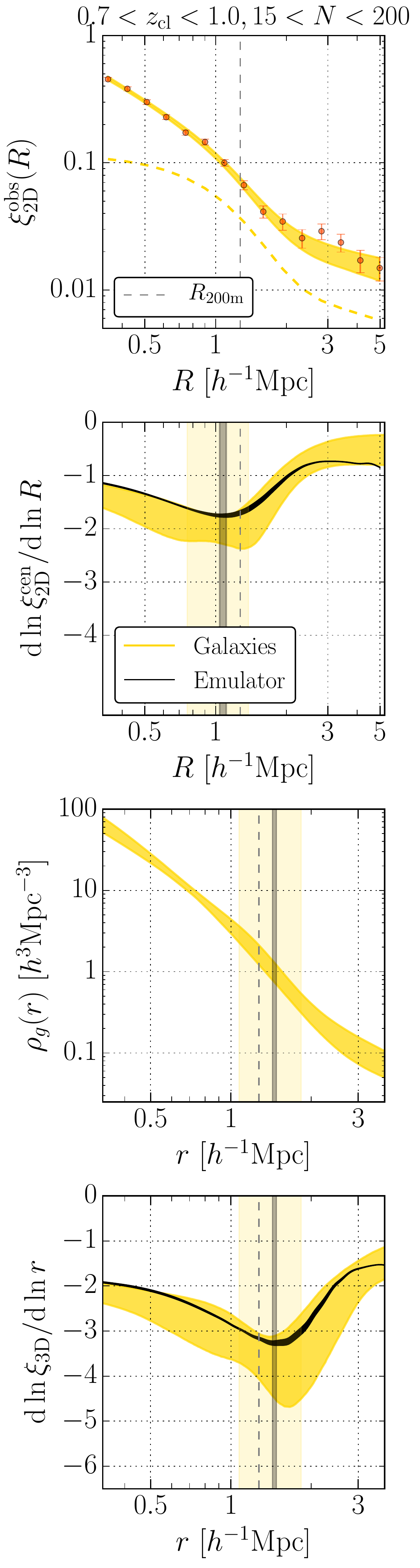}
    }
  \end{center}
  \caption{
    In the top row, we show observed projected cross-correlation measurements
    between the clusters and galaxies (points with error bars in red color) from left to right,
    for the Full, Low-$z$, Mid-$z$, and High-$z$ cluster samples from Table~\ref{tab:mainresults}
    with the fiducial absolute magnitude cut $M_{z}-5 \log_{10} h< -18.8$ for galaxy selections.
    Shaded colored regions show the 16th and 84th percentiles of the model predictions from the MCMC chains for each cluster sample.
    The dashed profile curves in the top row show contributions from off-centered clusters to projected cross-correlation measurements at the best-fit model parameters.
    The second row shows constraints on 
    the logarithmic derivative of the projected cross-correlation profiles without off-centering effects in equation~(\ref{eq:3Dto2D}).
    The third row presents constraints on the three-dimensional galaxy density profile calculated as $\rho_g(r)=\langle \Sigma_g \rangle \xi_{\rm 3D}(r)/(2 R_{\rm max})$
    (see Section~\ref{subsec:model} for more details).
    The last row shows the constraints on the logarithmic derivative of the three-dimensional profiles.
    The vertical black dashed lines denote mean values of $R_{\rm 200m}$ shown in Table~\ref{tab:sampleselection} for the cluster samples.
    Black vertical shaded regions show model predictions from the halo-matter cross-correlation (i.e., dark matter distributions around clusters)
    and mass-richness relation in Appendix~\ref{app:modelpredictions}
    for $R_{\rm sp}^{\rm 2D}$ in the second row and $r_{\rm sp}^{\rm 3D}$ in the third and last rows from Table~\ref{tab:sampleselection}, 
    whereas colored vertical shaded regions show their constraints from the MCMC chains in each cluster sample.
    Black shaded profile regions in the second and last rows show derivative profiles from the model predictions for each cluster sample.
  }
\label{fig:mainresults_full_z}
\end{figure*}

\begin{figure*}
  \begin{center}
    {
      \includegraphics[width=4.2cm]{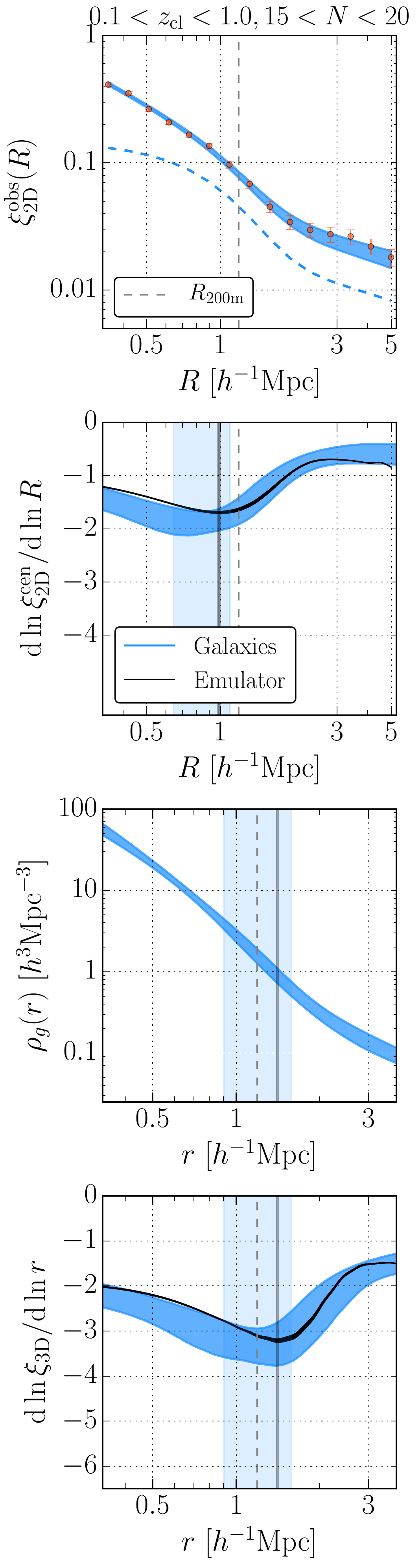}
      \includegraphics[width=4.2cm]{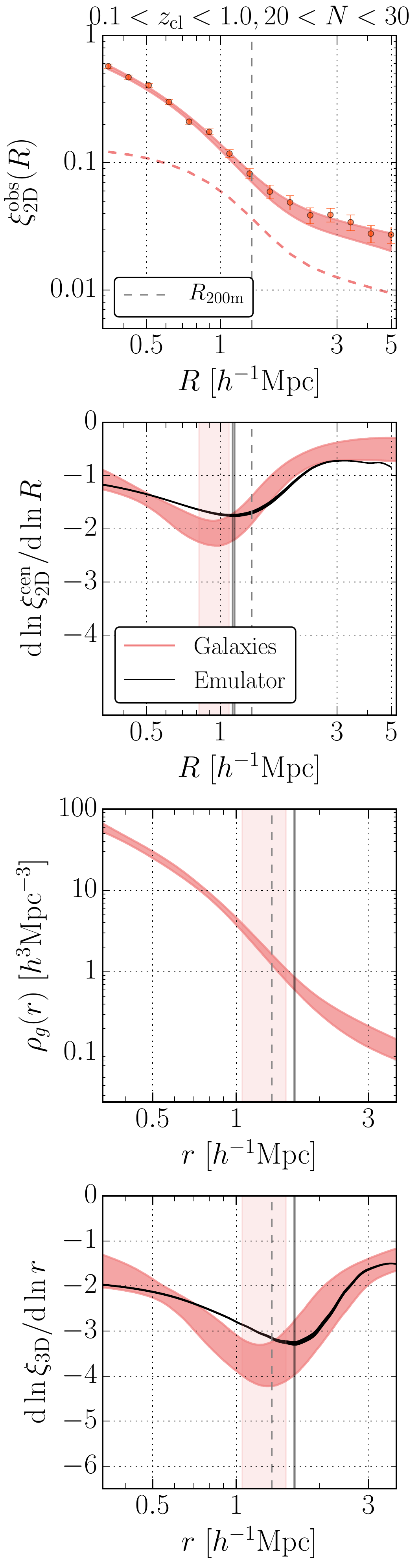}
      \includegraphics[width=4.2cm]{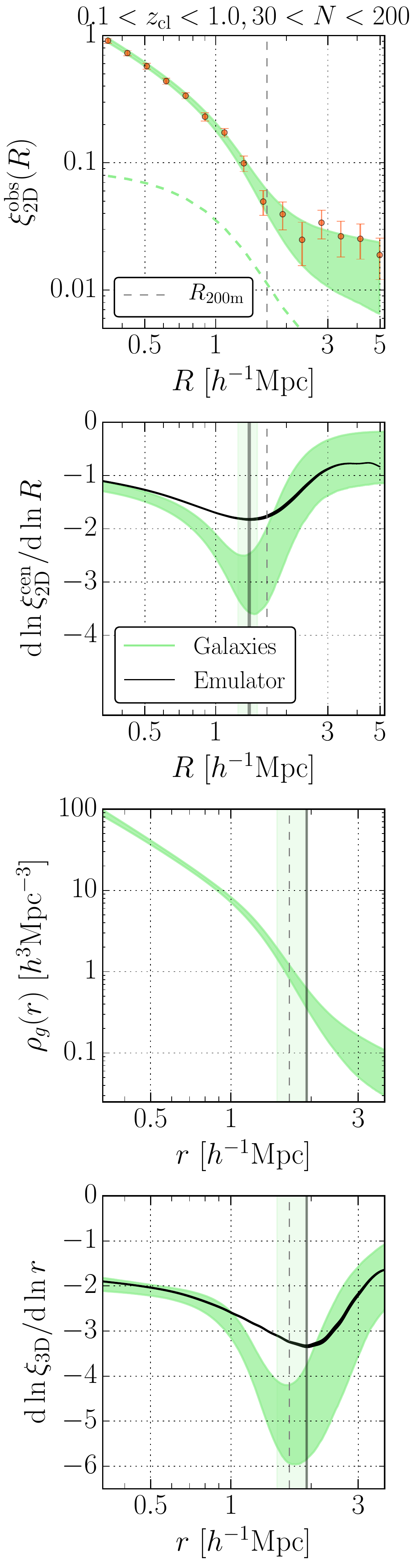}
    }
  \end{center}
  \caption{
    Same as Figure~\ref{fig:mainresults_full_z}, but for the Low-$N$, Mid-$N$, and High-$N$ cluster samples from Table~\ref{tab:mainresults}.
  }
\label{fig:mainresults_N}
\end{figure*}

\subsection{Model fitting}
We constrain the model parameters by comparing the model predictions
with the projected cross-correlation measurements.
We perform Bayesian parameter estimation assuming a Gaussian form for the likelihood,
$\mathcal{L}\propto |{\bf C}|^{-1/2} \exp(-\chi^2/2)$, with
\begin{equation}
 \chi^2 =
 \sum_{i, j}~ \Bigl[ {\bf D} - {\bf D}^{\rm model}\Bigr]_i\left({\bf C}^{-1}\right)_{ij}
    \Bigl[{\bf D} - {\bf D}^{\rm model} \Bigr]_j,
\end{equation}
where ${\bf D}$ is a data vector of the projected cross-correlation function
in the radial bins, ${\bf D}^{\rm model}$ is the model prediction,
and ${\bf C}^{-1}$ is the inverse of the jackknife covariance matrix.
We use the 15 radial bins for each sample selection.
The indices $i$ and $j$ run from $1$ to $15$.
We perform the parameter estimation with the affine-invariant Markov chain Monte Carlo (hereafter MCMC) sampler of \cite{Goodman&Weare2010}
as implemented in the python package \textsc{emcee} \citep{Foreman-Mackeyetal2013}.

\begin{table*}[t]
   \caption{ Same as Table~\ref{tab:mainresults}, but for the Low-$z$ cluster sample with different absolute magnitude cuts for galaxy selections. $^{*}$
            }
  \begin{center}
    \begin{tabular*}{1.0 \textwidth}{ lcccc } \hline\hline
      \centering
  Parameter                    &Low-$z$ & Low-$z$ & Low-$z$ & Low-$z$ \\ 
                               &$M_z-5\log_{10}h<-20.8$&$M_z-5\log_{10}h<-19.8$&$M_z-5\log_{10}h<-17.8$&$M_z-5\log_{10}h<-16.8$\\ \hline
  $\log_{10}\rho_s$            & $1.58^{+0.66}_{-0.92}$   &  $1.90^{+0.73}_{-0.86}$   &  $1.19^{+0.75}_{-0.97}$   &  $1.27^{+0.77}_{-1.22}$   \\ 
  $\log_{10}\alpha$            & $-0.82^{+0.57}_{-0.55}$   &  $-0.82^{+0.45}_{-0.52}$   &  $-0.80^{+0.50}_{-0.55}$   &  $-0.79^{+0.42}_{-0.53}$   \\ 
  $\log_{10}r_s$               & $-0.21^{+0.52}_{-0.33}$   &  $-0.43^{+0.44}_{-0.34}$   &  $-0.28^{+0.59}_{-0.36}$   &  $-0.44^{+0.54}_{-0.35}$   \\ 
  $\log_{10}\rho_o$            & $-0.06^{+0.18}_{-0.27}$   &  $-0.09^{+0.15}_{-0.22}$   &  $-0.26^{+0.16}_{-0.26}$   &  $-0.43^{+0.24}_{-0.42}$   \\ 
  $S_e$                        & $1.28^{+0.26}_{-0.33}$   &  $1.36^{+0.22}_{-0.27}$   &  $1.60^{+0.27}_{-0.35}$   &  $1.59^{+0.47}_{-0.53}$   \\ 
  $\log_{10}r_t$               & $0.08^{+0.12}_{-0.12}$   &  $0.10^{+0.14}_{-0.13}$   &  $0.01^{+0.21}_{-0.15}$   &  $0.12^{+0.29}_{-0.22}$   \\ 
  $\log_{10}\beta$             & $0.76^{+0.20}_{-0.19}$   &  $0.73^{+0.20}_{-0.19}$   &  $0.72^{+0.20}_{-0.19}$   &  $0.72^{+0.21}_{-0.20}$   \\ 
  $\log_{10}\gamma$            & $0.63^{+0.20}_{-0.20}$   &  $0.60^{+0.20}_{-0.20}$   &  $0.56^{+0.20}_{-0.19}$   &  $0.56^{+0.21}_{-0.20}$   \\ 
  $f_{\rm cen}$                & $0.69^{+0.08}_{-0.08}$   &  $0.70^{+0.08}_{-0.08}$   &  $0.69^{+0.08}_{-0.08}$   &  $0.70^{+0.08}_{-0.08}$   \\ 
  $R_{\rm off}$                & $0.39^{+0.08}_{-0.09}$   &  $0.40^{+0.09}_{-0.10}$   &  $0.39^{+0.09}_{-0.09}$   &  $0.39^{+0.09}_{-0.09}$   \\ \hline
  $r_{\rm sp}^{\rm 3D}\ [h^{-1}{\rm Mpc}]$        & $1.41^{+0.19}_{-0.17}$   &  $1.36^{+0.20}_{-0.20}$   &  $1.18^{+0.27}_{-0.21}$   &  $1.32^{+0.43}_{-0.36}$   \\ 
  $\frac{ {\rm d}\ln \xi_{\rm 3D} }{ {\rm d}\ln r }|_{r=r_{\rm sp}^{\rm 3D}}$         & $-4.14^{+0.46}_{-0.68}$   &  $-3.77^{+0.37}_{-0.53}$   &  $-3.40^{+0.32}_{-0.43}$   &  $-3.20^{+0.41}_{-0.51}$   \\ 
  $\frac{ {\rm d}\ln (\rho_{\rm in} f_{\rm trans}) }{ {\rm d}\ln r }|_{r=r_{\rm sp}^{\rm 3D}}$  & $-5.40^{+0.90}_{-1.23}$   &  $-5.03^{+0.79}_{-1.11}$   &  $-4.67^{+0.73}_{-0.93}$   &  $-4.55^{+0.76}_{-1.01}$   \\ 
  $\frac{ \rho_{\rm in}f_{\rm trans} }{ \rho_{\rm in}f_{\rm trans}+\rho_{\rm out} }|_{ r=r_{\rm sp}^{\rm 3D} }$ & $0.71^{+0.12}_{-0.11}$ & $0.66^{+0.12}_{-0.11}$ & $0.60^{+0.16}_{-0.14}$ & $0.56^{+0.20}_{-0.23}$ \\
  $R_{\rm sp}^{\rm 2D}\ [h^{-1}{\rm Mpc}]$      & $1.11^{+0.13}_{-0.14}$   &  $1.04^{+0.16}_{-0.15}$   &  $0.93^{+0.17}_{-0.14}$   &  $0.98^{+0.26}_{-0.21}$   \\ 
  $\frac{ {\rm d}\ln \xi_{\rm 2D} }{ {\rm d}\ln R }|_{R=R_{\rm sp}^{\rm 2D}}$      & $-2.22^{+0.19}_{-0.24}$   &  $-2.01^{+0.16}_{-0.19}$   &  $-1.88^{+0.17}_{-0.21}$   &  $-1.73^{+0.18}_{-0.22}$   \\ \hline
  $S/N$                        & $40.7$   &    $44.4$   &    $38.1$   &    $35.8$   \\ 
   $\langle \Sigma_{g} \rangle\ [h^{2}{\rm Mpc}^{-2}]$   &   8.4           &     24.2           &      139.5          &   300.0             \\
  $\chi_{\rm min}^2/{\rm dof}$                         &  $18.1/10$   &    $15.2/10$   &    $17.7/10$   &    $13.2/10$   \\ \hline
\end{tabular*}
\end{center}
\tabnote{
  $^{*}$
  We show the results for analyses in Section~\ref{subsec:dynamical} to investigate possible dynamical friction effects
  employing one or two magnitudes fainter and brighter photometric galaxies than the fiducial analysis in Table~\ref{tab:mainresults}
  with the Low-$z$ cluster sample described in Table~\ref{tab:sampleselection}.
}
\label{tab:dynamicalfriction}
\end{table*}

\section{Results} \label{sec:results}
In Section~\ref{subsec:result:fid},
we show results for the cluster samples in Table~\ref{tab:sampleselection}
with the fiducial absolute magnitude cut $M_{z}-5 \log_{10} h< -18.8$ for galaxy selections.
We present results for the Low-$z$ cluster sample with different absolute magnitude cuts for galaxy selections
in Section~\ref{subsec:dynamical} to investigate possible dynamical friction effects.
In Section~\ref{sec:redblue},
we also show results with separations for red and blue galaxies and 
the absolute magnitude cut $M_{z}-5 \log_{10} h< -19.3$
over $0.1 \leq z_{\rm cl} \leq 1.0$.

\subsection{Splashback features around HSC CAMIRA clusters with cluster redshift and richness dependences}
\label{subsec:result:fid}
We show our projected cross-correlation function measurements and the model predictions from the MCMC chains in the first row of Figures~\ref{fig:mainresults_full_z} and \ref{fig:mainresults_N} for each cluster sample in Table~\ref{tab:sampleselection} with the fiducial absolute magnitude cut $M_{z}-5 \log_{10} h< -18.8$ for galaxy selections. 
The second row shows constraints on the logarithmic derivative of the projected cross-correlation function based on our model fitting results. 
The third and fourth rows present constraints on the three-dimensional galaxy density profile and the logarithmic derivative of the three-dimensional profile, respectively. In the second and fourth rows, we also show the model predictions from the halo-matter cross-correlation function (i.e., dark matter distributions around clusters) in \cite{Nishimichietal2019}
and the mass-richness relation \citep{Murataetal2019} presented in Appendix~\ref{app:modelpredictions} for comparisons.

In Table~\ref{tab:mainresults}, we show constraints on the model parameters and derived splashback features. The constraints on $\log_{10} \alpha$, $\log_{10} \beta$, $\log_{10} \gamma$, $f_{\rm cen}$, and $R_{\rm off}$ parameters are determined mainly from the Gaussian priors, whereas those for the other parameters are well constrained by the measurements compared to their prior distributions. 
The $\chi^2$ values at the best-fit model parameters in Table~\ref{tab:sampleselection} indicate that our fitting results are acceptable. 
Following \cite{Baxteretal2017}, we list the values of the logarithmic derivative of the inner profiles (i.e., $\rho_{\rm in}(r) f_{\rm trans}(r)$) at the location of $r_{\rm sp}^{\rm 3D}$ in Table~\ref{tab:sampleselection}. These values are smaller than $-3$ with more than $\sim 2 \sigma$ significance for all the samples. 
This result suggests that classical fitting functions like the NFW profiles \citep{Navarroetal1996} have difficulty 
in fitting the observed cross-correlation measurements
and provides the evidence of the splashback features \citep{Baxteretal2017}.

The $1\sigma$ precisions of our constraints on $r_{\rm sp}^{\rm 3D}$ compared to the median values are approximately $12\%$, $16\%$, $17\%$, $25\%$, $28\%$, $18\%$, and $13\%$ for the Full, Low-$z$, Mid-$z$, High-$z$, Low-$N$, Mid-$N$, and High-$N$, respectively, after marginalizing over the off-centering effects with the fiducial absolute magnitude galaxy cut. These constraints are consistent with the model predictions from the halo-matter cross-correlation function in Appendix~\ref{app:modelpredictions} within $0.5 \sigma$, $1.9 \sigma$, $0.2 \sigma$, $0.2 \sigma$, $0.5 \sigma$, $1.6\sigma$, and $1.1 \sigma$ levels for the Full, Low-$z$, Mid-$z$, High-$z$, Low-$N$, Mid-$N$, and High-$N$, respectively, whereas our constraints are consistent with the $20\%$ smaller model predictions  (i.e., $0.8 \times r_{\rm sp, model}^{\rm 3D}$) within $1.2 \sigma$, $0.2 \sigma$, $1.0 \sigma$, $0.9 \sigma$, $0.3 \sigma$, $0.2 \sigma$, and $0.7 \sigma$ levels. 
Given the larger error bars due to the smaller survey area, our constraints are consistent with both the model predictions and their $20\%$ smaller values. 
However, we note that our constraints on $r_{\rm sp}^{\rm 3D}$ for the Low-$z$ sample at $0.1<z_{\rm cl}<0.4$ are on the lower side than the model prediction at the level of $1.9 \sigma$, which is in line with the constraints in the literature \citep{Moreetal2016, Baxteretal2017, Changetal2018} with clusters selected by another optical cluster finding algorithm redMaPPer. 
Further investigation is warranted given our larger error bars. 
We note that we compare constraints from red galaxies only with the model predictions again 
for the Low-$z$, Mid-$z$, and High-$z$ cluster samples in Section~\ref{sec:redblue}, 
since these constraints are more precise than those without separations of red and blue galaxies. 
Our constraints with the fiducial absolute magnitude galaxy cut in Figures~\ref{fig:mainresults_full_z} and \ref{fig:mainresults_N} suggest that the three-dimensional profiles are slightly steeper (i.e., smaller derivative values) around the splashback radius than the model predictions from dark matter. 
We discuss a possible origin for these results in Sections~\ref{subsec:dynamical} by comparing constraints from different absolute magnitude galaxy cuts.

\begin{figure*}
  \begin{center}
    {
      \includegraphics[width=0.495 \textwidth]{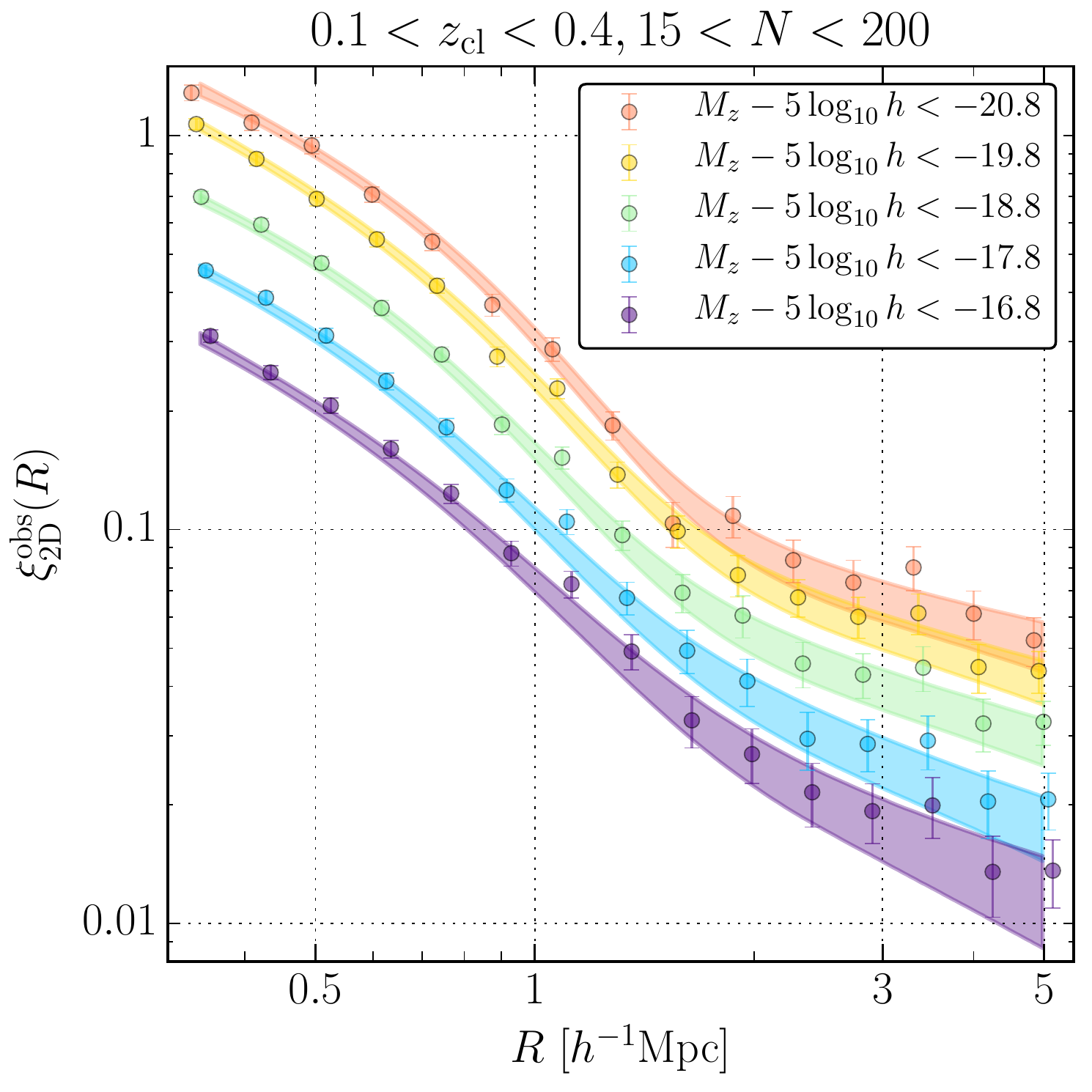}
      \includegraphics[width=0.495 \textwidth]{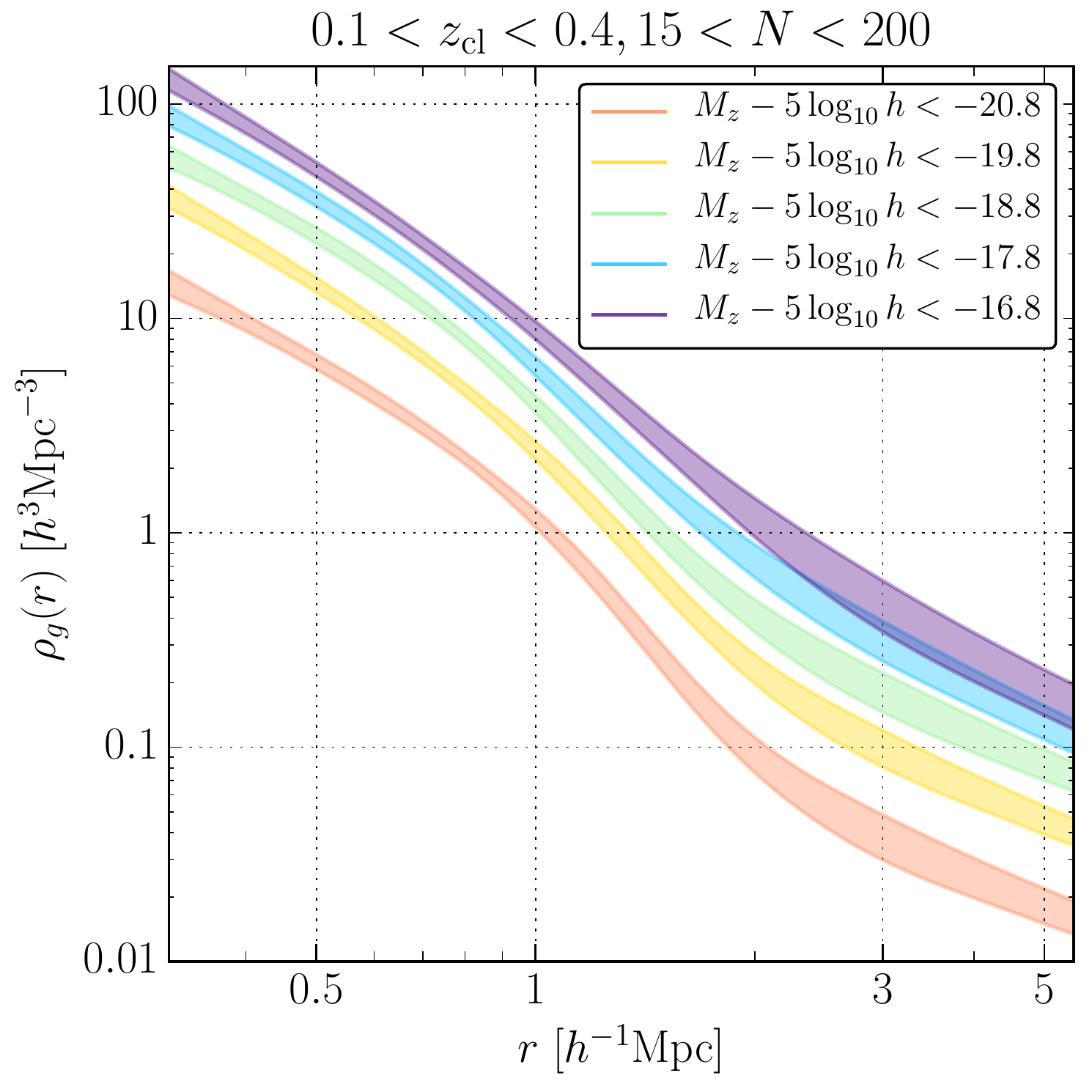}
    }
  \end{center}
  \caption{
    We show the results from Table~\ref{tab:dynamicalfriction} with the Low-$z$ cluster sample and the different absolute magnitude cuts for galaxy selections after marginalizing over the off-centering effects. 
    The left panel shows the projected cross-correlation measurements and the 16th and 84th percentiles of the model predictions from the MCMC chains for each selection.
    The right panel presents constraints on the three-dimensional galaxy density profile $\rho_g(r)=\langle \Sigma_g \rangle \xi_{\rm 3D}(r)/(2 R_{\rm max})$ (see Section~\ref{subsec:model} for more details)
    from the MCMC chains for each selection.
  }
\label{fig:fit_dynamicalfrictions}
\end{figure*}

\begin{figure*}
  \begin{center}
    \includegraphics[width=1.00 \textwidth]{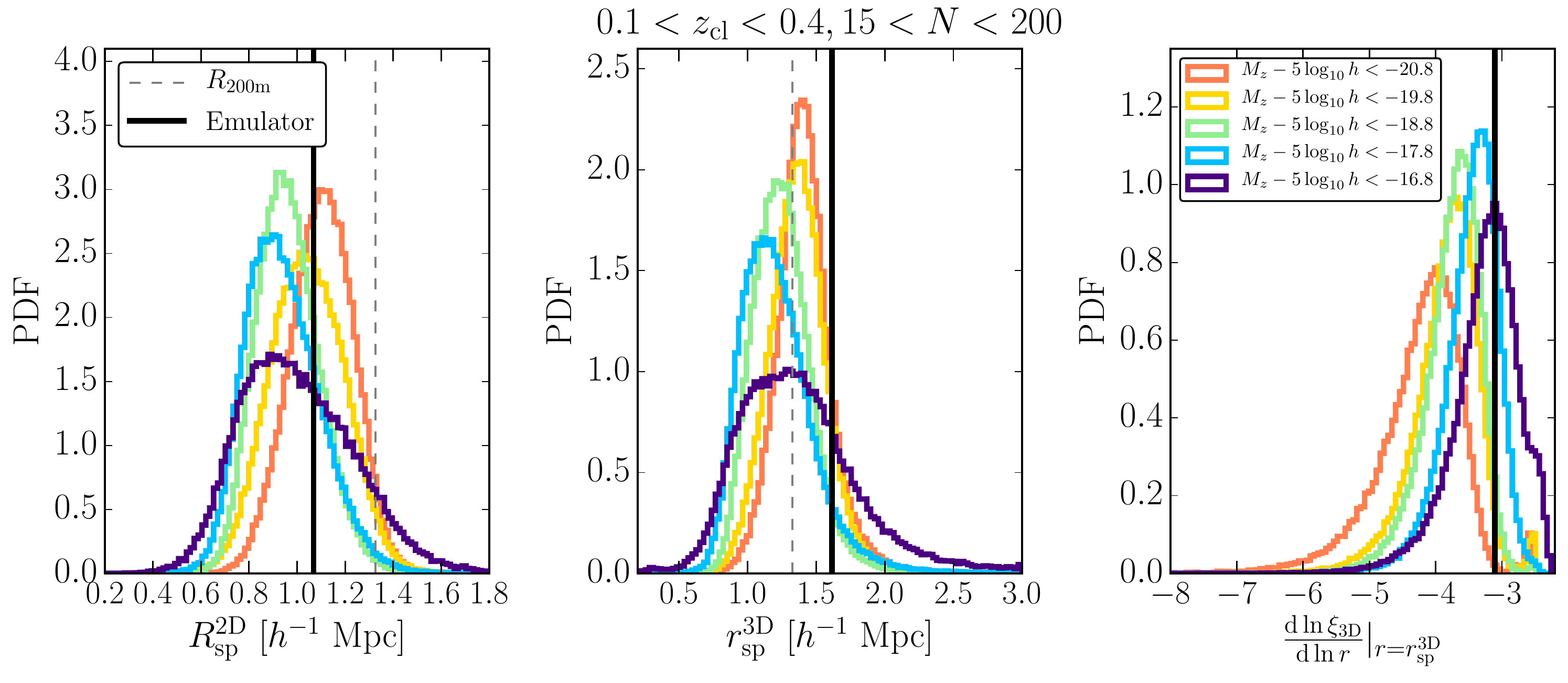}
  \end{center}
  \caption{
    We show the probability distribution functions of splashback features from the MCMC chains for the analyses with the Low-$z$ cluster sample and the different absolute magnitude cuts for galaxy selections in Table~\ref{tab:dynamicalfriction} and Figure~\ref{fig:fit_dynamicalfrictions}. The vertical black lines denote the model predictions from the halo-matter cross-correlation emulator and mass-richness relation in
    Appendix~\ref{app:modelpredictions} for comparisons.
    The vertical dashed lines show the mean value of $R_{\rm 200m}$.
  }
\label{fig:hist_dynamicalfrictions}
\end{figure*}

\subsection{Splashback features from different absolute magnitude galaxy cuts}\label{subsec:dynamical}
We employ the Low-$z$ cluster sample 
to derive constraints on splashback features with different absolute magnitude galaxy cuts, 
since we can use fainter absolute magnitude cuts for the galaxies with lower redshift clusters given a fixed apparent magnitude limit,
and the constraint on $r_{\rm sp}^{\rm 3D}$ for the Low-$z$ sample 
is better among the cluster samples 
as shown in Table~\ref{tab:mainresults} with the fiducial absolute magnitude cut.
Specifically, we shift the absolute magnitude cuts by one or two magnitudes to both fainter and brighter sides
with the {\it z}-band magnitude compared to the fiducial absolute magnitude cut 
$M_{z}-5 \log_{10} h< -18.8$ in Section~\ref{subsec:result:fid}.
The faintest absolute magnitude cut ($M_{z}-5 \log_{10} h< -16.8$) corresponds to an apparent {\it z}-band magnitude limit of 24.1 at $z_{\rm cl}=0.4$, which is shallower than the apparent magnitude cut for the galaxy catalog.

We show these constraints on the model parameters and splashback features 
in Table~\ref{tab:dynamicalfriction} after marginalizing over the off-centering effects.
We present our projected cross-correlation function measurements for the Low-$z$ sample 
with the different absolute magnitude galaxy cuts
and the model predictions from the MCMC chains in the left panel of Figure~\ref{fig:fit_dynamicalfrictions}.
The measurements with fainter absolute magnitude cuts have lower amplitudes,
indicating that brighter galaxies are more strongly clustered around clusters 
than fainter galaxies.
The right panel in Figure~\ref{fig:fit_dynamicalfrictions} 
shows constraints on the three-dimensional galaxy density profile from the MCMC chains.
We also show histogram comparisons of the constraints on splashback features
in Figure~\ref{fig:hist_dynamicalfrictions}.
Figure~\ref{fig:hist_dynamicalfrictions} indicates 
that the constraints on $R_{\rm sp}^{\rm 2D}$ and $r_{\rm sp}^{\rm 3D}$ 
are robust against the different absolute magnitude cuts within the error bars.
Specifically, 
the $1\sigma$ precisions of the constraints on $r_{\rm sp}^{\rm 3D}$
compared to the median values are $13\%$, $15\%$, $16\%$, $20\%$, and $30\%$ 
from the brightest to the faintest absolute magnitude cut including the fiducial cut,
respectively.
Within these error bars, we do not see any systematic trend consistent with
dynamical friction effects (i.e., smaller values for brighter galaxies), 
which is not surprising given expectations (a $\lesssim 20\%$ shift of $r_{\rm sp}^{\rm 3D}$ for our brightest galaxy sample and $\lesssim 5\%$ shifts for other galaxy samples) 
from subhalos in simulations \citep{Moreetal2016, Changetal2018} with simplistic abundance matching methods.
Note that the constraints on $r_{\rm sp}^{\rm 3D}$ are preferred to 
be 
smaller than that predicted by the model
across the five absolute magnitude range
within these error bars, although moderate correlations exist among these constraints.
Indeed, we confirm that there are moderate positive correlations in the jackknife cross-covariance for the projected cross-correlation 
measurements among these different magnitude cuts. 
Further investigation with reduced error bars from a larger survey area 
is warranted to constrain dynamical friction effects 
at a $\sim 5\%$ level 
for the CAMIRA clusters.

On the other hand, the logarithmic derivatives of the three-dimensional profiles at $r_{\rm sp}^{\rm 3D}$ 
are larger 
for fainter absolute magnitude cuts, as shown in the right panel of Figure~\ref{fig:fit_dynamicalfrictions} 
and Figure~\ref{fig:hist_dynamicalfrictions}.
\cite{Moreetal2016} shows larger logarithmic derivative values at $r_{\rm sp}^{\rm 3D}$ 
with decreasing maximum circular velocities of subhalos in simulations, which is consistent with our results.
We expect that these differences in the derivative values 
are mainly caused by different ratios between the inner and outer material number densities 
($=\rho_{\rm in}f_{\rm trans} / \rho_{\rm out} $) around the splashback radii,
since splashback radii mark a halo boundary between the multi-stream and single-stream regions.
We thus expect that derivative values would be smaller (i.e., steeper profiles) when the ratios are large at $r_{\rm sp}^{\rm 3D}$
whereas these values should be larger (i.e., shallower profiles) when the ratios are smaller.
We confirm this expectation by checking that constraints on 
$ \rho_{\rm in}f_{\rm trans} / (\rho_{\rm in}f_{\rm trans}+\rho_{\rm out} )
= (1+(\rho_{\rm in}f_{\rm trans}/\rho_{\rm out})^{-1} )^{-1}$ 
at $r_{\rm sp}^{\rm 3D}$ for the fainter galaxy samples are systematically smaller than those for the brighter galaxy samples 
as presented in Tables~\ref{tab:mainresults} and \ref{tab:dynamicalfriction} although the error bars are not very small.
Note that this correlation between the density ratios and the logarithmic derivative values at $r_{\rm sp}^{\rm 3D}$
also exists in Table~\ref{tab:mainresults} for the fiducial absolute magnitude cuts and 
Table~\ref{tab:redblue} for the color cuts presented in Section~\ref{sec:redblue}.

In particular, Figure~\ref{fig:hist_dynamicalfrictions} shows that
the constraint on the logarithmic derivative at $r_{\rm sp}^{\rm 3D}$
from the faintest absolute magnitude cut is consistent with the model prediction from the dark matter distribution,
suggesting that the ratio between the inner and outer materials for the faintest galaxies should be on average similar 
to the one for the dark matter distribution around splashback radii for the Low-$z$ cluster sample.
Also, our results might indicate that 
a larger fraction of fainter galaxies accrete into cluster regions ($r \lesssim r_{\rm sp}^{\rm 3D}$) 
compared to brighter galaxies at $0.1 < z_{\rm cl} < 0.4$,
or that fainter galaxies have been selectively destroyed or merged compared to brighter galaxies
to decrease number densities after infalls into cluster regions.
We note that blue fraction is higher at fainter magnitudes, 
thus the interpretation is also related 
to the results of Section~\ref{sec:redblue}, 
although the absolute magnitude limits are not the same.
It might be informative to compare these results 
with hydrodynamical simulations to constrain galaxy formation and evolution, possibly including relations between subhalo masses and 
galaxy luminosities.

\subsection{Splashback features from red and blue galaxies} \label{sec:redblue}
We employ a method in \cite{Nishizawaetal2018} to define
red (quenched) and blue (star-forming) galaxy populations for each cluster redshift bin 
with the help of red-sequence measurements by 
the CAMIRA red member galaxy catalog described in Section~\ref{sec:data:cluster}
without resorting to photometric redshifts of individual galaxies.
This method uses a color-magnitude diagram for the CAMIRA member galaxies to derive a linear red-sequence relation at each cluster redshift 
within the redshift interval $\Delta z_{\rm cl}=0.05$ (see Section~\ref{sec:measure}),
where we employ {\it g--r}, {\it r--i}, and {\it i--y} color 
for clusters at $0.1 < z_{\rm cl} < 0.4$, $0.4 < z_{\rm cl} < 0.7$, and $0.7 < z_{\rm cl} < 1.0$ respectively, 
and we use the {\it z}-band magnitude.
These color combinations are chosen to be sensitive to the 4000${\rm \AA}$ break from red galaxies at each redshift range.
This method defines red galaxies 
with colors above a linear color-magnitude relation, which is $2\sigma$ lower (bluer) than the derived linear red-sequence mean relation at each cluster redshift bin
and as a function of the {\it z}-band magnitude. 
We calculate these $\sigma$ error values by averaging over the absolute magnitude limit for each cluster redshift bin,
and these errors include the photometric errors as well as the intrinsic errors in the red-sequence.
With this method, we can define red and blue galaxy populations even outside the aperture size in the CAMIRA finder
($R \simeq 1\ h^{-1} {\rm Mpc}$ in physical coordinates).
\cite{Nishizawaetal2018} showed that number ratios between correlated red and blue galaxy populations around clusters marginally change by $\sim 0.1$ 
at the filter transitions for the color combinations (i.e., $z_{\rm cl}=0.4, 0.7$) at $R < 1h^{-1}{\rm Mpc}$.
We have confirmed that 
a difference of these selection criteria with the PDR2 is negligibly small
from those in \cite{Nishizawaetal2018} with the first-year HSC data.
In our analysis, we use the galaxy catalog after the same selection for the CAMIRA member catalog
with $\texttt{z\_cmodel\_mag - a\_z}<24.0$ and $\texttt{z\_cmodel\_magsigma}<0.1$, 
and we employ the absolute magnitude cut $M_{z}-5 \log_{10} h< -19.3$ as described in Section~\ref{sec:data:galaxy}.

We note that this method is different from those in the literature. 
Specifically, \cite{Baxteretal2017} and \cite{Shinetal2019} employed 
photometric redshifts of individual galaxies to select galaxies for deriving color-magnitude diagrams as a function of redshift 
and to infer red/blue separation criteria based on percentile ranges in color values.  
On the other hand, 
\cite{Zurcher&More2019} inferred {\it g-r} color cuts to separate red from blue galaxy population
as a function of redshift with spectroscopic samples 
although these spectroscopic samples are quite shallow and thus consist of galaxies that are brighter 
than their galaxies for cross-correlation measurements.

\begin{table*}[t]
   \caption{ Same as Table~\ref{tab:mainresults}, but for analyses with separations of red and blue galaxies. $^{*}$
            }
  \begin{center}
    \begin{tabular*}{0.86 \textwidth}{ lcccccc } \hline\hline
      \centering
  Parameter                    & Low-$z$ & Low-$z$ & Mid-$z$ & Mid-$z$ & High-$z$ & High-$z$    \\
                               & {\it red}     & {\it blue}    & {\it red}     & {\it blue}    & {\it red} & {\it blue} \\ \hline

  $\log_{10}\rho_{s}$ & $2.14^{+0.72}_{-0.82}$   &  $0.75^{+1.13}_{-1.68}$   &  $2.67^{+0.54}_{-0.58}$   &  $0.78^{+0.73}_{-0.87}$   &  $2.30^{+0.68}_{-0.68}$   &  $1.00^{+1.05}_{-1.57}$   \\
  $\log_{10}\alpha$  & $-0.91^{+0.44}_{-0.51}$   &  $-0.76^{+0.54}_{-0.53}$   &  $-0.75^{+0.28}_{-0.36}$   &  $-0.92^{+0.52}_{-0.52}$   &  $-0.80^{+0.38}_{-0.50}$   &  $-0.77^{+0.49}_{-0.55}$   \\
  $\log_{10}r_{s}$    & $-0.44^{+0.38}_{-0.34}$   &  $-0.35^{+0.60}_{-0.44}$   &  $-0.64^{+0.25}_{-0.24}$   &  $-0.14^{+0.48}_{-0.38}$   &  $-0.53^{+0.30}_{-0.31}$   &  $-0.45^{+0.60}_{-0.38}$   \\
  $\log_{10}\rho_{o}$ & $-0.14^{+0.18}_{-0.26}$   &  $0.10^{+0.11}_{-0.30}$   &  $-0.80^{+0.37}_{-0.37}$   &  $-0.82^{+0.39}_{-0.45}$   &  $-0.55^{+0.33}_{-0.46}$   &  $-0.32^{+0.21}_{-0.32}$   \\
  $S_{e}$             & $1.24^{+0.24}_{-0.31}$   &  $1.78^{+0.27}_{-0.44}$   &  $0.68^{+0.44}_{-0.38}$   &  $0.85^{+0.49}_{-0.48}$   &  $0.98^{+0.42}_{-0.50}$   &  $1.55^{+0.49}_{-0.44}$   \\
  $\log_{10}r_{t}$    & $0.09^{+0.10}_{-0.09}$   &  $0.10^{+0.41}_{-0.25}$   &  $0.14^{+0.10}_{-0.09}$   &  $0.19^{+0.14}_{-0.12}$   &  $0.16^{+0.09}_{-0.13}$   &  $0.11^{+0.34}_{-0.23}$   \\
  $\log_{10}\beta$   & $0.77^{+0.18}_{-0.17}$   &  $0.77^{+0.20}_{-0.20}$   &  $0.77^{+0.19}_{-0.19}$   &  $0.78^{+0.20}_{-0.19}$   &  $0.78^{+0.20}_{-0.20}$   &  $0.76^{+0.21}_{-0.21}$   \\
  $\log_{10}\gamma$  & $0.66^{+0.19}_{-0.18}$   &  $0.59^{+0.20}_{-0.20}$   &  $0.67^{+0.18}_{-0.18}$   &  $0.61^{+0.20}_{-0.20}$   &  $0.67^{+0.19}_{-0.19}$   &  $0.60^{+0.20}_{-0.20}$   \\
  $f_{\rm cen}$      & $0.70^{+0.08}_{-0.09}$   &  $0.70^{+0.08}_{-0.08}$   &  $0.64^{+0.06}_{-0.06}$   &  $0.64^{+0.06}_{-0.06}$   &  $0.59^{+0.11}_{-0.13}$   &  $0.61^{+0.10}_{-0.10}$   \\
  $R_{\rm off}$      & $0.41^{+0.09}_{-0.10}$   &  $0.37^{+0.09}_{-0.08}$   &  $0.56^{+0.07}_{-0.07}$   &  $0.56^{+0.10}_{-0.10}$   &  $0.58^{+0.08}_{-0.17}$   &  $0.54^{+0.13}_{-0.14}$   \\ \hline
  $r_{\rm sp}^{\rm 3D}\ [h^{-1}{\rm Mpc}]$  & $1.45^{+0.17}_{-0.16}$   &  $1.05^{+0.63}_{-0.33}$   &  $1.72^{+0.22}_{-0.21}$   &  $1.76^{+0.36}_{-0.31}$   &  $1.72^{+0.24}_{-0.31}$   &  $1.15^{+0.45}_{-0.45}$   \\
  $\frac{ {\rm d}\ln \xi_{\rm 3D} }{ {\rm d}\ln r }|_{r=r_{\rm sp}^{\rm 3D}}$ & $-4.55^{+0.53}_{-0.69}$   &  $-2.47^{+0.35}_{-0.47}$   &  $-5.62^{+0.79}_{-1.04}$   &  $-3.88^{+0.69}_{-0.93}$   &  $-5.13^{+0.76}_{-1.05}$   &  $-2.93^{+0.60}_{-0.77}$   \\
  $\frac{ {\rm d}\ln (\rho_{\rm in} f_{\rm trans} ) }{ {\rm d}\ln r }|_{r=r_{\rm sp}^{\rm 3D}}$  & $-5.76^{+0.88}_{-1.13}$   &  $-4.28^{+0.86}_{-1.07}$   &  $-6.51^{+1.02}_{-1.35}$   &  $-5.10^{+0.92}_{-1.24}$   &  $-6.22^{+1.06}_{-1.42}$   &  $-4.53^{+0.90}_{-1.29}$   \\
  $\frac{ \rho_{\rm in}f_{\rm trans} }{ \rho_{\rm in}f_{\rm trans}+\rho_{\rm out} }|_{ r=r_{\rm sp}^{\rm 3D} }$ &  $0.74^{+0.10}_{-0.09}$ & $0.31^{+0.25}_{-0.30}$ & $0.86^{+0.06}_{-0.09}$ & $0.73^{+0.12}_{-0.17}$ & $0.81^{+0.09}_{-0.12}$ & $0.50^{+0.21}_{-0.36}$ \\
  $R_{\rm sp}^{\rm 2D}\ [h^{-1}{\rm Mpc}]$    & $1.15^{+0.11}_{-0.12}$   &  $0.82^{+0.33}_{-0.23}$   &  $1.29^{+0.14}_{-0.12}$   &  $1.31^{+0.22}_{-0.22}$   &  $1.33^{+0.19}_{-0.26}$   &  $0.85^{+0.33}_{-0.30}$   \\
  $\frac{ {\rm d}\ln \xi_{\rm 2D} }{ {\rm d}\ln R }|_{R=R_{\rm sp}^{\rm 2D}}$ & $-2.50^{+0.24}_{-0.28}$   &  $-1.27^{+0.13}_{-0.18}$   &  $-3.05^{+0.35}_{-0.42}$   &  $-1.72^{+0.25}_{-0.32}$   &  $-2.84^{+0.33}_{-0.41}$   &  $-1.47^{+0.21}_{-0.30}$   \\ \hline
   $S/N$          & $43.9$   &    $23.6$   &    $42.3$   &    $20.9$   &    $52.6$   &    $19.5$   \\
   $\langle \Sigma_{g} \rangle\ [h^{2}{\rm Mpc}^{-2}]$   &  14.4  &    21.1      &   13.6     &   33.3      &  15.6     & 38.4    \\
   $\chi_{\rm min}^2/{\rm dof}$  &  $12.2/10$   &    $19.9/10$   &    $12.1/10$   &    $5.3/10$   &    $15.0/10$   &    $19.4/10$   \\ \hline
\end{tabular*}
\end{center}
\tabnote{
  $^{*}$
  We show the results for analyses in Section~\ref{sec:redblue} with separations of red and blue galaxies from a red-sequence based method in \cite{Nishizawaetal2018}
  and the cluster samples described in Table~\ref{tab:sampleselection}.
  We use an absolute magnitude cut $M_{z}-5 \log_{10} h< -19.3$ as described in Sections~\ref{sec:data:galaxy} and \ref{sec:redblue}.
}
\label{tab:redblue}
\end{table*}

\begin{figure*}
  \begin{center}
    {
    \includegraphics[width=5.63cm]{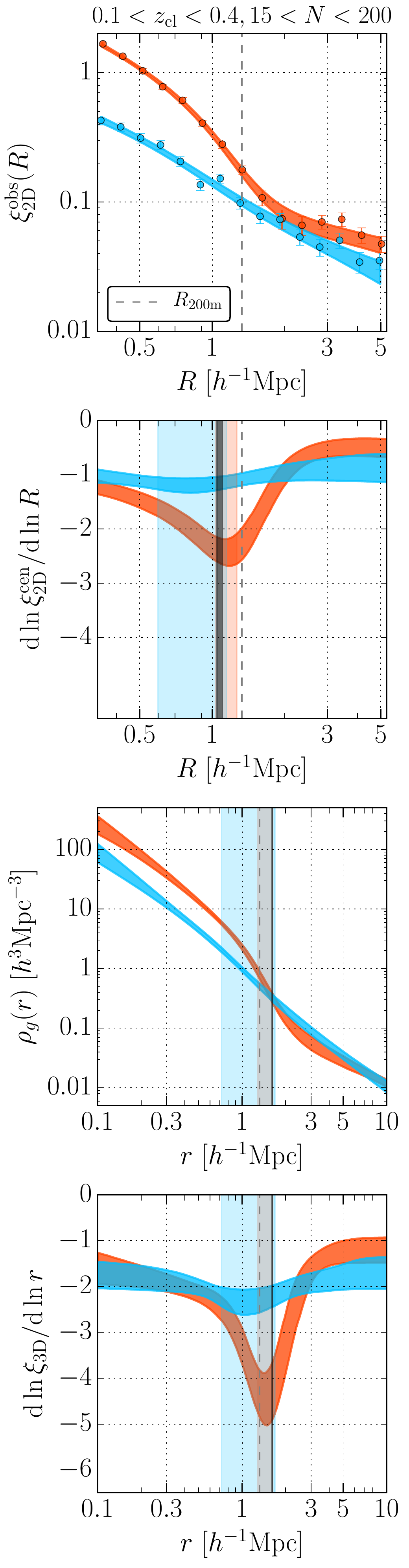}
    \includegraphics[width=5.63cm]{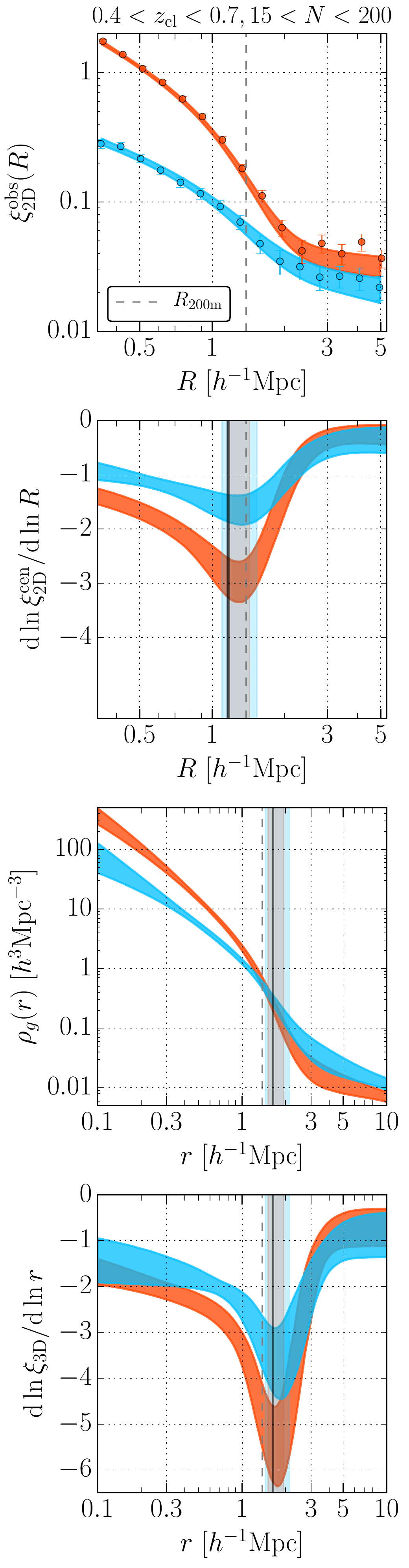}
    \includegraphics[width=5.63cm]{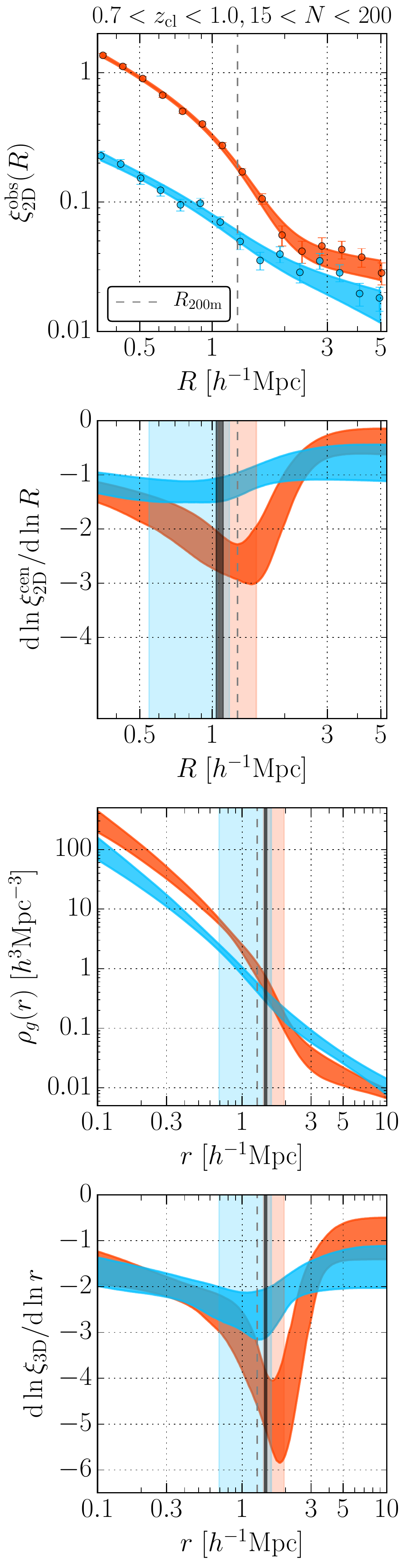}
    }
  \end{center}
  \caption{
    Same as Figure~\ref{fig:mainresults_full_z}, but for the analyses with separations of red (red color) and blue (blue color) galaxies
    and an absolute magnitude cut $M_{z}-5 \log_{10} h< -19.3$ for galaxy selections
    in Section~\ref{sec:redblue} and Table~\ref{tab:redblue}
    for the Low-$z$, Mid-$z$, and High-$z$ cluster samples after marginalizing over the off-centering effects.
  }
\label{fig:fit_redblue}
\end{figure*}

In Figure~\ref{fig:fit_redblue} and Table~\ref{tab:redblue},
we show the projected cross-correlation measurements and derived constraints.
Consistently with the literature \citep{Baxteretal2017, Nishizawaetal2018, Zurcher&More2019, Shinetal2019},
we find that red galaxies are more concentrated around cluster centers than blue galaxies in the projected measurements.
We detect splashback features for both the red and blue galaxy populations defined by our selection criteria over $0.1 < z_{\rm cl} < 1.0$,
although the significances from the derivative of the inner profiles \citep{Baxteretal2017} are smaller 
for blue galaxy populations of the Low-$z$ and High-$z$ cluster samples ($\sim 1.5 \sigma$). 
Within the error bars, the constraints on $r_{\rm sp}^{\rm 3D}$ are consistent 
among those for the fiducial absolute magnitude cuts without the red and blue separation in Section~\ref{subsec:result:fid}
and those for the red and blue galaxy populations at each redshift range.
Due to their sharper profiles around $r_{\rm sp}^{\rm 3D}$ (i.e., smaller derivative values),
constraints on $r_{\rm sp}^{\rm 3D}$ for the red galaxies only are more precise than those without the red and blue separation in Section~\ref{subsec:result:fid}.
Specifically, 
the $1\sigma$ precisions of the constraints on $r_{\rm sp}^{\rm 3D}$ 
are approximately $11 \%$, $13 \%$, and $16 \%$ for the Low-$z$, Mid-$z$, and 
High-$z$ cluster samples respectively when employing the red galaxies only, 
which are more precise than those in Section~\ref{subsec:result:fid} by $\sim 10 \%$ for the High-$z$ cluster sample.
These more precise constraints are consistent with the model predictions in Table~\ref{tab:sampleselection}
within $1.0 \sigma$, $0.4 \sigma$, and $1.0 \sigma$ levels, whereas
these constraints are consistent with the $20 \%$ smaller model predictions within
$1.0 \sigma$, $1.9 \sigma$, and $2.0 \sigma$ levels
for the Low-$z$, Mid-$z$, and High-$z$ samples, respectively.
We note that the constraints 
for the Mid-$z$ and High-$z$ cluster samples 
are more consistent with
the model predictions ($\lesssim 1\sigma$ levels) 
than the $20\%$ smaller values of the model predictions ($\sim 2\sigma$ levels).
We also note that
these analyses with the red galaxy populations approximately correspond to simulation results in \cite{Adhikarietal2014}
with selectively using simulation particles orbiting within clusters.
They find that using simulation particles with small radial motions, which
corresponds to preferentially selecting simulation particles within clusters,
does not significantly bias the splashback radius measurement
within the level of $\sim 5-10 \%$.
Also, 
based on the analysis of {\it N}-body simulations, 
\cite{Shinetal2019} argue that the splashback radius depends on accretion time
(that may correlate with galaxy colors) of subhalos used to
measure it, although the dependence appears to be weak except
for subhalos accreted very recently.
Given our larger error bars from the limited area, 
further investigations are warranted to constrain 
$r_{\rm sp}^{\rm 3D}$ values at the higher redshift ranges ($0.4<z_{\rm cl}<1.0$) 
with upcoming datasets more precisely.

We expect that the density profiles of blue galaxies would be consistent with a pure power-law without any splashback features
when these blue galaxies are still on their first infall passage
\citep{Baxteretal2017, Zurcher&More2019, Shinetal2019}.
We thus employ a pure power-law model only with marginalizing over the off-centering effects in analyses for blue galaxies 
instead of the DK14 model for the three-dimensional profile 
to check whether the power-law model can have reasonable $\chi_{\rm min}^2$ values or not.
We find that these analyses result in
$\chi^{2}_{\rm min}/{\rm dof}=29.2/13, 21.7/13, 31.9/13$ for the Low-$z$, Mid-$z$, 
and High-$z$ cluster samples, respectively.
Their low {\it p}-values suggest that the DK14 model (Table~\ref{tab:redblue}) 
can fit the measurements better 
than the power-law model given the difference 
in the minimum $\chi^2$ and the degree-of-freedom values over $0.1 < z_{\rm cl} < 1.0$
(i.e., $\Delta \chi^{2}_{\rm min} \gtrsim 10$ by adding the six parameters for the DK14 model with the Gaussian priors on three of them).
Note that, as shown in Figure~\ref{fig:fit_redblue},
the logarithmic derivative of the three-dimensional profile at $r=r_{\rm sp}^{\rm 3D}$ for the blue galaxy population at
$0.4<z_{\rm cl}<0.7$ is smaller than those at the other redshift bins at the level of $\sim 2 \sigma$, possibly due to the different combinations
of filters to define the blue galaxy populations in \cite{Nishizawaetal2018} as described above.

At the same time, we also expect red galaxies to scatter into the blue population, especially 
at fainter magnitudes and higher redshift due to larger photometric errors. 
We estimate the contamination fraction from the red to the blue galaxy population to be $f_{\rm red} \simeq 0.025$ 
since we employ the $2 \sigma$ lower limit for the separation criteria in \cite{Nishizawaetal2018}.
Compared to \cite{Zurcher&More2019},
we employ the same galaxy selection for the cross-correlation measurements as 
the one to define the red/blue separation criteria over all the redshift range 
to apply this contamination fraction more consistently for fainter galaxies at higher redshift.
This contamination could potentially alter the cross-correlation measurements 
for the blue populations to cause the observed splashback feature in the blue population.
We follow a method in Appendix~C of \cite{Zurcher&More2019} to assess such contamination.
First, we assume a pure power-law function for the blue galaxies with $S_{e}=1.5$ and the best-fit parameter value 
for $\log_{10}\rho_{o}$ in Table~\ref{tab:redblue}
at each redshift bin.
We then calculate a weighted sum of this power-law profile and the DK14 model profile for the red galaxy population with the best-fit parameters 
in Table~\ref{tab:redblue} with a contamination fraction from the red to the blue population at each redshift bin 
(i.e., $\rho_{g, \rm blue}+f_{\rm red} \rho_{g, \rm red}$).
We find that the $2.5 \%$ contamination fraction alters the logarithmic derivative of the three-dimensional profile around 
$r_{\rm sp}^{\rm 3D}$ of the red population by $0.1-0.2$, which is smaller than values needed
to explain the observed profiles ($\gtrsim 0.5$ differences in the logarithmic derivative of the three-dimensional profiles around $r_{\rm sp}^{\rm 3D}$)
in Figure~\ref{fig:fit_redblue}.
We would require $10-20 \%$ contamination fractions from the red to the blue galaxy population 
to be consistent with the observed profiles for the blue populations
when we assume a pure power-law function before such contaminations.

From these results, consistently with \cite{Zurcher&More2019} from the different selection method on the red and blue galaxy populations, 
we conclude that some fraction of blue galaxies 
stay blue even after reaching their first apocenters of orbits
within host clusters (i.e., splashback radii for these clusters)
from the presence of the splashback feature in the blue galaxy populations
when we define the blue galaxy populations from the method in \cite{Nishizawaetal2018}.
We note that \cite{Shinetal2019} found that their bluest galaxy sample defined by their criteria are
consistent with a pure power-law, which is consistent with them being on their first infall into their cluster samples. 
We expect that the difference from our results might come from the difference of 
the selection criteria of red and blue galaxy populations, as also discussed in \cite{Zurcher&More2019}. 

Also, we investigate how the constraints on three-dimensional galaxy density profiles of the red and blue galaxy populations change
around the locations of $r_{\rm sp}^{\rm 3D}$ more directly in Figure~\ref{fig:fit_redblue},
whereas \cite{Baxteretal2017} and \cite{Shinetal2019} showed that their red/blue fractions abruptly change
around $r_{\rm sp}^{\rm 3D}$ in their projected profiles.
Figure~\ref{fig:fit_redblue} shows that the three-dimensional galaxy density fractions 
between the red and blue galaxy populations change rapidly 
around the model predictions of $r_{\rm sp}^{\rm 3D}$ in Table~\ref{tab:sampleselection} for all the redshift bins 
and the blue galaxy populations in our separation criteria dominate at the outskirts of clusters over $0.1 < z_{\rm cl} < 1.0$ after the $15<N<200$ selection.
The constraints on three-dimensional galaxy densities in Figure~\ref{fig:fit_redblue}
do not evolve significantly within the error bars
over $0.1 < z_{\rm cl} < 1.0$ for both the red and blue galaxy populations around clusters when selected by $15<N<200$,
including concentrations for red galaxy populations possibly due to our off-centering marginalization compared to \cite{Nishizawaetal2018}.
In particular, 
we find that the three-dimensional galaxy densities
for the red and blue populations become the same value at 
$r=1.63^{+0.24}_{-0.15} h^{-1}{\rm Mpc}$, 
$1.50^{+0.29}_{-0.20} h^{-1}{\rm Mpc}$, and 
$1.76^{+0.28}_{-0.31} h^{-1}{\rm Mpc}$
for the Low-$z$, Mid-$z$, and High-$z$ cluster sample respectively, where we calculate these values from the MCMC chains for each analysis.
We note that we ignore positive cross-covariances between the projected cross-correlation measurements 
for the red and blue galaxy populations at each redshift bin for simplicity,
and thus our error bars on these radial values 
are conservatively larger than those with these positive cross-covariances.
These constraints are consistent with the model predictions of $r_{\rm sp}^{\rm 3D}$ in Table~\ref{tab:sampleselection}
within these error bars ($\lesssim 1\sigma$ level) for all the redshift bins.

Since the fraction of blue galaxy population starts to dominate outward at $r \simeq r_{\rm sp}^{\rm 3D}$ over all the redshift bins, 
we expect that the blue galaxy population consists of a larger fraction of the infalling materials around the splashback radius than the red galaxy population
over $0.1 < z_{\rm cl} < 1.0$ on average. 
However, the three-dimensional galaxy density profiles per comoving volumes for the blue galaxy populations 
do not increase at $r<r_{\rm sp}^{\rm 3D}$ within the error bars over $0.1 < z_{\rm cl} < 1.0$ as shown in Figure~\ref{fig:fit_redblue}
although the blue galaxy populations should fall into the cluster regions ($r<r_{\rm sp}^{\rm 3D}$) more frequently than the red galaxy populations.
These results suggest that cluster influences on infalling galaxies should increase when these galaxies 
enter cluster regions at $r<r_{\rm sp}^{\rm 3D}$ due to the multi-steaming materials, 
potentially leading to stronger quenching effects on blue galaxies into red galaxies \citep[e.g.,][]{Hamabataetal2019}
or more frequent merger rates with other galaxies 
to reduce the number of the blue galaxies selectively at $r<r_{\rm sp}^{\rm 3D}$ for the absolute magnitude cut.
Alternatively, these results would imply that the multi-streaming regions at $r<r_{\rm sp}^{\rm 3D}$ 
mix infalling galaxies with older galaxies which were accreted a longer time ago
to cause increased red galaxy fractions at splashback radii abruptly.

\section{Discussion} \label{sec:discussion}
In Section~\ref{subsec:offcenter},
we present how the off-centering effects cause biases and additional errors for constraints on splashback features.
In Section~\ref{subsec:projeffect}, 
we discuss the robustness of our results on the projection effects by employing mock catalogs 
with simplified cluster finders \citep{BuschandWhite2017, Sunayama&More2019}
and an HOD model to populate galaxies with halos phenomenologically
with matching the cluster abundance density and lensing profiles to observations approximately.

\subsection{Biases and errors on splashback features from off-centering effects} \label{subsec:offcenter}
We check how the off-centering effect causes biases and additional errors for the analysis of the splashback features 
with our prior distribution in Table~\ref{tab:paramspriors} 
from the stacked lensing measurements in \cite{Murataetal2019} for each cluster sample.
For this purpose, 
we repeat analyses without marginalizing over the off-centering effects 
for the Low-$z$, Mid-$z$, and High-$z$ cluster samples
with the fiducial setups in Section~\ref{subsec:result:fid}
and with the red galaxy population only in Section~\ref{sec:redblue}.
We note that the prior distributions on the off-centering 
for higher redshift clusters have larger uncertainties due to lower signal-to-noise ratios of 
the stacked lensing measurements presented in \cite{Murataetal2019}. 
We conservatively use the prior distributions from the stacked lensing measurements rather than those from a comparison of CAMIRA centers 
with X-ray selected clusters presented in \cite{Ogurietal2018a} 
since the number of X-ray selected clusters that were matched to CAMIRA clusters ($\sim 50$)
is limited, especially for higher redshift and lower richness clusters.

We show constraints on the splashback features for the analyses without marginalizing over the off-centering effects in Table~{\ref{tab:withoutoffcentering}.
We also present constraints on the logarithmic derivative of the three-dimensional profile with and without the off-centering model
for the Low-$z$ cluster sample with smaller off-centering effects and the High-$z$ cluster sample with larger off-centering effects
in Figure~\ref{fig:offcenter}.
These results show that the median values for $r_{\rm sp}^{\rm 3D}$ in the constraints without the marginalization 
are biased to $\sim 10\%$ higher values compared to the fiducial results in Tables~\ref{tab:mainresults} and \ref{tab:redblue}
with marginalizing over the off-centering effects \citep[see also][]{Moreetal2016, Baxteretal2017, Zurcher&More2019}.
The widths of the 16th and 84th percentile region for $r_{\rm sp}^{\rm 3D}$ after marginalizing over the off-centering effects 
increase by $21\%$, $36\%$, and $105\%$ for the Low-$z$, Mid-$z$, and High-$z$ cluster samples respectively 
with the fiducial absolute magnitude cut only
compared to the analyses without the marginalization.
On the other hand, the widths 
increase by $22\%$, $30\%$, and $62\%$ for the Low-$z$, Mid-$z$, and High-$z$ cluster samples, respectively, when employing the red galaxy populations only, 
which is smaller degradation compared to that for the absolute magnitude cut only
mainly because the steeper profiles (i.e., smaller derivative values)
for the red galaxy population only determine $r_{\rm sp}^{\rm 3D}$ better even with the off-centering model.
Thus we obtain the tighter constraints on $r_{\rm sp}^{\rm 3D}$ when employing the red galaxy population only,
especially for the higher redshift cluster samples in Section~\ref{sec:redblue}.
Figure~\ref{fig:offcenter} also shows 
that the larger uncertainties in the off-centering increase errors for the derivative profile at inner radii 
for the High-$z$ cluster sample, whereas the difference between the two results
for the Low-$z$ sample is smaller due to the smaller uncertainty of the offset distribution
for the prior distribution.

Our result indicates that constraints on splashback features can be improved further by 
reducing the uncertainties of offset distributions, especially for clusters at higher redshifts.
We note that upcoming datasets of lensing measurements with higher signal-to-noise ratios
should reduce uncertainties of the off-centering that we adopt for the prior distribution,
leading to tighter constraints on splashback features.
We could instead use upcoming X-ray datasets for the high redshift clusters
to constrain offset distributions 
by checking offsets between X-ray and CAMIRA cluster centers for a large number of clusters.

\begin{table*}[t]
   \caption{ Constraints on splashback features without marginalizing over off-centering effects. $^{*}$
            }
  \begin{center}
    \begin{tabular*}{0.99 \textwidth}{ lccccccccccccc } \hline\hline
      \centering
  Parameter                    & Full & Low-$z$ & Mid-$z$ & High-$z$ & Low-$z$   & Mid-$z$   & High-$z$                  \\ 
                               &      &         &         &          & {\it red} & {\it red} & {\it red}                 \\\hline
  $r_{\rm sp}^{\rm 3D}\ [h^{-1}{\rm Mpc}]$ &  $1.63^{+0.12}_{-0.11}$   &  $1.37^{+0.17}_{-0.16}$   &  $1.76^{+0.21}_{-0.18}$   &  $1.68^{+0.21}_{-0.16}$   &  $1.52^{+0.14}_{-0.13}$   &  $1.86^{+0.17}_{-0.16}$   &  $1.87^{+0.18}_{-0.16}$   \\ 
  $\frac{ {\rm d}\ln \xi_{\rm 3D} }{ {\rm d}\ln r }|_{r=r_{\rm sp}^{\rm 3D}}$ & $-4.29^{+0.40}_{-0.51}$   &  $-3.67^{+0.33}_{-0.44}$   &  $-4.44^{+0.53}_{-0.74}$   &  $-4.39^{+0.56}_{-0.72}$  &  $-4.68^{+0.50}_{-0.65}$   &  $-5.64^{+0.63}_{-0.83}$   &  $-5.38^{+0.69}_{-0.97}$   \\ 
  $\frac{ {\rm d}\ln ( \rho_{\rm in}f_{\rm trans} ) }{ {\rm d}\ln r }|_{r=r_{\rm sp}^{\rm 3D}}$  &  $-5.90^{+0.83}_{-1.06}$   &  $-4.97^{+0.82}_{-0.93}$   &  $-5.54^{+0.81}_{-1.10}$   &  $-5.82^{+0.99}_{-1.27}$   &  $-5.96^{+0.88}_{-1.11}$   &  $-6.52^{+0.92}_{-1.15}$   &  $-6.59^{+1.03}_{-1.35}$   \\ 
  $\frac{ \rho_{\rm in}f_{\rm trans} }{ \rho_{\rm in}f_{\rm trans}+\rho_{\rm out} }|_{ r=r_{\rm sp}^{\rm 3D} }$ & $0.65^{+0.10}_{-0.10}$ & $0.64^{+0.14}_{-0.11}$ & $0.77^{+0.11}_{-0.12}$ & $0.69^{+0.15}_{-0.15}$ &  $0.73^{+0.10}_{-0.09}$   &  $0.86^{+0.06}_{-0.09}$   &  $0.79^{+0.10}_{-0.11}$   \\ 
  $R_{\rm sp}^{\rm 2D}\ [h^{-1}{\rm Mpc}]$  & $1.36^{+0.08}_{-0.08}$   &  $1.07^{+0.11}_{-0.10}$   &  $1.37^{+0.10}_{-0.10}$   &  $1.42^{+0.11}_{-0.10}$                                          &  $1.24^{+0.09}_{-0.08}$   &  $1.46^{+0.08}_{-0.08}$   &  $1.53^{+0.09}_{-0.09}$   \\ 
  $\frac{ {\rm d}\ln \xi_{\rm 2D} }{ {\rm d}\ln R }|_{R=R_{\rm sp}^{\rm 2D}}$  & $-2.18^{+0.15}_{-0.19}$   &  $-1.96^{+0.14}_{-0.16}$   &  $-2.27^{+0.21}_{-0.27}$   &  $-2.30^{+0.23}_{-0.29}$   &  $2.54^{+0.21}_{-0.25}$   &  $-3.09^{+0.25}_{-0.29}$   &  $-2.94^{+0.29}_{-0.37}$   \\ \hline
  $\chi_{\rm min}^2/{\rm dof}$ & $9.1/10$   &    $19.4/10$   &    $6.0/10$   &    $17.0/10$   &    $11.7/10$   &    $12.5/10$   &    $13.2/10$   \\ \hline
\end{tabular*}
\end{center}
\tabnote{
  $^{*}$
  We show the results without marginalizing over off-centering effects in the model fitting described in Section~\ref{subsec:offcenter}.
  We use the fiducial absolute magnitude cut $M_{z}-5 \log_{10} h< -18.8$ for galaxy selections in the constraints without {\it red} in the second row,
  whereas we employ the absolute magnitude cut $M_{z}-5 \log_{10} h< -19.3$ and the red galaxy criteria in Section~\ref{sec:redblue} for the constraints with {\it red}.
  In Section~\ref{subsec:offcenter},
  we discuss biases and errors in the constraints from off-centering effects
  by comparing with the fiducial results after marginalizing over off-centering
  effects in Tables~\ref{tab:mainresults} and \ref{tab:redblue}.
}
\label{tab:withoutoffcentering}
\end{table*}

\begin{figure*}
  \begin{center}
    {
    \includegraphics[width=0.495 \textwidth]{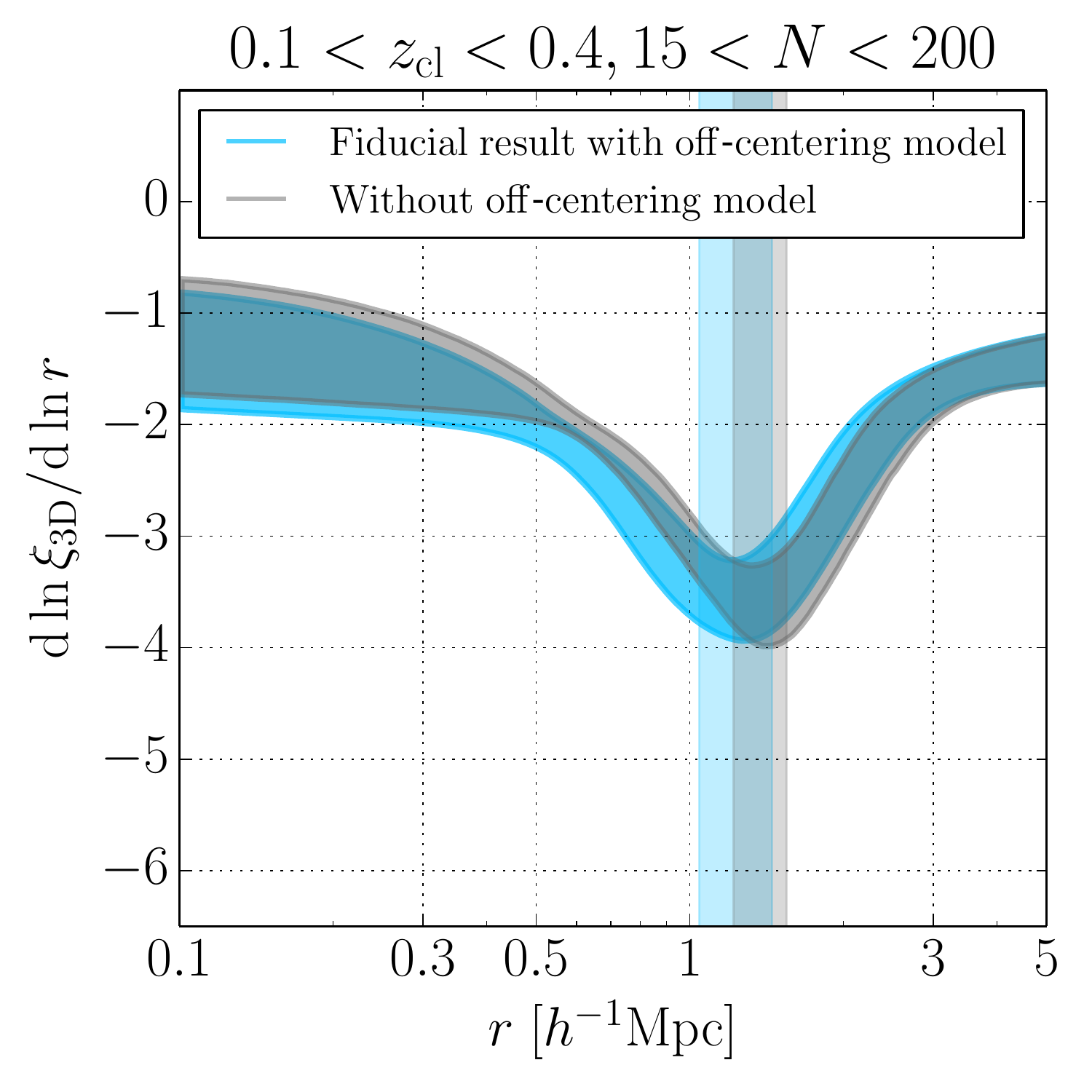}
    \includegraphics[width=0.495 \textwidth]{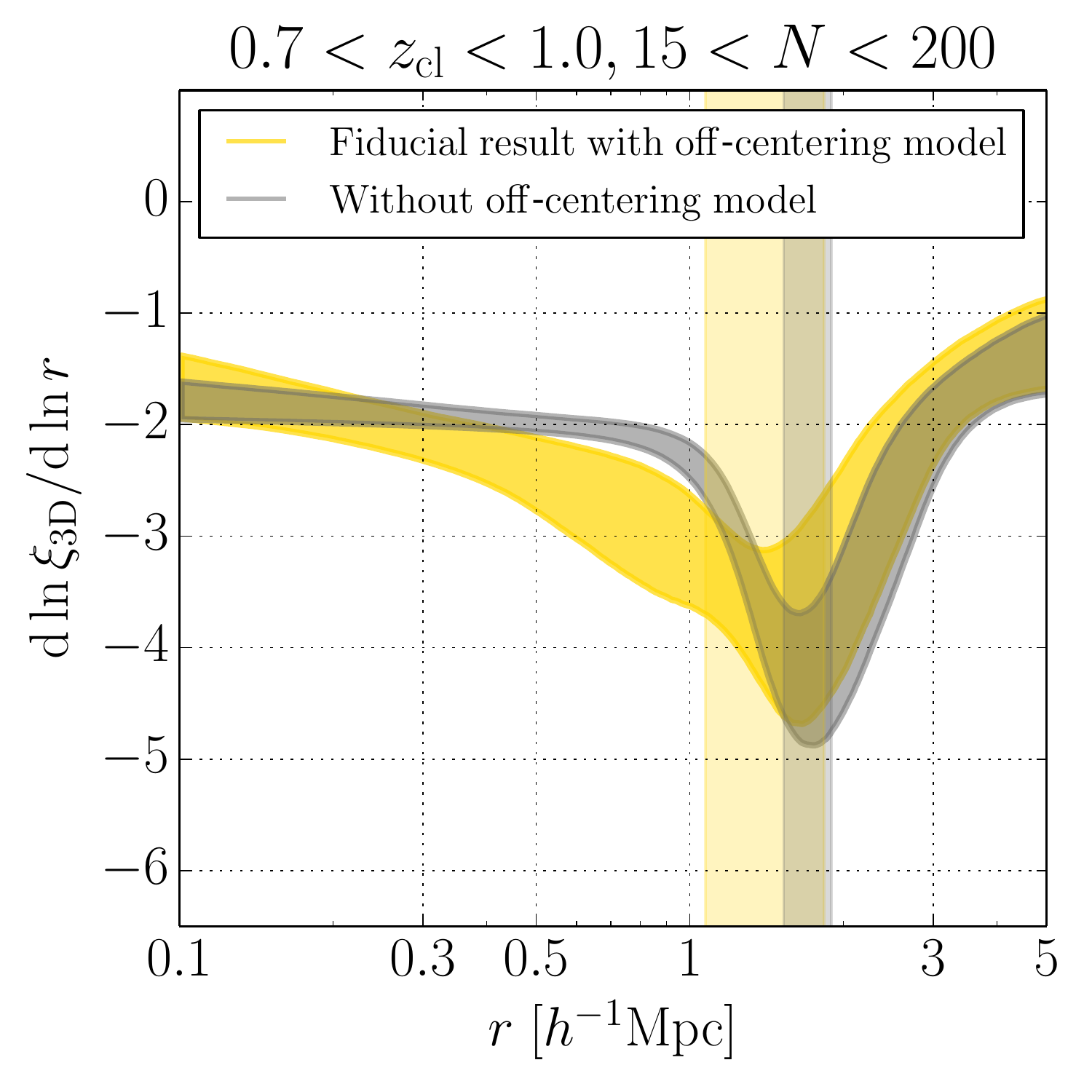}
    \includegraphics[width=0.495 \textwidth]{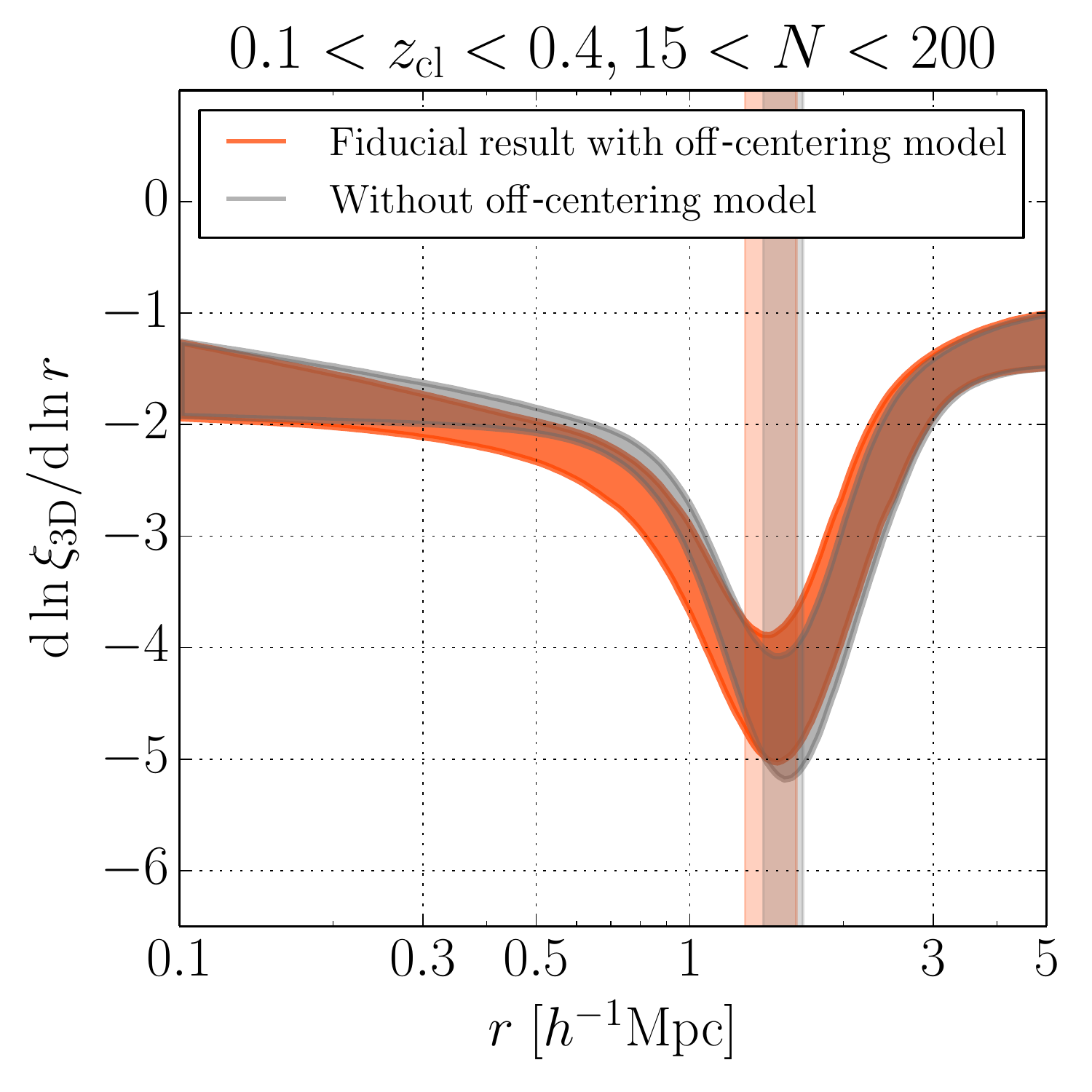}
    \includegraphics[width=0.495 \textwidth]{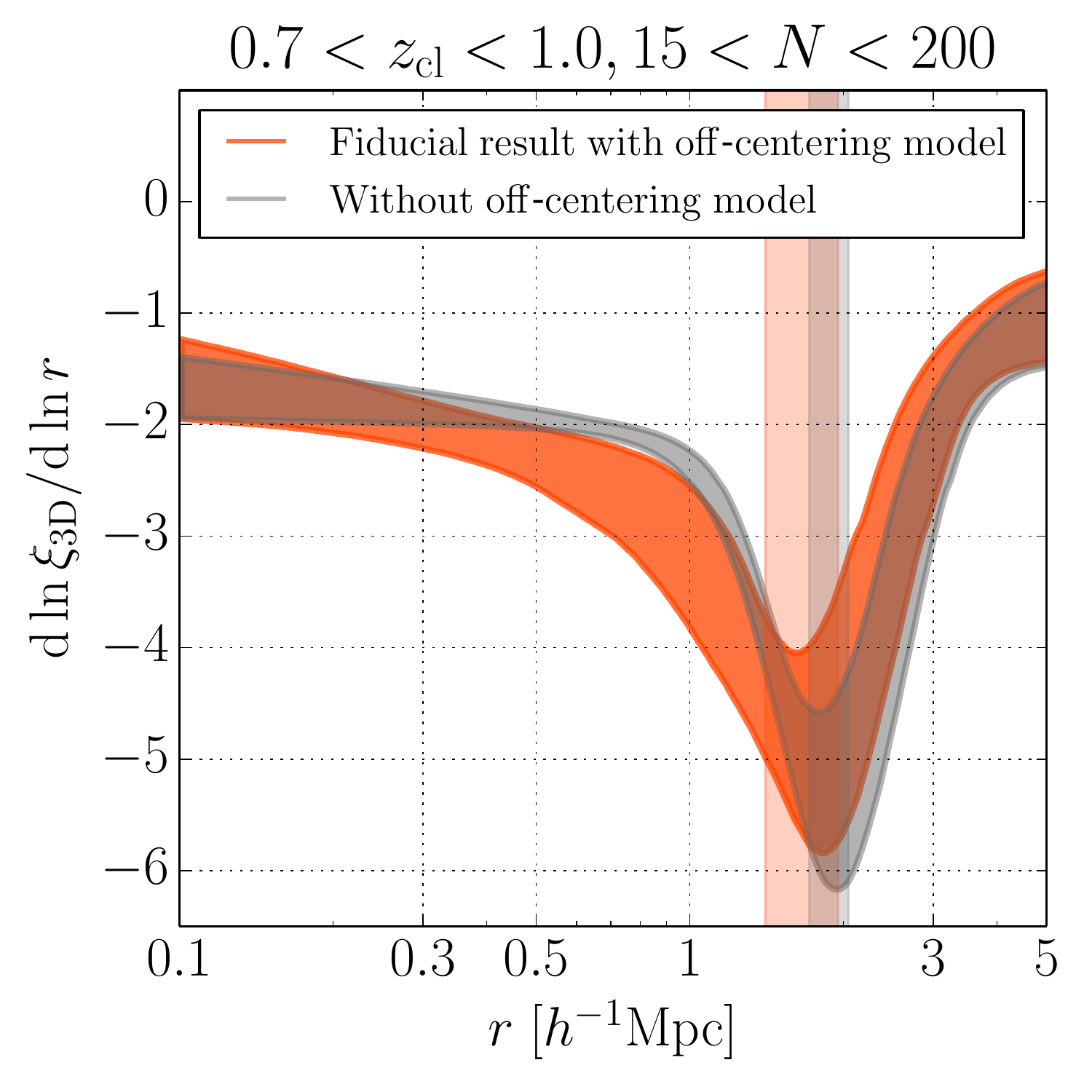}
    }
  \end{center}
  \caption{
    {\it Top panels:} We show how the off-centering effects change constraints on splashback features from Tables~\ref{tab:mainresults} and
    \ref{tab:withoutoffcentering} 
    for the Low-$z$ cluster sample with smaller off-centering effects and the High-$z$ cluster sample with larger off-centering effects
    with the fiducial absolute magnitude cut $M_{z}-5 \log_{10} h< -18.8$ for galaxy selections.
    We note that these fiducial results come from Figure~\ref{fig:mainresults_full_z}.
    Vertical shaded regions show the 16th and 84th percentiles of the model predictions on $r_{\rm sp}^{\rm 3D}$
    from the MCMC chains for each model, whereas
    shaded profile regions denote those on the logarithmical derivative profiles for the three-dimensional profile.
    {\it Bottom panels:} Same as the top panels, but for analyses with the red galaxy populations only 
    and the absolute magnitude cut $M_{z}-5 \log_{10} h< -19.3$ in Tables~\ref{tab:redblue} and \ref{tab:withoutoffcentering}.
    We also show these fiducial results in Figure~\ref{fig:fit_redblue}.
  }
\label{fig:offcenter}
\end{figure*}

\subsection{Potential biases from projection effects}\label{subsec:projeffect}
Optical cluster-finding algorithms such as CAMIRA and redMaPPer
could preferentially detect clusters with filamentary structure along the line-of-sight direction
as galaxies in the filament boosts observed richness.
Such projection effects
possibly cause some biases in a non-trivial way 
to observed splashback features based on the DK14 fitting 
in which the three-dimensional spherical symmetry is implicitly assumed,
since the optical cluster selections are based on cluster richness values.
Also, aperture sizes in optical cluster-finding algorithms could imprint 
some effects on the observed splashback features
since 
cluster finders 
preferentially select clusters
with higher galaxy density fluctuations within their aperture size,
which is comparable to the scale of splashback features in projected space \citep{BuschandWhite2017, Changetal2018, Sunayama&More2019}.
We investigate possible projection effects 
with mock observations
and simplified cluster-finding algorithms
for CAMIRA and redMaPPer cluster-finding algorithms (hereafter CAMIRA-like and redMaPPer-like, respectively),
partly following methods presented
in \cite{BuschandWhite2017} and \cite{Sunayama&More2019} as detailed below.

For this purpose, we use a suite of cosmological {\it N}-body simulations 
generated by the \textsc{Dark Quest} campaign \citep{Nishimichietal2019}
with a box size of $1h^{-1}{\rm Gpc}$ on a side and mass resolution of $\simeq 10^{10} h^{-1}M_{\odot}$
for the {\it Planck} cosmology at $z=0.25$ (see Appendix~\ref{app:modelpredictions} for more details).
We select one realization of the simulations at $z=0.25$ 
since this redshift is close to that for 
one of the most precise measurements in \cite{Moreetal2016} with the SDSS redMaPPer clusters, 
and projection effects are expected to be pronounced 
at lower redshift for a fixed cluster redshift error
due to the larger non-linear structure growth.
In particular, we employ halo catalogs based on \textsc{Rockstar} \citep{Behroozietal2013} 
to populate galaxies within halos,
and dark matter to measure stacked lensing profiles around clusters 
to check an approximate consistency with the observation
for the CAMIRA-like cluster finder \citep{Murataetal2019}
and the redMaPPer-like cluster finder \citep{Murataetal2018}.

We use an HOD model 
\citep{Jingetal1998, Peacock&Smith2000, Seljak2000, Scoccimarroetal2001, Zhengetal2005}
to populate red galaxies within the halos, 
which are used to select clusters by the simplified optical cluster-finding algorithms
and to measure projected cross-correlation functions with detected clusters to infer splashback features below.
Thus, these HOD galaxies correspond to the red galaxy populations analyzed in Section~\ref{sec:redblue} approximately.
We adopt the HOD model with central and satellite galaxies as
\begin{equation}
\langle N \rangle_M = \langle N_c \rangle_M + \langle N_s \rangle_M,
\end{equation}
where the mean halo occupation distribution for central galaxies is given by
\begin{equation}
\langle N_c \rangle_M = \frac{1}{2} \left[ 1 + {\rm erf}\left( \frac{ \log_{10}M_{\rm 200m} - \log_{10}M_{\rm min} }{ \sigma_{\log_{10}M }} \right) \right]
\end{equation}
and that for satellite galaxies is given by
\begin{equation}
\langle N_s \rangle_M = \langle N_c \rangle_M \left[ \frac{ M_{\rm 200m}-\kappa M_{\rm min}}{ M_1 } \right]^{\alpha}
\end{equation}
when $M_{\rm 200m}>\kappa M_{\rm min}$ and zero otherwise. 
Note that $\langle \cdots \rangle_M$ denotes averages for halos with mass $M(=M_{\rm 200m})$ and 
we use all halos with $M_{\rm 200m}>10^{12} h^{-1}M_{\odot}$ in the simulation. 
We assume that the distribution of central galaxies follows the Bernoulli distribution.
We populate satellite galaxies 
by the Poisson distribution to a halo only when a central galaxy exists.
Note that our HOD model depends only on halo masses and does not depend on accretion rates for halos.
For the model parameters in this HOD model for the CAMIRA-like finder, 
we employ 
$M_{\rm min}=10^{11.7}h^{-1}M_{\odot}$, 
$\sigma_{\log_{10}M}=0.1$, 
$\kappa=1$, 
$\alpha=1.15$, 
and 
$M_{1}=10^{12.96}h^{-1}M_{\odot}$,
so that the resultant cluster catalog after applying the CAMIRA-like cluster finder
approximately reproduces cluster abundance densities and lensing profiles 
in several richness bins that are similar to observations, as we will confirm below.
For redMaPPer-like cluster finders, we use
$M_{\rm min}=10^{11.7}h^{-1}M_{\odot}$,
$\sigma_{\log_{10}M}=0.1$,
$\kappa=1$,
$\alpha=1$,
and
$M_{1}=10^{12.84}h^{-1}M_{\odot}$.
With each set of HOD parameters, all the halos with $M_{\rm 200m}>10^{12}h^{-1}M_{\odot}$ have a central galaxy.
We set central HOD galaxies at identified central positions of each halo.
For the radial spatial distributions of satellite HOD galaxies around each halo, 
we employ the halo-matter cross-correlation function $\xi_{\rm hm}(r; M, z)$
from the \textsc{Dark Emulator} 
\citep[][see Appendix~\ref{app:modelpredictions} for more details]{Nishimichietal2019}, 
assuming a spherical symmetry around each halo for simplicity. 
We assign radial distances from each halo center 
to satellite HOD galaxies probabilistically,
following a probability distribution as $\propto \xi_{\rm hm}(r; M, z)$ within $r<r_{\rm sp}^{\rm 3D}(M,z)$
where $M$ is halo mass of each halo, $z=0.25$, and 
$r_{\rm sp}^{\rm 3D}(M,z)$ is splashback radius calculated from $\xi_{\rm hm}(r; M, z)$ in the emulator.
On the other hand, at each radius, we randomly distribute satellite HOD galaxies in the angular directions.
Since satellite galaxies are populated up to $r=r_{\rm sp}^{\rm 3D}(M, z)$ for each halo with a sharp galaxy density drop, 
we expect that the locations of splashback radius after stacking the satellite HOD galaxies around halos 
match the splashback radius in stacking profiles of $\xi_{\rm hm}(r; M,z)$ from the emulator over halos 
when we select halos by halo mass (i.e., without projection effects) as we will confirm below. 
Note that we do not account for redshift space distortion in our mock galaxy catalog since we do not assign velocities 
for our HOD galaxies.
The resultant galaxy catalogs have 
7392897 and 7824298
HOD galaxies in cluster selections and projected cross-correlation measurements
for CAMIRA-like and redMaPPer-like cluster finders, respectively.

Given the HOD galaxy catalog for cluster selections, 
we use a simplified mock cluster-finding algorithm for CAMIRA
accounting for cluster member probabilities, 
following a methodology presented in \cite{Sunayama&More2019} for redMaPPer.
We employ a top-hat filter along the line-of-sight direction
around each halo center in the simulation to select member galaxies 
and then estimate an optical richness for each cluster
to mimic photometric redshift uncertainties in observations,
following methods presented in \cite{BuschandWhite2017} and \cite{Sunayama&More2019},
since our mock galaxies do not have realistic galaxy color information 
(i.e., we instead assume all the HOD galaxies are red-sequence galaxies used for cluster selections).
We adopt an effective projection length of $d_{\rm eff}=40 h^{-1}{\rm Mpc}$,
based on a typical standard deviation of the differences
between cluster redshift and spectroscopic redshift of BCGs (when available) 
for the HSC CAMIRA clusters ($\equiv \Delta z_{\rm cl}$) 
through $d_{\rm eff} \simeq \Delta z_{\rm cl}/H(z)$,
which is roughly a constant value over $0.1 \le z_{\rm cl} \le 1.0$ 
within a precision of $\sim 5 h^{-1}{\rm Mpc}$ around $40~h^{-1}{\rm Mpc}$.
Note that we differentiate central and satellite galaxies in the mock galaxy catalog,
and we only consider central galaxies as potential cluster centers (i.e., we ignore off-centering effects for simplicity).
\begin{figure*}
  \begin{center}
    {
    \includegraphics[width=0.508 \textwidth]{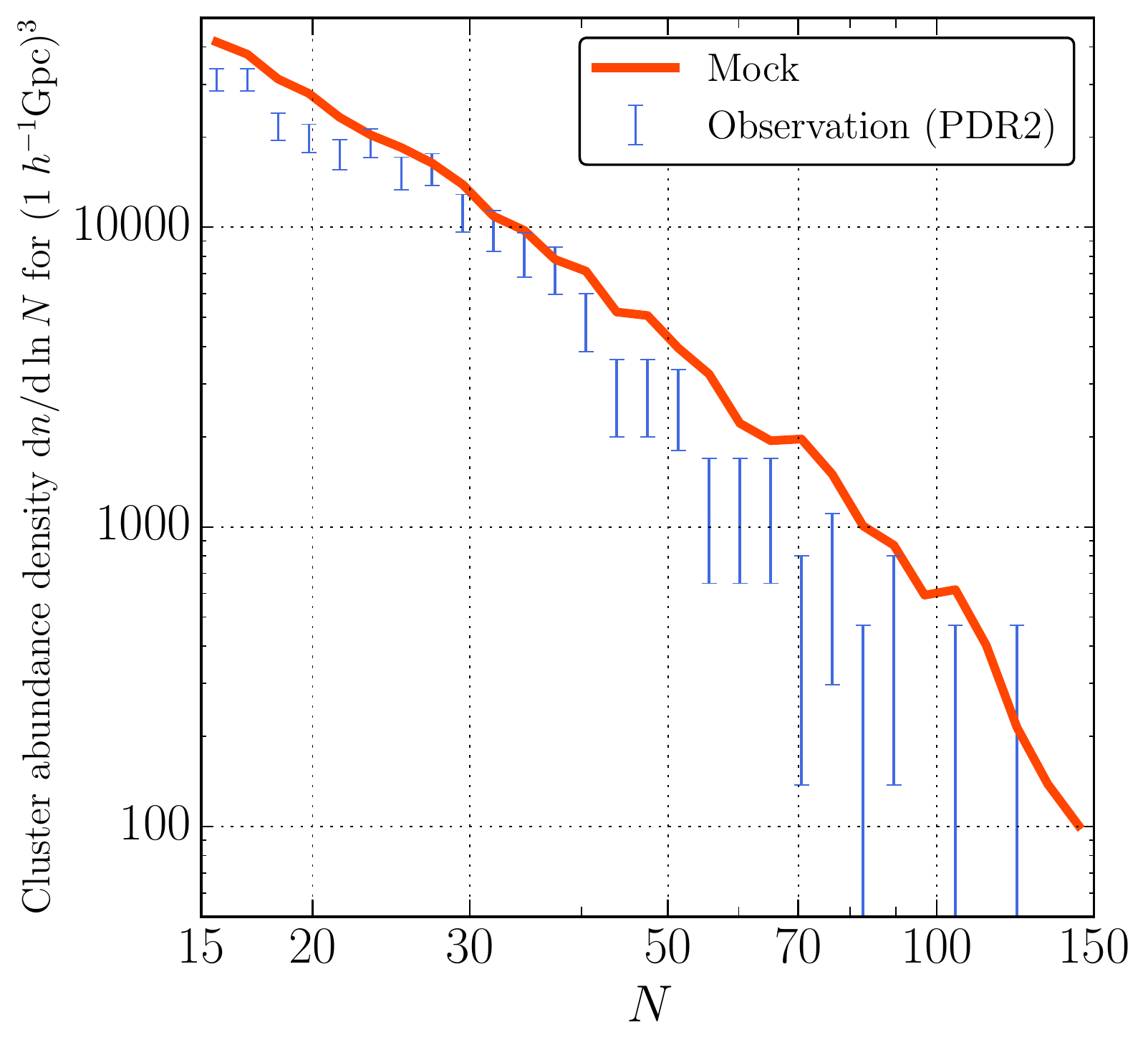}
    \includegraphics[width=0.475 \textwidth]{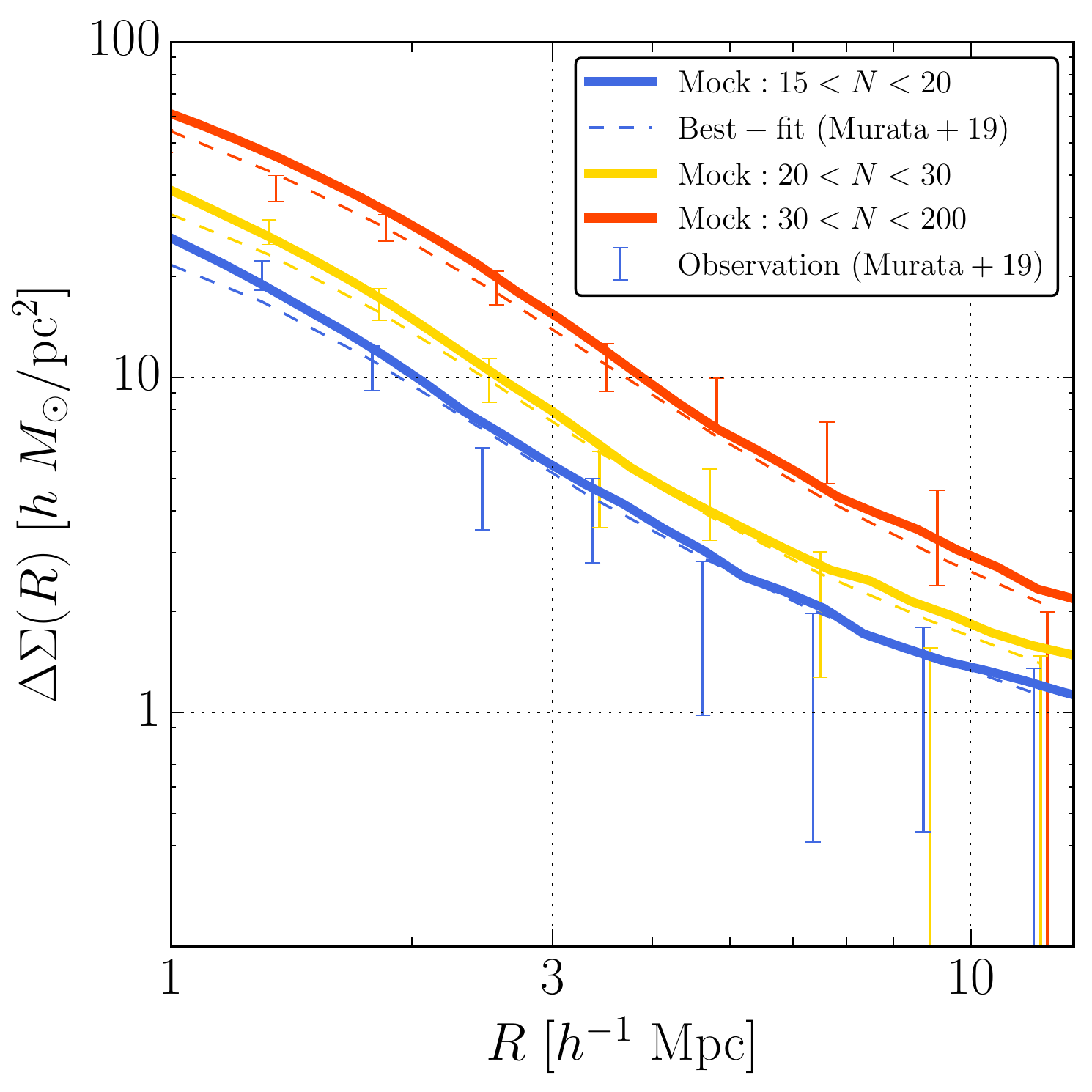}
    }
  \end{center}
  \caption{
    A consistency check for the CAMIRA-like cluster-finding algorithm in our mock catalog.
    {\it Left panel:} 
    We compare cluster abundance densities between the observation from the CAMIRA cluster catalog 
    based on PDR2 used for this paper and the mock catalog.
    Error bars for the observation is based on the Poisson errors for simplicity. 
    With the sample covariance contribution, the diagonal errors increase 
    by $\sim 1.3$ compared to the Poisson errors with a large positive correlation among different richness bins
    \citep[see Figures~1 and 2 in][for more details]{Murataetal2019}.
    Note that the volume of the mock catalog is $\sim 20$ times larger than that of the observation at $0.1<z_{\rm cl}<0.4$.
    {\it Right panel:} We compare lensing profiles from the observations at $0.1<z_{\rm cl}<0.4$ in \cite{Murataetal2019}
    and our mock cluster catalog in three richness bins. Solid lines show the lensing profiles from the mock catalog. 
    Dashed lines denote the model predictions at the best-fit parameters in \cite{Murataetal2019}, 
    and points with error-bars are the observed lensing measurements from data.
    Note that there are large positive correlations of errors at larger radii, as shown in Figure~2 in \cite{Murataetal2019}, 
    and that the model and data are affected by the off-centering effects at $R \lesssim 1.5h^{-1}{\rm Mpc}$ \citep{Murataetal2019}.
  }
\label{fig:mock_abundance_lens}
\end{figure*}

We start our percolation of galaxy clusters by making a ranked list of all the central 
galaxies according to their host halo masses in a descending order.
With this ranked list, we compute richness $N$ from a more massive halo as follows
based on a radial spatial filter $F_{R}(R)$ in equation~(9) of \cite{Oguri2014}
for each candidate central galaxy by taking all the mock galaxies 
(both central and satellite galaxies) within $|d| < d_{\rm eff}$, 
where $|d|$ is a distance along the line-of-sight direction from each halo center.
Note that we assign $p_{\rm free}=1$ 
as priors that a galaxy does not belong to any other more massive galaxy cluster
to mitigate double counting of galaxies in richness estimations
for all the HOD galaxies before staring the percolation process.
The membership probability for a galaxy, $g$, with projected distance $R$ from each halo
is given by
\begin{eqnarray}
p_{ {\rm mem}, g}(R) &=&  F_{R}(R) \nonumber \\
                     &\propto& \Gamma[2, (R/R_0)^2] -(R/R_0)^4 \exp{ \left( -(R/R_0)^2 \right) }
\end{eqnarray}
with $R_0=0.8~h^{-1}{\rm Mpc}$ in physical coordinate and $F_{R}(R=0)=1$, 
where we only include the radial spatial filter $F_{R}(R)$ in \cite{Oguri2014}
since for simplicity we assume all the HOD galaxies are red-sequence galaxies used for cluster selections.
As shown in Figure~2 of \cite{Oguri2014}, a typical aperture size for $F_{R}(R)$ is 
$\sim 1~h^{-1}{\rm Mpc}$ in physical coordinates, and the radial filter subtracts background galaxy level locally 
with estimates from $1~h^{-1}{\rm Mpc} \lesssim R \lesssim 2.5~h^{-1}{\rm Mpc}$ in physical coordinates
to account for a local background level from large-scale structure.
The resulting richness value for each halo is calculated from all the galaxies within $|d| < d_{\rm eff}$ as
\begin{equation}
\widehat{N} = \sum_{g} p_{ {\rm free}, g} p_{ {\rm mem}, g}(R),
\label{eq:lambda}
\end{equation}
and we then update $p_{ {\rm free}, g}$ 
with $p_{ {\rm free}, g} (1-p_{ {\rm mem}, g})$ when $p_{ {\rm mem}, g}>0$ for all the galaxies
to mitigate double counting of galaxies similarly to \cite{Oguri2014}
for subsequent less massive halos in the ranked list during the percolation process.
We estimate richness values for all the halos in the ranked list only once for the CAMIRA-like cluster-finder
since its aperture size does not depend on richness, unlike redMaPPer.
Note that we skip richness estimation for a halo when there is a more massive halo within 
a projected distance of $0.8~h^{-1}{\rm Mpc}$ in physical coordinates.

We also employ the redMaPPer-like cluster-finding algorithm with the different HOD galaxy catalog.
We refer its algorithm
to Section~3.2 in \cite{Sunayama&More2019} for more details, which is similar to that for the CAMIRA-like above.
One difference from \cite{Sunayama&More2019} is that
we use the projection length of $d_{\rm eff}=40~h^{-1}{\rm Mpc}$ following the CAMIRA-like algorithm.
We only consider central galaxies as cluster centers, and
we start our percolation by rank ordering central galaxies
by their halo masses in a descending order
to assign richness values in the first step, as described in \cite{Sunayama&More2019}.
After the first richness assignments,
we repeat richness estimations a few times to converge
since the aperture size depends on richness values $\lambda$ for redMaPPer\footnote{We use $\lambda$ to denote richness for the redMaPPer-like cluster-finding algorithms.},
as $R_c(\lambda)=1.0 \times (\lambda/100)^{0.2}~h^{-1}{\rm Mpc}$ in physical coordinates,
where its spatial filter is based on the normalized projected NFW profile and an error function with a smooth truncation at $R=R_c$.
Note that the aperture size of redMaPPer is smaller than that of CAMIRA at $\lambda<100$.
Furthermore, redMaPPer estimates and subtracts the background galaxy level in cluster selections
assuming a constant value (i.e., a global value) in projected spaces,
whereas CAMIRA does it locally to account for large-scale structure as described above.
In order to investigate aperture size effects on splashback features below,
we also apply a redMaPPer-like algorithm with a larger aperture size of
$1.35 \times (\lambda/100)^{0.2}~h^{-1}{\rm Mpc}$ in physical coordinates.
In this case, the aperture size at $\lambda=20$ is comparable to that for CAMIRA.

\begin{figure*}
  \begin{center}
    {
    \includegraphics[width=0.508 \textwidth]{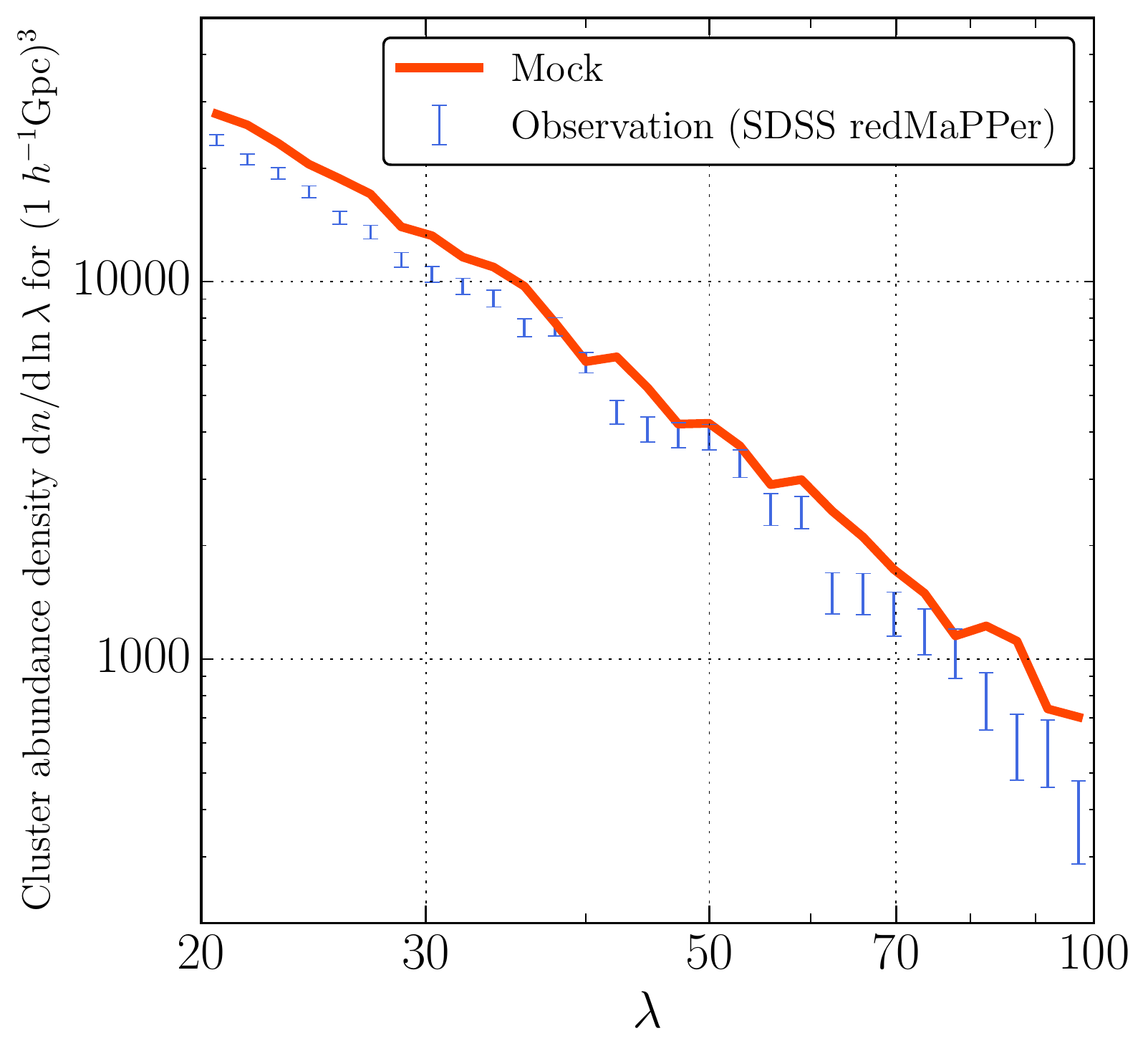}
    \includegraphics[width=0.475 \textwidth]{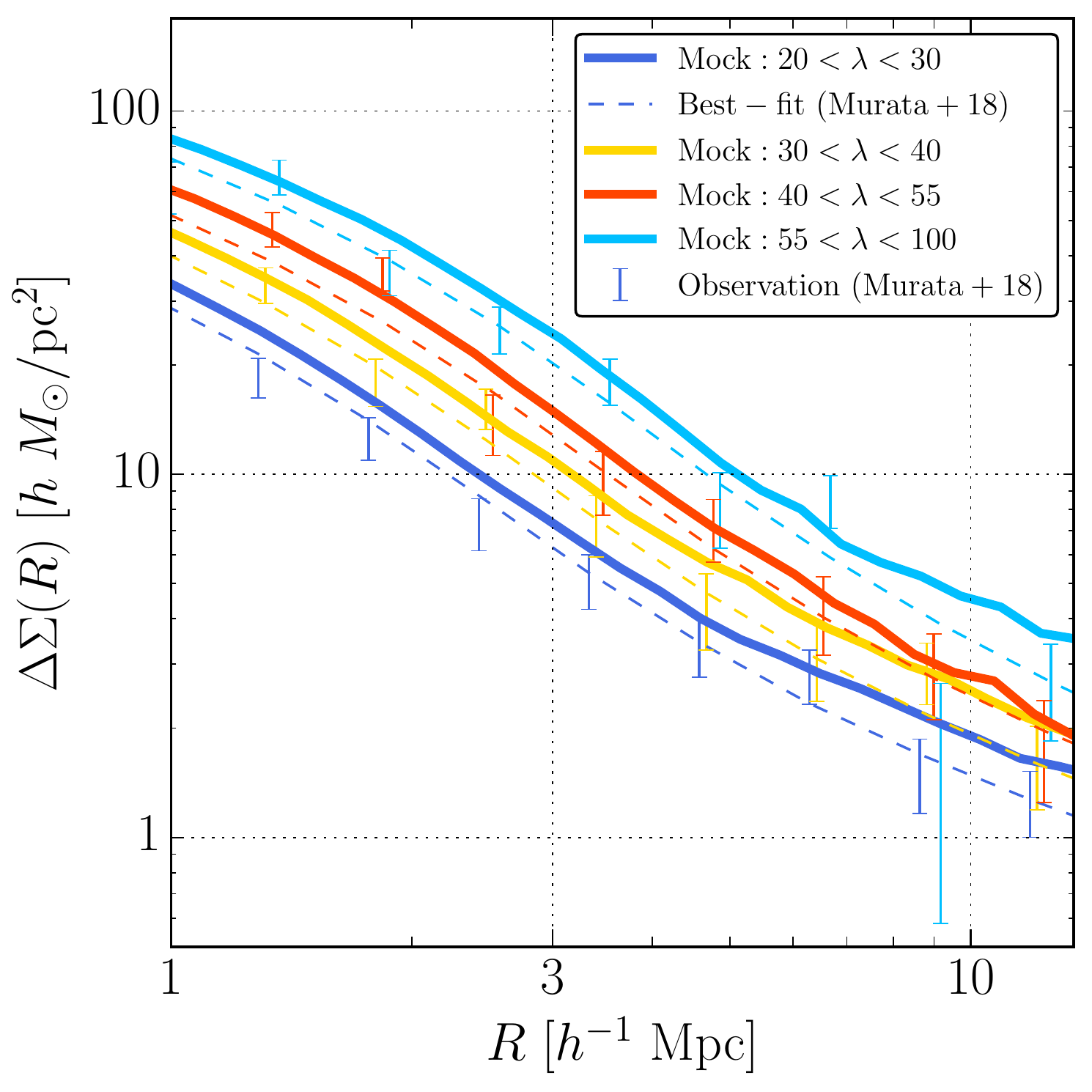}
    }
  \end{center}
  \caption{
    Same as Figure~\ref{fig:mock_abundance_lens},
    but for the redMaPPer-like cluster-finding algorithm with the fiducial aperture size in our mock catalog.
    We use observational measurements of the abundance and lensing profiles
    for the SDSS redMaPPer clusters at $0.1<z_{\rm cl}<0.33$ in \cite{Murataetal2018}.
    We also use the best-fit model for the lensing profiles in \cite{Murataetal2018} in the right panel.
  }
\label{fig:mock_abundance_lens_redmapper}
\end{figure*}

For a consistency check on the setups of the HOD galaxy catalogs and the simplified cluster-finding algorithms,
we investigate 
whether cluster abundance densities and lensing profiles from the dark matter around the clusters
match approximately 
to the observations.
For the CAMIRA-like cluster-finding algorithm, we employ the abundance measurements
from the CAMIRA catalog based on PDR2 in Section~\ref{sec:data:cluster} at $0.1<z_{\rm cl}<0.4$
with the corresponding survey volume, 
and the lensing measurements in \cite{Murataetal2019} for the Low-$z$ cluster sample ($0.1<z_{\rm cl}<0.4$) 
in several richness bins.
In Figure~\ref{fig:mock_abundance_lens},
we show that the cluster abundance density from our mock catalog and simplified cluster-finder 
is larger than observations, but is consistent within a $\sim 30\%$ level 
especially for a large number of low-richness clusters.
We note that 
observed abundances in different richness bins are positively correlated
especially at lower-richness bins
\citep[see Figures~1 and 2 in][for more details]{Murataetal2019}, 
and that these levels of matching is much better than 
those from the Millennium Simulation 
in \cite{BuschandWhite2017} and \cite{Sunayama&More2019},
where their cluster abundance densities are $\sim 3$ times larger than observations.
The overestimation of cluster number is due to different ways to populate red galaxies.
We also show in Figure~\ref{fig:mock_abundance_lens}
that the lensing profiles from our mock catalog
at $15<N<20$, $20<N<30$, and $30<N<200$ 
reproduce the observed lensing profiles in \cite{Murataetal2019} for the Low-$z$ cluster sample 
quite well within the error-bars in observations. 
They are also consistent with the best-fit model in \cite{Murataetal2019} at a $\sim 10\%$ level.
In particular, 
a mean mass averaged over 
halos with identified central galaxies at $15<N<200$ in the mock catalog for the CAMIRA-like finder
is 
$1.8 \times 10^{14} h^{-1}M_{\odot}$,
which is consistent with the one 
in Table~\ref{tab:sampleselection} 
within the $1\sigma$ uncertainty
for the Low-$z$ cluster sample
derived from the mass-richness relation in \cite{Murataetal2019}. 
The total number of clusters in the simulation for the CAMIRA-like
is 23486 at $15<N<200$.
For the redMaPPer-like cluster-finding algorithm with the fiducial aperture size, 
we also check a consistency with the observation of the SDSS redMaPPer at $0.1<z_{\rm cl}<0.33$ in \cite{Murataetal2018}.
In Figure~\ref{fig:mock_abundance_lens_redmapper} we show the comparison, suggesting that the cluster abundance density and lensing profiles 
for the redMaPPer-like cluster-finder also match to the observation approximately.
\begin{figure*}
  \begin{center}
    {
    \includegraphics[width=0.495 \textwidth]{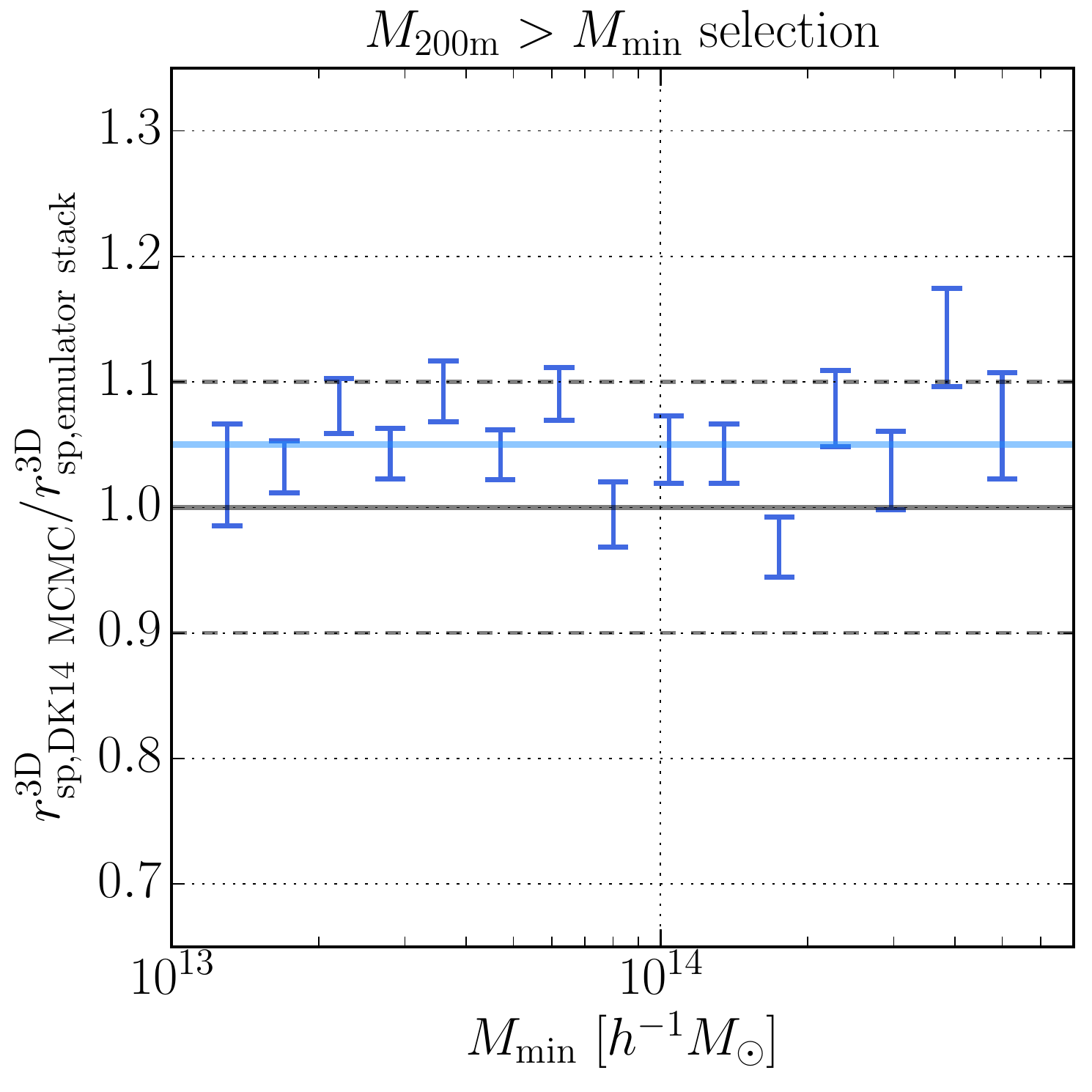}
    \includegraphics[width=0.495 \textwidth]{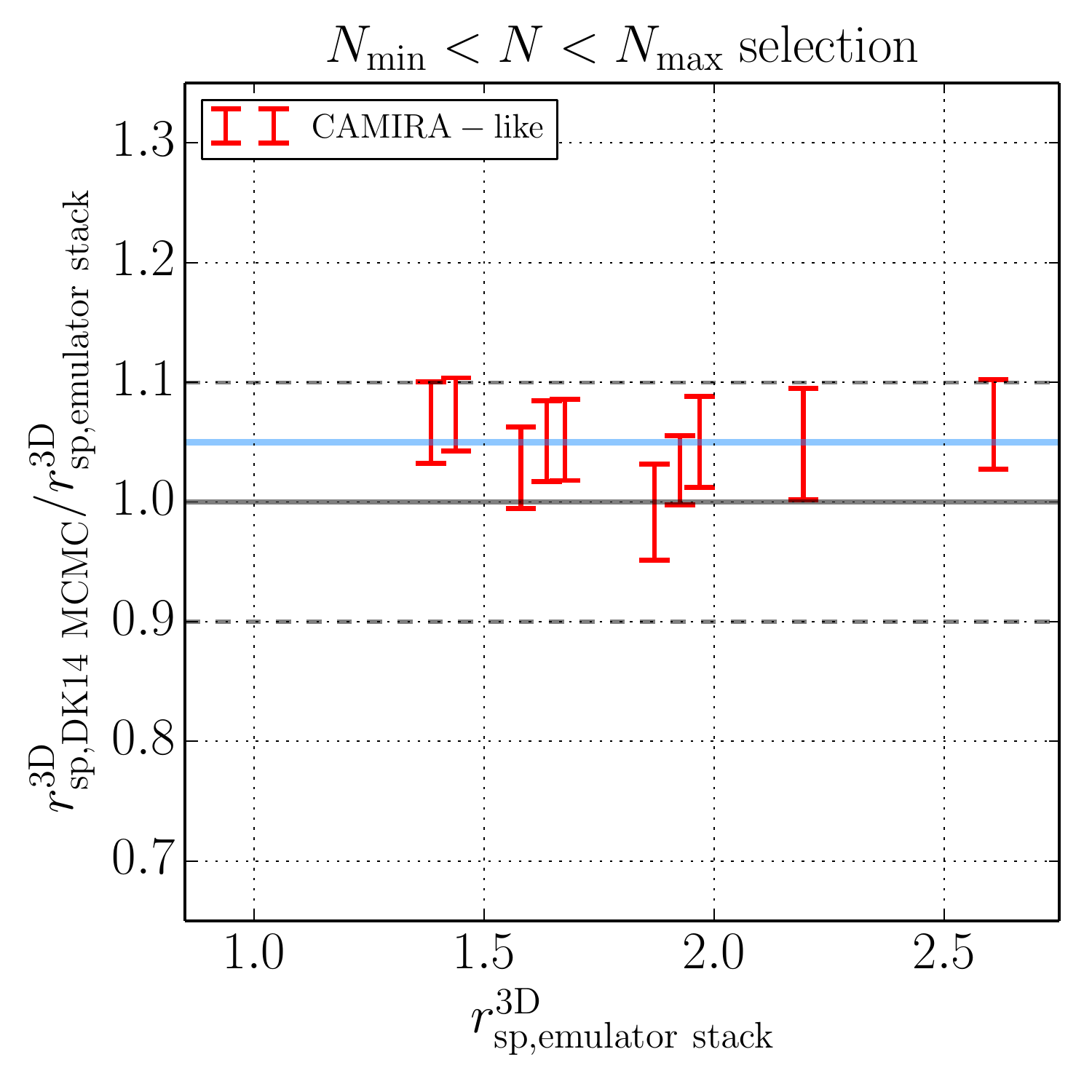}
    }
  \end{center}
  \caption{
    We show results for the CAMIRA-like cluster-finding algorithm in our mock catalog.
    {\it Left panel:} We show a comparison of estimated locations of $r_{\rm sp}^{\rm 3D}$ based on the DK14 profile fitting
    assuming the spherical symmetry
    to the projected cross-correlation functions between halos and the HOD galaxies in the simulation,
    and those from the emulator stack when selecting halos based on halo masses
    (i.e., without projection effects related to richness values).
    As expected from the construction of the HOD galaxy catalog, the two values match quite well
    (see texts for more details).
    The blue horizontal line denotes the average value of the ratio, $1.05$.
    We estimate the error-bars from the MCMC chains with half widths of the $68\%$ percentile region.
    {\it Right panel:}
    Similarly to the left panel, we compare the observed splashback radii from the DK14 fitting to the projected cross-correlation functions
    with those from the emulator stack,
    but here we select halos based on richness values with projection effects.
    We use ten richness bins with bin edges of $N=[15.0, 17.6, 20.6, 24.2, 28.4, 33.3, 39.1, 45.9, 53.8, 74.0, 200.0]$.
    The constraint for the CAMIRA-like finder is $r_{\rm sp, DK14\ MCMC}^{\rm 3D}/r_{\rm sp, emulator\ stack}^{\rm 3D} = 1.06 \pm 0.03 $ 
    for the sample of $15<N<200$, which
    is consistent with $1.05$ from the mass selection in the left panel without projection effects,
    and its uncertainty is smaller than our statistical errors in the HSC data analysis. 
    Note that we estimate this ratio by repeating the procedure for the cluster sample of $15<N<200$.
  }
\label{fig:mock_mainresults}
\end{figure*}
\begin{figure*}
  \begin{center}
    {
    \includegraphics[width=0.495 \textwidth]{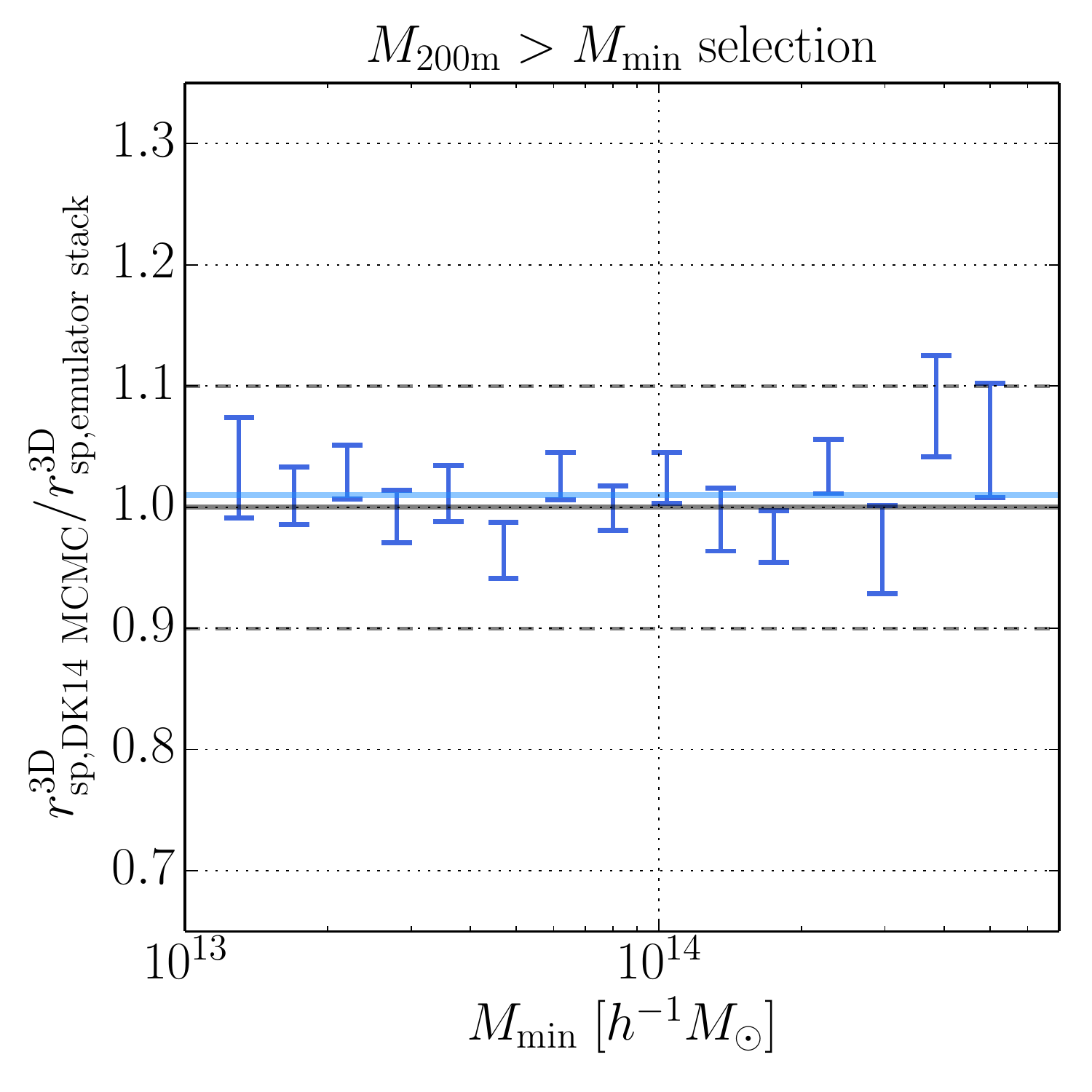}
    \includegraphics[width=0.495 \textwidth]{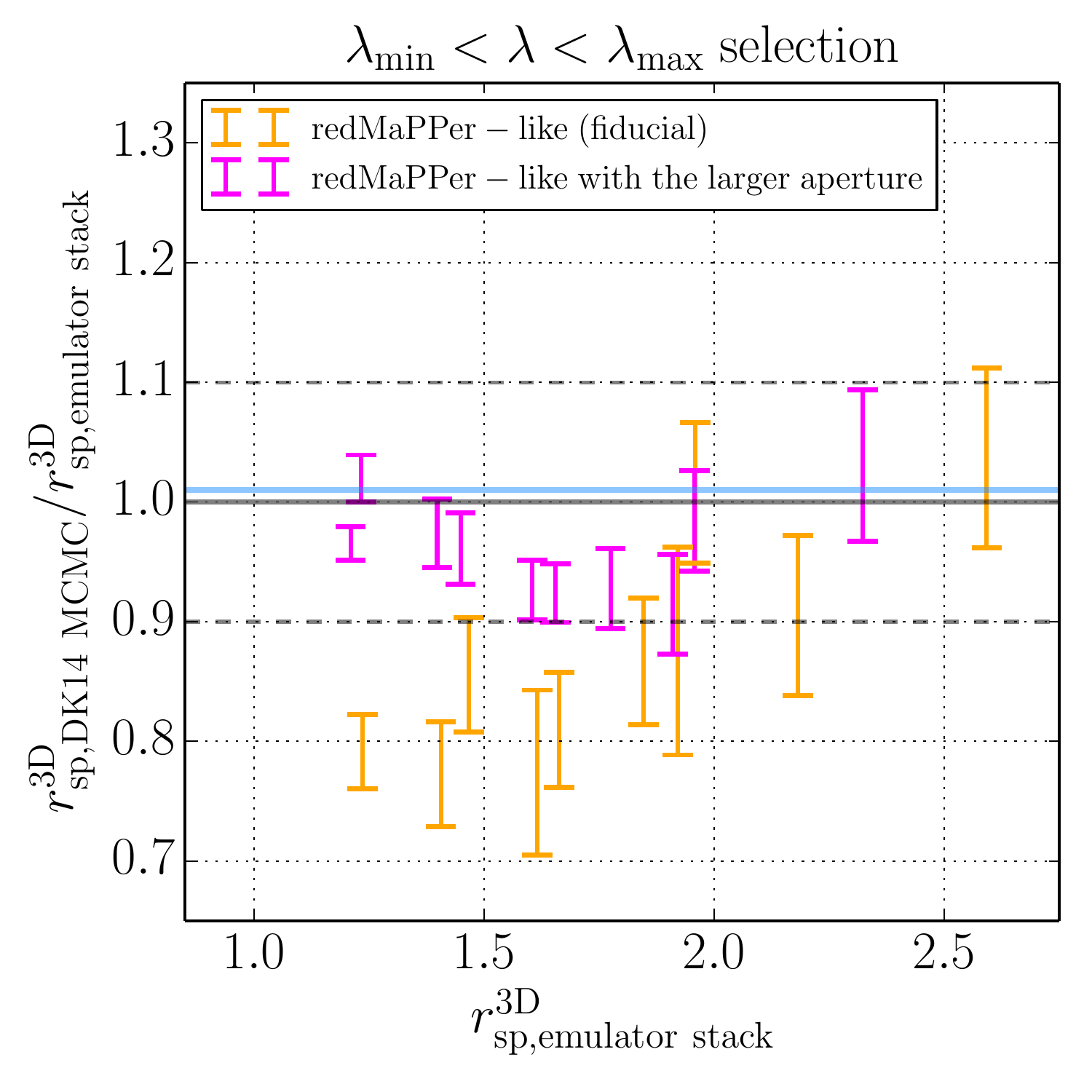}
    }
  \end{center}
  \caption{
    Same as Figure~\ref{fig:mock_mainresults},
    but for the redMaPPer-like cluster-finding algorithms with the fiducial and larger aperture sizes 
    in our mock catalog.
    The blue horizontal line denotes the average value of the ratio in the left panel, $1.01$.
    We use the same richness bins as Figure~\ref{fig:mock_mainresults} 
    in the right panel.
    The constraint for the redMaPPer-like with the fiducial aperture size
    is $r_{\rm sp, DK14\ MCMC}^{\rm 3D}/r_{\rm sp, emulator\ stack}^{\rm 3D} = 0.86 \pm 0.03 $ for the sample of $20<\lambda<100$,
    which is smaller than $1.01$ from the mass cuts by $\sim 15\%$ and is consistent with the trend of the smaller value
    in observation \citep{Moreetal2016, Changetal2018}. 
    The biases are smaller at higher richness bins, which is also consistent with the literature \citep{Shinetal2019}.
    Specifically, the constraints for the samples of $15<\lambda<40$ and $40<\lambda<100$ are
    $r_{\rm sp, DK14\ MCMC}^{\rm 3D}/r_{\rm sp, emulator\ stack}^{\rm 3D} = 0.72 \pm 0.05$ and
    $r_{\rm sp, DK14\ MCMC}^{\rm 3D}/r_{\rm sp, emulator\ stack}^{\rm 3D} = 0.91 \pm 0.04$, respectively.
    In contrast, with the redMaPPer-like with the larger aperture ($\times 1.35$), the biases
    are significantly reduced even at lower richness bins within a $\sim 5 \%$ level compared to the mass cuts.
  }
\label{fig:mock_mainresults_redmapper}
\end{figure*}

Given the cluster catalog with halo mass and richness values and the HOD galaxy catalog,
we conduct mock observations with the DK14 fitting
and mock model calculations with the emulator
by closely following our procedure in the data analysis with real data 
to investigate 
whether these two values match within statistical errors for each mock cluster-finding algorithm,
or not due to the projection effects.
First, we estimate the splashback radius in the three-dimensional space $r_{\rm sp}^{\rm 3D}$ 
from the projected cross-correlation functions in the mock catalogs.
We compute these projected cross-correlation functions 
for each cluster sample after selection and the HOD galaxy catalog,
where we use the maximum integral scale along the line-of-sight direction
of $100~h^{-1}{\rm Mpc}$.
We use $64$ jackknife regions in order to compute the covariance in these measurements.
We then repeat the MCMC fitting procedure with the DK14 profile assuming the spherical symmetry
as in Section~\ref{sec:model} to the mock projected cross-correlation functions
in order to estimate $r_{\rm sp}^{\rm 3D}(\equiv r_{\rm sp, DK14\ MCMC}^{\rm 3D})$.  
Note that we do not include the off-centering model for simplicity
since our mock cluster-finding algorithms do not account for the off-centering effects.
Second, we calculate the mock model calculations with the emulator following Appendix~\ref{app:modelpredictions}.
For each cluster sample after selection, 
we calculate the average profiles of $\xi_{\rm hm}(r; M, z)$ at $z=0.25$ 
by stacking over the halo masses for massive main halos with identified central galaxies for each cluster sample 
similarly to Appendix~\ref{app:modelpredictions} to calculate 
the splashback radius as model predictions 
$(\equiv r_{\rm sp, emulator\ stack}^{\rm 3D})$.
These model predictions in the mock catalogs 
correspond to those for the real data with the observed mass-richness relation used in Appendix~\ref{app:modelpredictions},
as long as the observed mean mass given a fixed richness value
via the mass-richness relation from weak lensing measurements 
is sufficiently close to 
true masses of such massive main halos.
As discussed in \cite{Shinetal2019},
the location of splashback radius scales as $\sim M^{1/3}$ 
and thus we require a roughly $30 \%$ biases in the observed mass-richness relation 
compared to the massive main halo masses 
to explain a $10\%$ deviation in model predictions for $r_{\rm sp}^{\rm 3D}$, but
this level of mass bias would be unlikely
\citep{Sunayamaetal2020}.
Thus, our mock model predictions closely resemble the procedure for the real data in Appendix~\ref{app:modelpredictions}.
These procedures in the mock catalog are different from those in \cite{Sunayama&More2019} 
with the Millennium Simulation, 
where the Abel transformation is employed to evaluate the asymmetry of galaxies around clusters
after richness selections in projected spaces.
We instead investigate the biases in $r_{\rm sp}^{\rm 3D}$ 
by following the procedures for the real data more directly.

We first show comparisons for cluster samples with mass threshold selections as $M_{\rm 200m}>M_{\rm min}$. 
These selections are not related to richness values, 
and hence this comparison shows the difference between 
$r_{\rm sp, DK14\ MCMC}^{\rm 3D}$
and $r_{\rm sp, emulator\ stack}^{\rm 3D}$ for cases without projection effects. 
We expect that these values match quite well since
we construct the HOD galaxy catalog based on $\xi_{\rm hm}(r; M, z)$ 
for the radial distribution of satellite galaxies
in $r<r_{\rm sp}^{\rm 3D}$
with a sharp drop at $r=r_{\rm sp}^{\rm 3D}$.
For the HOD catalog with the CAMIRA-like finder,
we find that $r_{\rm sp, DK14\ MCMC}^{\rm 3D}/r_{\rm sp, emulator\ stack}^{\rm 3D} \simeq 1.05$
over a wide mass range above $M_{\rm 200m}=10^{13} h^{-1}M_{\odot}$
as shown in the left panel of Figure~\ref{fig:mock_mainresults}.
The values for $r_{\rm sp, DK14\ MCMC}^{\rm 3D}$ are slightly larger than 
$r_{\rm sp, emulator\ stack}^{\rm 3D}$
since our HOD model catalog employs $\alpha=1.15$ 
for the mass dependence in the number of HOD galaxies 
as $\propto M_{\rm 200m}^{\alpha}$
with a larger weight on more massive halos,
whereas the approximately corresponding number for the emulator stack is
$\int_{0}^{ r_{\rm sp}^{\rm 3D} } r^2 {\rm d} r \xi_{\rm hm}(r; M_{\rm 200m}, z)$, 
which is roughly propotional to $M_{\rm 200m}$.
For the different HOD catalog with the redMaPPer-like cluster finders,
we find that $r_{\rm sp, DK14\ MCMC}^{\rm 3D}/r_{\rm sp, emulator\ stack}^{\rm 3D} \simeq 1.01$
over the wide mass range in the left panel of Figure~\ref{fig:mock_mainresults_redmapper},
since this HOD model catalog instead employs $\alpha=1$.
We use $r_{\rm sp, DK14\ MCMC}^{\rm 3D}/r_{\rm sp, emulator\ stack}^{\rm 3D} = 1.05\ {\rm or}\ 1.01$
for the CAMIRA-like and redMaPPer-like cluster finders, respectively, 
as a baseline when we discuss the significance of projection effects in comparisons
between $r_{\rm sp, DK14\ MCMC}^{\rm 3D}$ and $r_{\rm sp, emulator\ stack}^{\rm 3D}$ below.

\begin{figure*}
  \begin{center}
    {
      \includegraphics[width=0.995\textwidth]{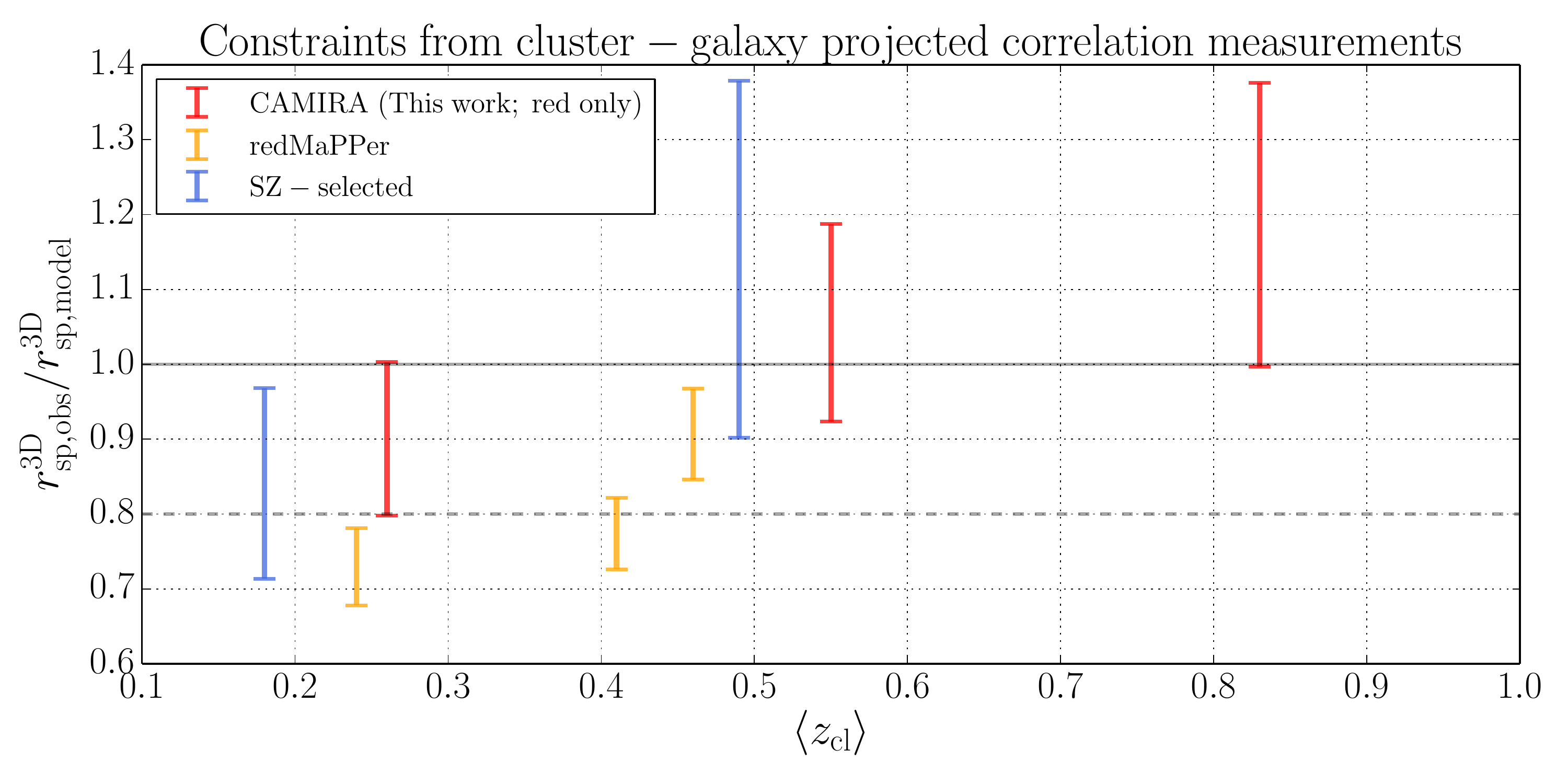}
    }
  \end{center}
  \caption{
    We show a comparison of ratios between
    the observed $r_{\rm sp}^{\rm 3D}$ and its model prediction for fiducial results
    among the literature with our results as a function of mean redshift in cluster samples.
    We refer to table~3 in \cite{Shinetal2019} as a summary table.
    Red error-bars show our results with the CAMIRA clusters
    from the red galaxy populations only in Section~\ref{sec:redblue}
    for the Low-$z$, Mid-$z$, and High-$z$ cluster samples.
    We employ median values with errors calculated from a half width of the $68\%$ percentile region in Table~\ref{tab:redblue}
    for simplicity.
    Orange error-bars show constraints with the redMaPPer clusters
    from SDSS \citep{Moreetal2016} and
    DES \citep{Changetal2018, Shinetal2019}, from low to high-redshift values.
    The average mass scale for \cite{Moreetal2016} and \cite{Changetal2018} is comparable to that for our constraints in red color,
    but the constraint in \cite{Shinetal2019} is based on a more massive mass scale
    with $r_{\rm sp, model}^{\rm 3D}=1.60, 1.46, 2.07\ h^{-1}{\rm Mpc}$ for these constraints, respectively.
    Blue error-bars denote constraints with clusters selected by the Sunyaev–Zel’dovich (SZ) effects
    in the {\it Planck} data \citep{Zurcher&More2019}
    and the South Pole Telescope data \citep{Shinetal2019} from low to high-redshift values.
    The horizontal dashed line shows the $20\%$ smaller values compared to the model predictions, as suggested by the literature.
  }
  \label{fig:summaryplot}
\end{figure*}

Next, we show comparisons when selecting the halos with richness values for each cluster-finding algorithm
in the right panel of Figures~\ref{fig:mock_mainresults} or \ref{fig:mock_mainresults_redmapper}
to investigate projection effects on the splashback radius.
For the CAMIRA-like cluster-finding algorithm, 
we find that $r_{\rm sp, DK14\ MCMC}^{\rm 3D}/r_{\rm sp, emulator\ stack}^{\rm 3D}$ is 
consistent with that (i.e., 1.05) 
without projection effects at all richness bins above $N=15$
at a precision 
of $5\%$, which is well below our statistical errors.
In particular, 
the constraint for the sample of $15<N<200$ 
is $r_{\rm sp, DK14\ MCMC}^{\rm 3D}/r_{\rm sp, emulator\ stack}^{\rm 3D} = 1.06 \pm 0.03 $ 
with $r_{\rm sp, emulator\ stack}^{\rm 3D} = 1.6~h^{-1}{\rm Mpc}$.
This result suggests that the biases of projection effects on $r_{\rm sp}^{\rm 3D}$ 
estimated from the DK14 fitting should be negligible for CAMIRA in our HSC data analysis.
Since CAMIRA employs the constant aperture size independent of richness values, 
our mock analysis confirms that we indeed detect the physical splashback radii
of halos rather than
artifact due to optical cluster finding that selects overdensities of galaxies within a given radius.

In the right panel of Figure~\ref{fig:mock_mainresults_redmapper},
we also show the results for the redMaPPer-like algorithms.
For the redMaPPer-like finder with the fiducial aperture size,
we find that the observed splashback radii from the DK14 fitting
are smaller than the model predictions,
especially at lower richness values.
Specifically,
the constraint for the sample of $20<\lambda<100$
is $r_{\rm sp, DK14\ MCMC}^{\rm 3D}/r_{\rm sp, emulator\ stack}^{\rm 3D} = 0.86 \pm 0.03$
with $r_{\rm sp, emulator\ stack}^{\rm 3D} = 1.7~h^{-1}{\rm Mpc}$,
suggesting a $\sim 15\%$ bias compared to that without projection effects (i.e., $1.01$).
This trend is consistent
with the results of the smaller observed splashback radii
in the literature with redMaPPer clusters
\citep{Moreetal2016, Changetal2018, Shinetal2019}.
In particular, the mass scales for \cite{Moreetal2016} and \cite{Changetal2018} are
comparable to that for $20<\lambda<100$ above, whereas
the mass scale for \cite{Shinetal2019} is more massive with
$r_{\rm sp, model}^{\rm 3D} \simeq 2.1~h^{-1}{\rm Mpc}$.
As shown in Figure~\ref{fig:summaryplot},
the result of \cite{Shinetal2019} is more consistent with their model prediction,
which is also consistent
the right panel of Figure~\ref{fig:mock_mainresults_redmapper}
with the smaller biases at higher richness values.

In addition, we show that
the biases are significantly reduced even at lower richness bins
for the redMaPPer-like finder with the larger ($\times 1.35$) aperture size
as
$r_{\rm sp, DK14\ MCMC}^{\rm 3D}/r_{\rm sp, emulator\ stack}^{\rm 3D} \simeq 0.95$.
These results suggest
that the smaller aperture size in the fiducial redMaPPer-like finder
than the splashback feature scale of $R_{\rm sp, model}^{\rm 2D}\simeq 1.1~h^{-1}{\rm Mpc}$ in comoving coordinates
(see Table~\ref{tab:sampleselection}) induces selection bias effects on the observed splashback radii.
Specifically, the fiducial redMaPPer finder uses
$R=0.72~h^{-1}{\rm Mpc}$ at $\lambda=20$ in physical coordinates
(i.e., $R=0.90~h^{-1}{\rm Mpc}$ in comoving coordinates at $z=0.25$)
which is smaller than the scale of $R_{\rm sp, model}^{\rm 2D}$ above.
On the other hand,
the redMaPPer with the larger aperture at $\lambda=20$
and CAMIRA with the smaller biases on the observed splashback radii
employs $R=1.0~h^{-1}{\rm Mpc}$ in physical coordinates
($R=1.25~h^{-1}{\rm Mpc}$ in comoving coordinates at $z=0.25$),
which is larger than the scale of $R_{\rm sp, model}^{\rm 2D}$.
The biases for the fiducial redMaPPer-like finder are reduced at larger richness values
in Figure~\ref{fig:mock_mainresults_redmapper},
probably because its aperture size increases as $\propto \lambda^{0.2}$
with $R = R_{\rm sp, model}^{\rm 2D}$ around $\lambda=50$ at $z=0.25$.
However,
$r_{\rm sp, DK14\ MCMC}^{\rm 3D}/r_{\rm sp, emulator\ stack}^{\rm 3D}$
for the redMaPPer-like finder with the larger aperture
is smaller than those for the mass cuts by a $\sim 5 \%$ level,
which could be due to the difference between their radial filters.
In particular, the CAMIRA finder employs the local background subtraction
in richness estimations to account for large-scale structure contributions,
which might provide another reason for the smaller projection effects on splashback radii.

It is important to confirm these results 
with more realistic mock galaxy populations with galaxy colors by applying more realistic cluster-finding algorithms for them
when available,
since we here empirically employ the HOD model and the simplified cluster-finding algorithms.
Nevertheless,
our results are informative to 
clarify the different effects of CAMIRA and redMaPPer on splashback radii
at least qualitatively.
It is also interesting to
directly compare splashback features for these two
cluster-finding algorithms by using the same datasets
for clusters, galaxies, and weak lensing calibrations as a fairer comparison.
%
\section{Conclusion} \label{sec:conclusion}
In this paper, 
we present the results of analyses on the splashback feature for 3316 HSC CAMIRA clusters
in a wide redshift range ($0.1 < z_{\rm cl} < 1.0$) with a richness range of $N>15$ 
by fitting with the DK14 model profile 
to projected cross-correlation measurements 
between the clusters and photometric galaxies from the HSC-SSP second public data release 
($\sim 427~{\rm deg}^2$ for the cluster catalog).
Compared to previous results using optically-selected clusters from different survey data \citep{Moreetal2016, Changetal2018, Shinetal2019}, 
we measure the projected cross-correlation functions 
for higher redshift clusters even at $0.7<z_{\rm cl}<1.0$ 
with fainter apparent magnitude limits 
for the galaxies ($\sim 2~ {\rm mag}$ deeper) 
thanks to the depth of HSC images
to investigate the splashback features around the clusters,
which allows us to study splashback features in great detail even for such a small survey area.
In this paper, we employ clusters selected by the CAMIRA cluster-finding algorithm,
whereas \cite{Moreetal2016}, \cite{Changetal2018}, and \cite{Shinetal2019}
used optically-selected clusters detected by a different finding algorithm redMaPPer. 
We employ \textsc{Dark Emulator} \citep{Nishimichietal2019} 
for the halo-matter cross-correlation function
and the mass-richness relation \citep{Murataetal2019} 
from the stacked lensing and abundance measurements for the model predictions 
to compare with observations.
We marginalize over the off-centering effects on the projected cross-correlation measurements
with the constraints in \cite{Murataetal2019} from the stacked lensing measurements.

We summarize our results as follows.
\begin{itemize}
\item We detect the splashback features around the CAMIRA clusters from the galaxies with $M_{z}-5 \log_{10} h< -18.8$ 
      in different redshift and richness bins. 
      We infer the location of the splashback radius for each cluster sample 
      and find that these radii are consistent with the model predictions within $2\sigma$ levels for all the cluster samples.
\item We investigate how the splashback features change with different absolute magnitude limits 
      for low-redshift clusters at $0.1 < z_{\rm cl} < 0.4$
      using galaxy samples fainter than those used in the literature by $\sim 2~ {\rm mag}$ 
      to constrain possible dynamical friction effects.
      We find that the constraints on $r_{\rm sp}^{\rm 3D}$ do not change significantly within a $\sim 20\%$ level over different magnitude limits. 
      We also find that the logarithmic derivative of the three-dimensional profile at $r_{\rm sp}^{\rm 3D}$ becomes smaller (shaper profiles) 
      for brighter absolute magnitude limits. 
      We attribute this dependence to a magnitude dependence of ratios 
      between multi-streaming and infalling materials around $r_{\rm sp}^{\rm 3D}$.
\item We detect splashback features by separating galaxies into red and blue populations 
      with the red-sequence method in \cite{Nishizawaetal2018} based on the color-magnitude diagram.
      We find that the constraints on $r_{\rm sp}^{\rm 3D}$ from the red galaxy populations are more precise 
      than those without these red/blue separations due to shaper profiles for the red galaxies only.
      The $1\sigma$ precisions on $r_{\rm sp}^{\rm 3D}$ from these analyses are
      $11 \%$, $13 \%$, and $16 \%$ for clusters 
      at $0.1 < z_{\rm cl}< 0.4$, $0.4 < z_{\rm cl} < 0.7$, and $0.7 < z_{\rm cl} < 1.0$, respectively.
      These constraints for $0.4 < z_{\rm cl} < 0.7$ and $0.7 < z_{\rm cl} < 1.0$
      are more consistent with the model predictions ($\lesssim 1\sigma$ levels) 
      than their $20 \%$ smaller values ($\sim 2 \sigma$ levels).
      We note that our precision for $r_{\rm sp}^{\rm 3D}$ at $0.4 < z_{\rm cl} < 0.7$
      is comparable to that at $0.55 < z_{\rm cl}<0.7$ ($14\%$) in \cite{Changetal2018},
      whereas our constraint on $r_{\rm sp}^{\rm 3D}$ at $0.7 <z_{\rm cl} < 1.0$ 
      provides the first constraint on splashback features at such high redshift,
      which is made possible
      thanks to the deep HSC images.
      In Figure~\ref{fig:summaryplot},
      we show a comparison between these results and those in the literature 
      from projected cross-correlation measurements between clusters and galaxies.
\item Also, we show that the blue galaxy populations can be fitted better with the DK14 model 
      than a pure power-law model, 
      suggesting that some fraction of the blue galaxy populations defined by the red-sequence method  
      stay blue even after reaching their first apocenters of orbits within host clusters over $0.1<z_{\rm cl}<1.0$. 
      We also show the red and blue fraction in the three-dimensional profiles
      changes abruptly around the model predictions of $r_{\rm sp}^{\rm 3D}$ 
      over $0.1 < z_{\rm cl} < 1.0$. 
      The model predictions of $r_{\rm sp}^{\rm 3D}$ are consistent with radii 
      where the three-dimensional galaxy density of red and blue galaxy populations become the same values within their error bars
      over $0.1 < z_{\rm cl} < 1.0$.
\item We show that the uncertainties in the prior distributions for the off-centering effects 
      increase the errors on $r_{\rm sp}^{\rm 3D}$ significantly for clusters at higher redshifts. 
\item We create mock galaxy catalogs from the HOD model with
      matching cluster abundance densities and lensing profiles to observations approximately. 
      We then investigate how the cluster finders could affect measurements of $r_{\rm sp}^{\rm 3D}$ from the DK14 fitting
      by employing simplified cluster finding algorithms for CAMIRA and redMaPPer with the mock catalogs
      following methods presented in \cite{BuschandWhite2017} and \cite{Sunayama&More2019}.
      We closely follow the procedure for analyses with real data in the mock observations and mock model predictions.
      We find that 
      the biases for the CAMIRA finder are insignificant compared to our statistical errors.
      Also, for the redMaPPer-like cluster finder, we find that
      the observed splashback radius from the DK14 fitting with the mock catalogs
      is biased to smaller values at the level of $\sim 15\%$
      than the model predictions especially at lower richness,
      which is a consistent trend in
      the smaller splashback radius with the redMaPPer cluster-finding algorithm in the literature \citep{Moreetal2016, Changetal2018, Shinetal2019}.
      These biases for the redMaPPer-like finder are significantly reduced with the larger aperture size,
      which is comparable to that for CAMIRA at $\lambda=20$.
      Since the fiducial aperture size of redMaPPer is smaller than the splashback features in the projected space ($R_{\rm sp}^{\rm 2D}$),
      these results suggest that these smaller aperture sizes induce selection bias effects on the observed splashback radius locations.
\item In Appendix~\ref{app:compmodel}, 
      we confirm that our model predictions with accounting for the scatter in the mass-richness relation 
      in Appendix~\ref{app:modelpredictions}
      are quite consistent with those without these scatter effects used in the literature.
\end{itemize}

In the future, we can improve precisions on measurements of splashback features 
by employing the full HSC survey data covering more than $1000~{\rm deg}^2$
or other survey data such as the Large Synoptic Survey Telescope \citep[LSST;][]{Ivezicetal2008},
which reduce the error bars with the current datasets
especially for higher redshift clusters 
to compare with model predictions from weak lensing masses more precisely.
Also, we could improve constraints on the splashback feature 
for higher redshift clusters
by adding more precise prior distributions 
for the off-centering distribution 
from upcoming lensing measurements with higher signal-to-noise ratios
or upcoming X-ray images for the clusters.

\bigskip
\begin{ack} \label{sec:ack}
We thank the anonymous referee for helpful comments that improved the quality of this work.

RM acknowledges Dominik Zürcher for sharing his codes developed in collaboration with SM
for his published work before further developments by ourselves for this work.

RM acknowledges financial support from
the University of Tokyo-Princeton strategic partnership grant,
Advanced Leading Graduate Course for Photon Science (ALPS),
Research Fellowships of the Japan Society for the Promotion of Science
for Young Scientists (JSPS), and JSPS Overseas Challenge Program for Young Researchers.
KO is supported by JSPS Overseas Research Fellowships.
This work was supported by JSPS KAKENHI Grant Numbers JP15H05892, JP17J00658, JP17K14273, JP18K03693, and JP19H00677.
This work was supported by World Premier International Research Center Initiative (WPI Initiative), MEXT, Japan.
This work was also supported by Japan Science and Technology Agency CREST JPMHCR1414.

The Hyper Suprime-Cam (HSC) collaboration includes the astronomical communities of Japan and Taiwan, and Princeton University. The HSC instrumentation and software were developed by the National Astronomical Observatory of Japan (NAOJ), the Kavli Institute for the Physics and Mathematics of the Universe (Kavli IPMU), the University of Tokyo, the High Energy Accelerator Research Organization (KEK), the Academia Sinica Institute for Astronomy and Astrophysics in Taiwan (ASIAA), and Princeton University. Funding was contributed by the FIRST program from Japanese Cabinet Office, the Ministry of Education, Culture, Sports, Science and Technology (MEXT), the Japan Society for the Promotion of Science (JSPS), Japan Science and Technology Agency (JST), the Toray Science Foundation, NAOJ, Kavli IPMU, KEK, ASIAA, and Princeton University.

This paper makes use of software developed for the Large Synoptic Survey Telescope. We thank the LSST Project for making their code available as free software at http://dm.lsst.org.

The Pan-STARRS1 Surveys (PS1) have been made possible through contributions of the Institute for Astronomy, the University of Hawaii, the Pan-STARRS Project Office, the Max-Planck Society and its participating institutes, the Max Planck Institute for Astronomy, Heidelberg and the Max Planck Institute for Extraterrestrial Physics, Garching, The Johns Hopkins University, Durham University, the University of Edinburgh, Queen's University Belfast, the Harvard-Smithsonian Center for Astrophysics, the Las Cumbres Observatory Global Telescope Network Incorporated, the National Central University of Taiwan, the Space Telescope Science Institute, the National Aeronautics and Space Administration under Grant No. NNX08AR22G issued through the Planetary Science Division of the NASA Science Mission Directorate, the National Science Foundation under Grant No. AST-1238877, the University of Maryland, and Eotvos Lorand University (ELTE) and the Los Alamos National Laboratory.

Based on data collected at the Subaru Telescope and retrieved from the HSC data archive system, which is operated by Subaru Telescope and Astronomy Data Center, National Astronomical Observatory of Japan.
\end{ack}

\appendix
\section{Model predictions for splashback features from halo-matter cross-correlation and richness-mass relation} 
\label{app:modelpredictions}
We model splashback radius locations and derivatives at these locations for each cluster redshift and richness selection 
by using three-dimensional cross-correlation functions between halos and dark matter from a halo emulator 
in \cite{Nishimichietal2019} 
and the richness-mass relation $P(N|M,z)$ in \cite{Murataetal2019}
constrained from the stacked weak gravitational lensing and abundance measurements.

Cosmological $N$-body simulations are one of the methods commonly used in the literature to model predictions of halo statistics.
We use the database generated by the \textsc{Dark Quest} campaign \citep{Nishimichietal2019} to predict the halo mass function 
and the halo-matter cross-correlation,\footnote{We define the halo-matter cross-correlation function $\xi_{\rm hm}^{\rm 3D}(r; M, z)$ 
as $\rho(r; M, z)=\bar{\rho}_{\rm m0}[1+\xi_{\rm hm}^{\rm 3D}(r; M, z)]$, 
where $\rho(r; M, z)$ 
is the average matter density profile around halos with mass $M$ at redshift $z$ 
and $\bar{\rho}_{\rm m0}$ is the present-day mean matter density. 
We note that we use the present-day mean matter density since we use the comoving coordinate for the radius and density in this paper.}
both of which are important ingredients for our model predictions.
\cite{Nishimichietal2019} developed a scheme called \textsc{Dark Emulator} to predict statistical quantities of halos, 
including the mass function and the halo-matter cross-correlation function as a function of halo mass, redshift, separation length, and cosmological parameters, 
based on a series of high-resolution, cosmological $N$-body simulations. 
The simulation suite is composed of cosmological $N$-body simulations for 101 cosmological models within a flat $w$CDM framework.
The simulations trace the nonlinear evolution of $2048^3$ particles in a box size of $1$ or $2 h^{-1}{\rm Gpc}$ on a side with mass resolution
of $\sim 10^{10} h^{-1}M_{\odot}$ or $\sim 8\times10^{10} h^{-1}M_{\odot}$, respectively. 
There are $21$ redshift slices for each simulation realization across the range $0 \leq z \leq 1.47$.
\cite{Nishimichietal2019} employed \textsc{Rockstar} \citep{Behroozietal2013}
to identify dark matter halos.
The halo mass definition in the simulations includes all particles within a distance of $R_{\rm 200m}$ from the halo center.
The minimum halo mass of the emulator is $10^{12}h^{-1}M_{\odot}$. 
For the model prediction below, 
we set $M_{\rm min}=10^{12} h^{-1}M_{\odot}$ and $M_{\rm max}=2\times 10^{15}h^{-1}M_{\odot}$
for the minimum and maximum halo masses, respectively, to evaluate the halo mass integration
consistently with \cite{Murataetal2019}.
We note that we also use a halo catalog
from one of these simulations to investigate projection effects on
the observational constraints for the splashback features
in Section~\ref{subsec:projeffect}.
\textsc{Dark Emulator} relies on data compression based on
Principal Component (PC) Analysis followed by Gaussian Process Regression
for each PC coefficient from a subset of 80 cosmological models. 
\cite{Nishimichietal2019} used the other subsets as a validation set 
to assess the performance of the emulator. 
We note that \cite{Nishimichietal2019} did not use the realizations for the {\it Planck} cosmology in the regression.

In the model prediction of splashback features to compare 
with the observations 
of projected cross-correlation functions between clusters and photometric galaxies,
we employ the following assumptions that are also used 
in the literature \citep{Moreetal2016, Baxteretal2017, Changetal2018, Zurcher&More2019, Shinetal2019}.
First, we assume that the galaxies are a reasonably good tracer of dark matter
and thus their distributions around halos can well be modeled with dark matter distribution 
around halos to a first approximation.
In particular, \cite{Moreetal2016} and \cite{Changetal2018} 
showed that subhalos, which should host galaxies, are good tracers with almost same location of splashback radius 
in $r_{\rm sp}^{\rm 3D}$ 
and slightly smaller derivative values (i.e., steeper profiles) for more massive subhalos
at $r_{\rm sp}^{\rm 3D}$ compared to dark matter distributions around halos when selected by halo masses in simulations
with matching subhalo abundances to observed galaxy abundances.
Second, we also assume that the observed cross-correlation signals can be approximately 
modeled employing statistical isotropy in the halo emulator, which was
constructed from the halo-matter cross-correlation function after the spherical averages in the simulations.
We note that we check how projection effects affect the observed splashback radius assuming the statistical isotropy
with the mock catalogs in Section~\ref{subsec:projeffect}.
Third, while splashback features depend on accretion rate of halos even at fixed halo mass and redshift, 
we assume that the correlation between richness and accretion rates at fixed halo mass and redshift are negligibly small
for our cluster samples 
and thus
we use the mass-richness relation 
to model the splashback features 
with the halo-matter cross-correlation function as a function of mass and redshift 
after averaging over accretion rates.
Finally, we additionally assume that the halo mass and redshift dependence on the number of host galaxies for halos can well be approximated by amplitudes of halo-matter cross-correlation functions.
Since we use the photometric galaxies without spectroscopic redshifts, we cannot account for this dependence in the measurements and the fitting analyses.
We could account for this effect with a large number of spectroscopic galaxies around clusters with upcoming spectroscopic surveys.

Under these assumptions, 
we calculate the model prediction of the halo-matter cross-correlation function
in each cluster redshift and richness sample
after averaging over the redshift range with volume weight ${\rm d}^2 V/{\rm d}z {\rm d}\Omega = \chi^2(z)/H(z)$
as
\begin{eqnarray}
&&\xi_{ \rm hm }^{\rm 3D}(r; z_{\rm min} \leq z_{\rm cl} \leq z_{\rm max}, N_{\rm min}\leq N \leq N_{\rm max})\nonumber \\
&=&
\frac{ 1 }{ N_{\rm cl} }
\int_{ z_{\rm min} }^{ z_{\rm max} } {\rm d}z~ \frac{ \chi^2(z) }{ H(z) } \int_{ M_{\rm min} }^{ M_{\rm max} } {\rm d}M
\frac{ {\rm d} n }{ {\rm d} M } \nonumber \\
&&\times S(M, z| N_{\rm min},  N_{\rm max} ) \xi_{\rm hm}^{\rm 3D}(r; M, z), 
\label{eq:xihmstack}
\end{eqnarray}
where 
${\rm d}n/{\rm d}M$ is the halo mass function at redshift $z$,
$\xi_{\rm hm}^{\rm 3D}(r; M, z)$ 
is the halo-matter cross-correlation function for halo mass $M$ at redshift $z$,
and we calculate the cluster number count per unit steradian for the normalization factor as 
\begin{eqnarray} 
N_{\rm cl}&=&\int_{ z_{\rm min} }^{ z_{\rm max} }{\rm d}z~\frac{ \chi^2(z) }{ H(z) }\int_{ M_{\rm min} }^{ M_{\rm max} } {\rm d}M \frac{ {\rm d} n }{ {\rm d} M } 
S(M, z| N_{\rm min},  N_{\rm max} ).\nonumber\\
\label{eq:xihmstack_normalization}
\end{eqnarray}   
The selection function of halo mass at a fixed redshift in the richness range is calculated 
by integrating the richness-mass relation over the richness range as
\begin{equation}
S(M, z| N_{\rm min},  N_{\rm max} ) = \int_{ N_{\rm min} }^{ N_{\rm max} } {\rm d} N P(N|M, z),
\end{equation}
where $P(N|M,z)$ is the probability distribution of the richness for a given halo mass and redshift
constrained from the stacked lensing and abundance measurements in \cite{Murataetal2019}.
Here we use MCMC chains for a fiducial result of the richness-mass relation under the {\it Planck} cosmology in \cite{Murataetal2019}.
We note that this integration can be expressed analytically with the error function 
since \cite{Murataetal2019} assumed the log-normal distribution for $P(N|M,z)$ 
with model parameters in mean and scatter relations.
We calculate model predictions for the projected cross-correlation $\xi_{\rm 2D}(R)$ based on equation~(\ref{eq:3Dto2D})
with the results in equation~(\ref{eq:xihmstack})
and the same maximum integration length value.
In Table~\ref{tab:sampleselection},
we show the resulting model prediction values for the splashback radius locations
and the derivatives at these locations for each cluster redshift and richness selection.

\section{Comparisons of model predictions from different methods for splashback radius}\label{app:compmodel}
We compare our model predictions in equation~(\ref{eq:xihmstack}) for splashback features accounting 
for the scatter in mass-richness relation 
in \cite{Murataetal2019}
with those from different methods in the literature without accounting for the scatter 
to check whether there are significant differences or not.

In particular, \cite{Moreetal2016}, \cite{Changetal2018}, and \cite{Shinetal2019} model splashback radius locations 
for their cluster sample by employing halos in {\it N}-body simulations with a mass threshold cut $M_{\rm 200m}>M_{\rm min}$ 
at their redshift in the simulations
to match mean cluster mass from their observational mass-richness or other mass-observable relations.
On the other hand, \cite{Zurcher&More2019} models the splashback location for their cluster sample 
by employing mean halo mass and redshift value such as $\xi_{\rm hm, model}^{\rm 3D}=\xi_{\rm hm}^{\rm 3D}(R; M=\langle M \rangle, z=\langle z \rangle)$
without accounting for the scatter between halo mass and observable explicitly.

We compare these model predictions with our fiducial model prediction in  equation~(\ref{eq:xihmstack}) 
by using the halo emulator in \cite{Nishimichietal2019} and 
the best-fit mass-richness relation parameters in \cite{Murataetal2019} for the clusters at $0.1 \leq z_{\rm cl} \leq 0.4$ with $15 \leq N \leq 200$.
We show a model comparison of the logarithmic derivative of the three-dimensional cross-correlation function in Figure~\ref{fig:model_diff}.
We note that we use the halo mass function in \cite{Nishimichietal2019} with
the mean redshift value of the sample for the mass threshold cut to match the mean mass for the method in \cite{Moreetal2016}, 
and that here we do not account for the redshift weights in equations~(\ref{eq:xihmstack}) and (\ref{eq:xihmstack_normalization}) 
over the redshift range, but we fix the redshift to the mean value.
Figure~\ref{fig:model_diff} shows that the differences for $r_{\rm sp}^{\rm 3D}$ from these different methods are $\sim 1 \%$ level. 
We find that results are similar for different redshift and richness selections shown in Table~\ref{tab:sampleselection}.
We have confirmed that the mass dependence of $\xi_{\rm hm}(R; M_{\rm 200m}, z)$ is approximately 
$\propto M_{\rm 200m}$ at $M_{\rm 200m} \gtrsim 10^{14} h^{-1}M_{\odot}$ with a small redshift 
dependence around $R \sim 1{\rm -}2 h^{-1}{\rm Mpc}$
from the halo emulator,
and thus averaging accounting for the scatter results in $\sim \xi_{\rm hm}(R; M_{\rm 200m}=\langle M_{\rm 200m} \rangle, z=\langle z \rangle)$ to a first approximation 
around splashback radius locations
as shown in Figure~\ref{fig:model_diff}.
Our result indicates that previous approaches to ignore the scatter in the mass-richness relation are sufficiently accurate and therefore are justified, although in this paper we adopt a model that fully takes account of the scatter.

We note that \cite{Shinetal2019} checked the scatter effect by simulated halos with their selection accounting for scatters in their mass-observable relations 
and found results that are similar to our results shown in Figure~\ref{fig:model_diff},
albeit for more massive halos.

\begin{figure}[t]
  \begin{center}
    \includegraphics[width=0.5 \textwidth]{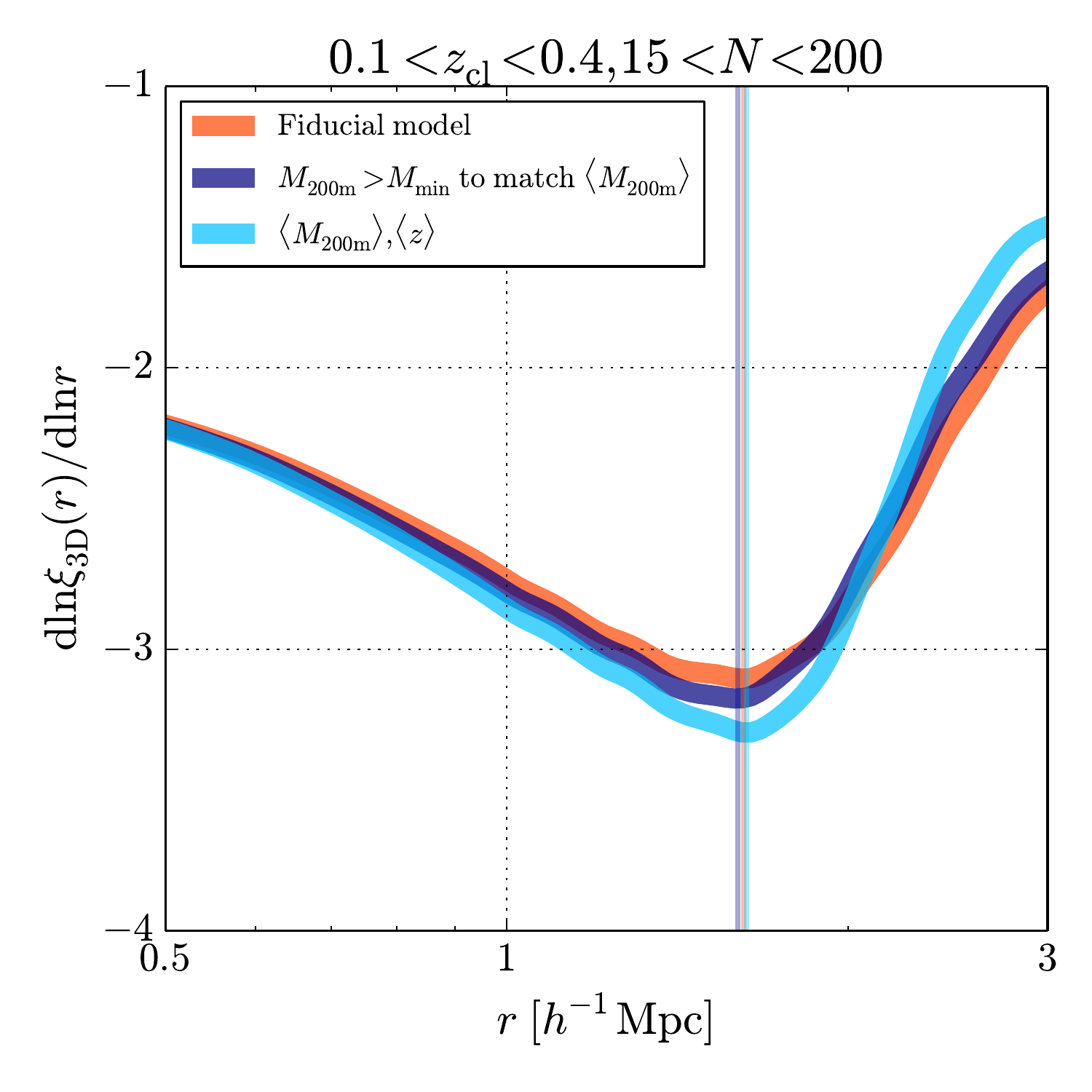}
  \end{center}
  \caption{ 
    We show a comparison of model predictions for the logarithmic derivative of the three-dimensional cross-correlation function 
    with the best-fit mass-richness relation parameters for the clusters at $0.1 \leq z_{\rm cl} \leq 0.4$ with $15 \leq N \leq 200$ in \cite{Murataetal2019}
    from the fiducial analysis under the {\it Planck} cosmology.
    ``Fiducial model'' shows our fiducial model in equation~(\ref{eq:xihmstack}) that fully accounts for the scatter in the mass-richness relation, 
    ``$M_{\rm 200m}>M_{\rm min}$ to match $\langle M_{\rm 200m} \rangle$'' for the method in \cite{Moreetal2016}, 
    and ``$\langle M_{\rm 200m} \rangle, \langle z \rangle$'' for the method in \cite{Zurcher&More2019} (see the text for more details).
    The vertical lines show the three-dimensional splashback radius location $r_{\rm sp}^{\rm 3D}$ for each method.
  }
  \label{fig:model_diff}
\end{figure}

\section{Model parameter constraint contours}\label{appendix:MCMCcontour}
We show the model parameter constraint contours in Figure~\ref{fig:MCMCcontourfid} 
from the fiducial analysis with the full sample of $0.1 \leq z_{\rm cl} \leq 1.0$ and $15 \leq N \leq 200$ in Table~\ref{tab:sampleselection}
to show the marginalized one-dimensional posterior distributions from each parameter and the 68$\%$ and 95$\%$ credible levels contours for each two-parameter subspace from the MCMC chains.

\begin{figure*}
  \begin{center}
    \includegraphics[width=17.0cm]{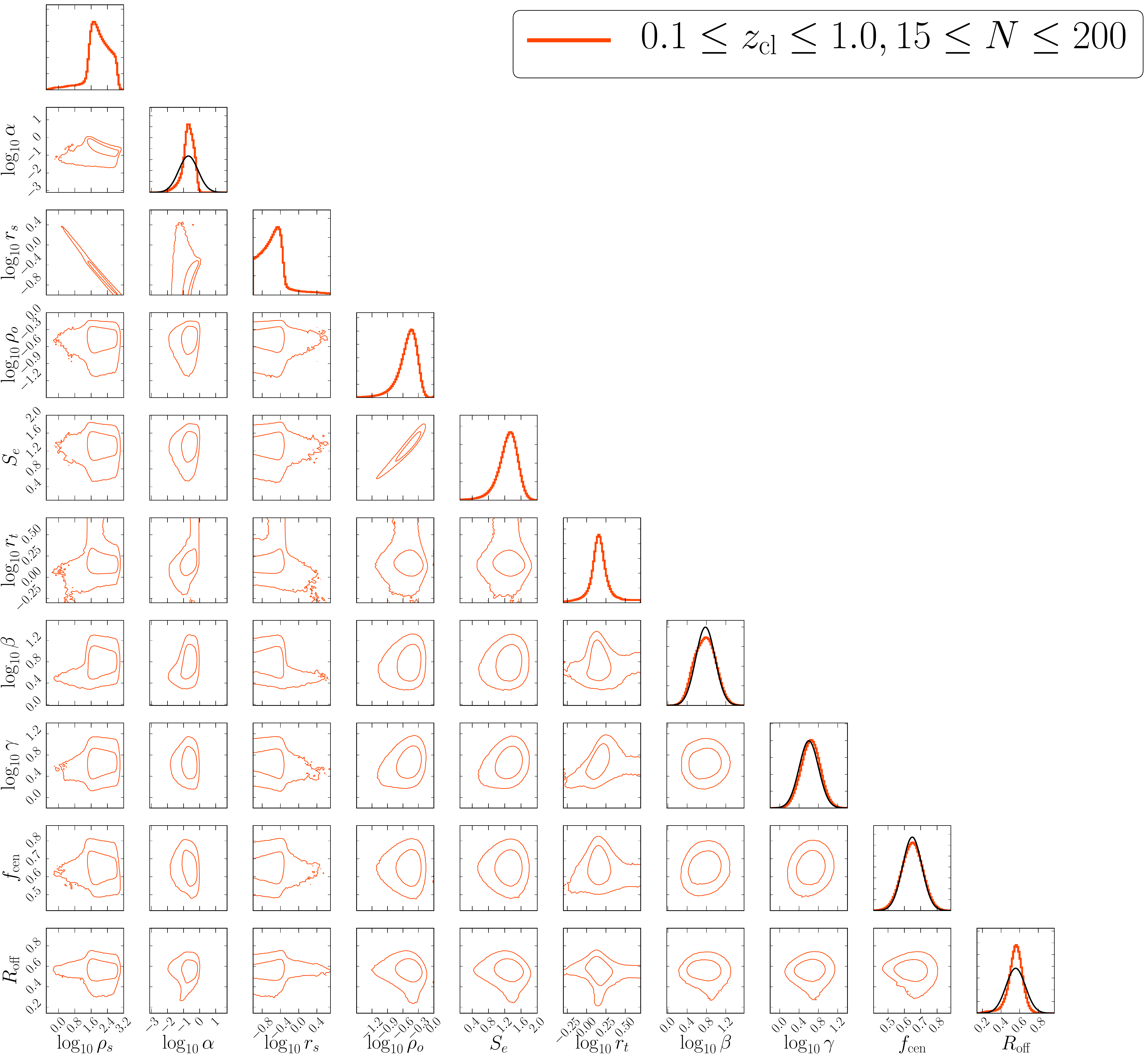}
  \end{center}
  \caption{ 
            Model parameter constraints in the fiducial analysis for the full sample 
            with $0.1 \leq z_{\rm cl} \leq 1.0$ and $15 \leq N \leq 200$ in Table~\ref{tab:mainresults}.
            Diagonal panels show the posterior distributions of the model parameters, and the other panels show the 68$\%$ and 95$\%$ credible levels contours in each two-parameter subspace from the MCMC chains.
            Black lines show the Gaussian prior distributions in Table~\ref{tab:paramspriors}.
          }
\label{fig:MCMCcontourfid}
\end{figure*}

%
\bigskip
\noindent
%

\label{lastpage}
\end{document}